\documentclass[ reprint,superscriptaddress,amsmath,amssymb,aps,showpacs, longbibliography]{revtex4-2}

\usepackage[utf8]{inputenc}
\usepackage{graphicx}
\usepackage{dcolumn}
\usepackage{bm}
\usepackage{xcolor}
\usepackage{tabularx}
\usepackage{amssymb}
\usepackage{pgfplots}
\pgfplotsset{compat=newest}
\usepackage{rotating}
\usepackage{hyperref}
\usepackage{todonotes}
\usepackage{float}
\usepackage{dsfont}
\usepackage{amsmath}
\usepackage{array}

\begin{document}

\title{Interpenetration of fractal clusters drives elasticity \\ in colloidal gels formed upon flow cessation}

\author {No{\'e}mie Dag{\`e}s}
\affiliation{Univ Lyon, Ens de Lyon, Univ Claude Bernard, CNRS, Laboratoire de Physique, F69342 Lyon, France\looseness=-1}%
\author {Louis V. Bouthier}
\affiliation{Groupe CFL, CEMEF, Mines Paristech, 1 Rue Claude Daunesse, 06904 Sophia Antipolis, France}%
\author {Lauren Matthews}
\affiliation{ESRF - The European Synchrotron,  38043 Grenoble Cedex, France}
\author {S{\'e}bastien Manneville}
\affiliation{Univ Lyon, Ens de Lyon, Univ Claude Bernard, CNRS, Laboratoire de Physique, F69342 Lyon, France\looseness=-1}
\author {Thibaut Divoux}
\affiliation{Univ Lyon, Ens de Lyon, Univ Claude Bernard, CNRS, Laboratoire de Physique, F69342 Lyon, France\looseness=-1}%
\author {Arnaud Poulesquen}
\affiliation{CEA, DES, ISEC, DE2D, SEAD, LCBC, Université of Montpellier, Marcoule, France}%
\author {Thomas Gibaud}
\email{Corresponding author, thomas.gibaud@ens-lyon.fr}
\affiliation{Univ Lyon, Ens de Lyon, Univ Claude Bernard, CNRS, Laboratoire de Physique, F69342 Lyon, France\looseness=-1}

\date{\today}

\begin{abstract}
Colloidal gels are out of equilibrium soft solids composed of attractive Brownian particles that form a space-spanning network at low volume fractions. The elastic properties of these systems result from the network microstructure, which is very sensitive to shear history. Here, we take advantage of such sensitivity to tune the viscoelastic properties of a colloidal gel made of carbon black nanoparticles. Starting from a fluidized state under an applied shear rate $\dot \gamma_0$, we use an abrupt flow cessation to trigger a liquid-to-solid transition. We observe that the resulting gel is all the more elastic when the shear rate $\dot \gamma_0$ is low and that the viscoelastic spectra can be mapped on a master curve. Moreover, coupling rheometry to small angle X-ray scattering allows us to show that the gel microstructure is different from gels solely formed by thermal agitation where only two length scales are observed: the dimension of the colloidal and the dimension the fractal aggregates. Competition between shear and thermal energy leads to gels with three characteristic length scales. Such gels structure in a percolated network of fractal clusters that interpenetrate each other. Experiments on gels prepared with various shear histories reveal that cluster interpenetration increases with decreasing values of the shear rate $\dot \gamma_0$ applied before flow cessation. These observations strongly suggest that cluster interpenetration drives the gel elasticity, which we confirm using a structural model. Our results, which are in stark contrast with previous literature, where gel elasticity was either linked to cluster connectivity or to bending modes, highlight a novel local parameter controlling the macroscopic viscoelastic properties of colloidal gels. 
\end{abstract}

\pacs{xxx}
                             
\maketitle

 \tableofcontents
 \newpage

\section{Introduction}
Colloidal gels are out of equilibrium amorphous soft solids composed of attractive Brownian particles that aggregate to form a space-spanning network at low concentrations~\cite{trappe2001}. These viscoelastic materials are ubiquitous both in nature and in industrial applications as diverse as flow batteries, food products and cementitious materials \cite{Lu:2013,Ioannidou:2016,Parant:2017,Cao:2020}.
The scenario underlying their formation, namely the sol-gel transition, governs the vast majority of their structural and mechanical properties. In practice, Brownian motion is the driving force that allows colloids to encounter each other, whereas the colloid concentration $\phi$ and the interaction potential $U$ set up the aggregation path and the final gel properties~\cite{trappe2004, Zaccarelli:2007}. In the limit of low concentrations and high interparticle attraction strength, diffusion-limited or reaction-limited cluster aggregation  take place and lead to the formation of fractal gels \cite{weitz1984,schaefer1984} characterised by two length scales the particle radius $r_0$ and the cluster size $\xi_c$ of fractal dimension $d_f$. At intermediate volume fractions and for moderate attraction strength, the sol-gel transition corresponds to an arrested phase separation \cite{cardinaux2007,lu2008,Zia:2014}. 
In the former situation, the gel mechanical properties are captured by fractal scaling \cite{shih1990,Krall:1998}, whereas in the latter case, the gel properties are set by the connectivity of glassy clusters \cite{zaccone2009,whitaker2019} or by the stretching and bending rigidity of the glassy network~\cite{gibaud2013}. 

\begin{figure}[t!]
\centering
\includegraphics[width=8.5 cm]{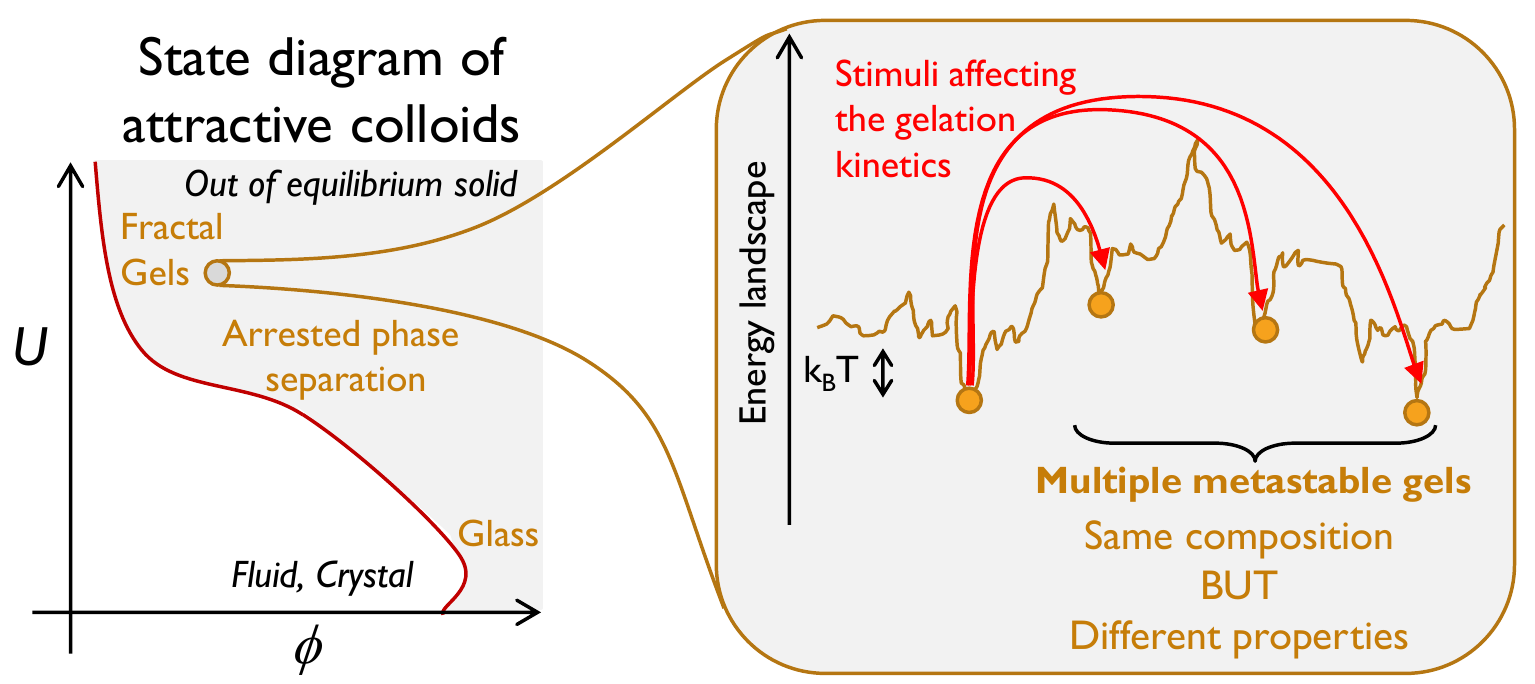}
\caption{Schematic principle of the concept of multiple metastable gels in attractive colloidal dispersions.
}
\label{fig:mmg}
\end{figure}

Because colloidal gels are out-of-equilibrium, additional parameters play a key role in controlling their properties. External fields that act on the colloids dynamics and compete with Brownian motion may disrupt the direct correspondence between the gel properties and its coordinates ($\phi$, $U$) in the state diagram, leading to gels with a broad variety of structural and mechanical properties from a single colloidal dispersion. 
Indeed, the energy landscape of suspensions of attractive colloids is complex and strewn with multiple local minima (Fig.~\ref{fig:mmg}). Some of those minima are only accessible with an additional input of energy, larger than the thermal energy $k_BT$. This is precisely the role of external stimuli: leading gelation into local minima inaccessible via Brownian motion. Therefore, the properties of the multiple metastable colloidal gels are a function of the external stimuli intensity. At the fundamental level, an external stimulus is  ideal for exploring the interplay between microstructure and mechanics: as the gel originates from the same dispersion, the colloid volume fraction and interactions remain unchanged, while variations in the gel mechanical properties only result from microstructural changes. 
In practice, different external control parameters such as temperature and external shear have been used as external stimulus. For instance, the microstructure of globular protein gels can be tuned by varying the quench rate during the sol-gel transition: the faster the temperature quench, the smaller the characteristic length of the gel network~\cite{gibaud2009}. In colloidal gels build from depletion interaction, flow cessation from a shear rate of intensity $\dot\gamma_0$ to 0~s$^{-1}$ leads to gels, whose structure and rheology are governed by the intensity of the preshear $\dot\gamma_0$: a high preshear $\dot\gamma_0$ leads to homogeneous and strong gels whereas a low preshear $\dot\gamma_0$ leads to heterogeneous and weak gels~\cite{koumakis2015}. In dispersions of fractal-like particles with attractive interparticle interactions, stress-controlled flow cessation at various cessation rates yields gels whose strength and connectivity increase for increasing cessation rate~\cite{helal2016}.  

Despite multiple evidence that the mechanical properties of colloidal gels can be tuned by an external stimulus, a quantitative correspondence between the gel microstructure and its mechanical properties is still lacking. 
We have chosen to take up this challenge in the case of mechanical shear applied to particulate colloidal gels with non-covalent interactions for three reasons. 
First, such colloidal gels can be rejuvenated by shear, i.e., their microstructure can be reconfigured by an external shear of large enough magnitude for a long enough duration, before the gel reforms upon flow cessation \cite{Utz:2000,Bonn:2004}. Shear rejuvenation allows us to conveniently explore different gelation scenarii on the same sample. 
Second, it is already well established that shear may interfere with the gelation pathway of particulate colloidal gels, thus giving the opportunity to tune gels in term of microstructure~\cite{koumakis2015, moghimi2017}, connectivity~\cite{helal2016} or yield stress~\cite{ovarlez2013}. 
Third, on the application level, such an interplay between shear and gelation is involved in numerous industrial processes, and especially in additive manufacturing  where external fields such as an additional shear coupled with 3D printing allows tuning the microstructure and the properties of the printed materials
\cite{Raney:2018}. 

In practice, we choose to work with gels of carbon black nanoparticles whose properties can be tuned using shear history \cite{ovarlez2013,helal2016, dages2021}. Here we influence the gelation pathway of these gels as follows: starting from a fluidized state under an applied shear rate $\dot\gamma_0$, an abrupt flow cessation triggers a liquid-to-solid transition. Varying the shear rate intensity $\dot\gamma_0$ allows us to generate gels whose viscoelastic properties spans over a decade in stress units. Specifically, lower shear intensities yield  more elastic gels upon flow cessation, while the viscoelastic spectrum for different $\dot \gamma_0$ shows a robust frequency dependence that can be rescaled onto a master curve.  
Using rheometry coupled to small angle X-ray scattering (SAXS), we further show that the gel microstructure is composed of clusters of size $\xi_c$ and fractal dimension $d_f$ separated by a cluster center to center distance $\xi_s$. Those structural parameters depends on $\dot \gamma_0$. More importantly, we show that $\xi_s<\xi_c$ meaning that adjacent clusters interpenetrate each others. The degree of interpenetration defined by the ratio $\xi_c/\xi_s$  decreases for increasing values of
$\dot\gamma_0$. The degree of interpenetration is crucial, for it controls the gel elasticity and captures the impact of $\dot \gamma_0$ on the gel viscoelastic properties, as confirmed by a fractal scaling model. 

The outline of the paper is as follows. We first introduce carbon black gels as well as our experimental toolbox in Section~\ref{sec:MaMe}. Second, in Section~\ref{sec:Re}, we present our experimental results. We show how shear history allows tuning the gel viscoelastic properties, which can be rescaled onto a master curve. We then establish that the gel structures in fractal clusters that interpenetrate each other. Third, in Section~\ref{sec:Disc}, after dismissing a superposition principle to account for the scaling of the viscoelastic properties of the gel, we derive a fractal scaling model establishing a direct link between the gel microstructure and the gel network elasticity as a function of the shear rate intensity $\dot\gamma_0$ applied before flow cessation.

\section{Materials and methods} \label{sec:MaMe}
\subsection{Carbon black dispersions}
Carbon black (CB) particles  are fractal carbonated colloids that result from the partial combustion of hydrocarbon oils \cite{lahaye1994,xi2006,sztucki2007}. These particles are widely used in the industry for mechanical reinforcement or to enhance the electrical conductivity of plastic and rubber materials \cite{wang2018}. Among the large variety of carbon black particles \cite{herd1992,martinez2017,dages2021, richards2017, hipp2021}, we choose to work with Vulcan PF particles (Cabot, density $d_{cb}=2.26\pm 0.03$). The density of Vulcan PF particles is $d_{cb}=2.26\pm 0.03$ and we estimate their radius of gyration to $r_g= 35$~nm with a 20\% polydispersity, and their fractal dimension to $d_{f0}=2.9$ (see Appendix~\ref{Apd:caracCB} for details). 

When dispersed in mineral oil (RTM17 Mineral Oil Rotational Viscometer Standard, Paragon Scientific, viscosity $\eta=354$~mPa.s at $T=20^{\circ}$C, density $d_{bck}=0.871$), CB particles are weakly attractive. The depth $U$ of the interparticle  potential depends on the type of CB particles, the solvent, and the presence of dispersant, and falls typically in the range $U\sim 10-30 k_BT$  \cite{trappe2007,prasad2003}. At a working weight concentration in CB particles of $c_w=4$~\%, the particles aggregate to form a gel, i.e., a space-spanning network, which behaves as a viscoelastic soft solid. 
Indeed, at rest, the elastic modulus $G'$ dominates the viscous modulus $G''$ in the limit of low frequencies, whereas the sample displays a solid-to-liquid transition beyond a critical strain $\gamma_y\sim 10$~\%. Moreover, under steady shear, the flow curve that links the shear stress $\sigma$ to the shear rate $\dot \gamma$ is well fitted by the Herschel-Bulkley model, $\sigma=\sigma_y+K\dot \gamma^n$ \cite{herschel1926}, with a dynamical yield stress $\sigma_y=4.5$~Pa, a consistency index $K=1.0$~Pa.s$^{1/0.83}$, and a fluidity index $n=0.83$ (see  Fig.~\ref{fig:flowcurve} in appendix~\ref{Apd:HB}).

\begin{figure}[t!]
\centering
\includegraphics[width=7 cm]{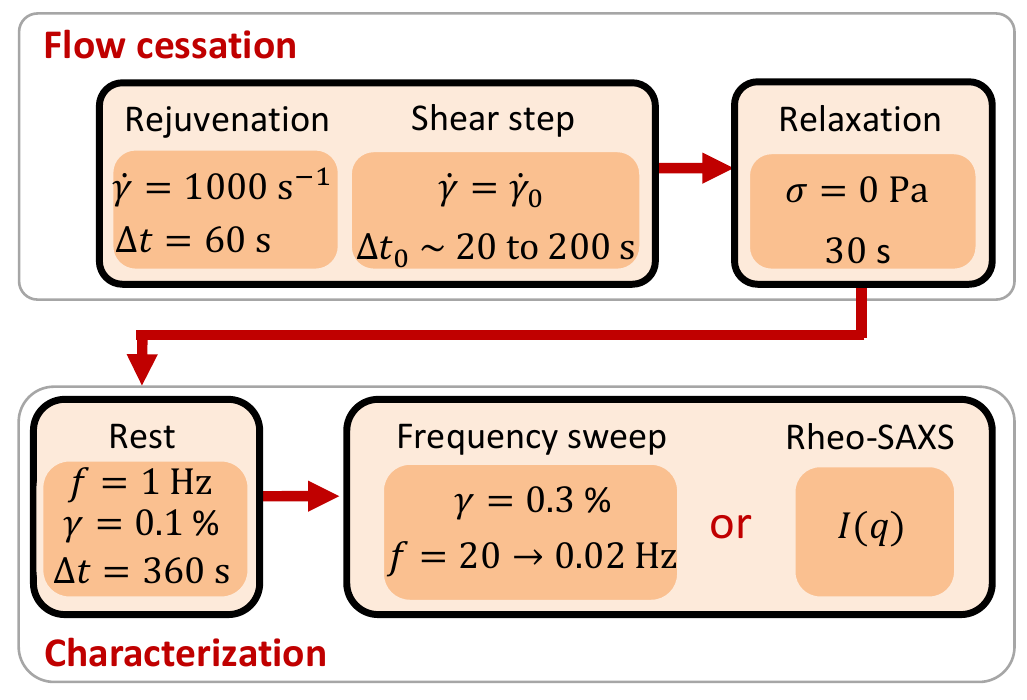}
\caption{Experimental protocol: schematic of the flow cessation and characterization sequences.
}
\label{fig:protocol}
\end{figure}

\subsection{Rheology}
In the present work, we use a rheometer both to measure the mechanical properties of CB gels and to shape up their microstructure. We carry out our experiments with two stress-controlled rheometers: ($i$) a MCR301 (Anton Paar) equipped with a rough cone (radius 40~mm, angle $1^{\circ}$) and a smooth bottom plate both made of steel, and ($ii$) a Haake RS6000 (Thermo Scientific) equipped with a Couette geometry composed of concentric polycarbonate cylinders (inner diameter 20 mm, outer diameter 22 mm, and height 40 mm) for rheo-SAXS experiments. Both apparatuses give identical results provided that the shear rate does not exceed 500~s$^{-1}$ in the Couette geometry due to the Taylor-Couette instability \cite{fardin2014}. 

\subsection{Small angle X-ray scattering}
The microstructural properties of the carbon black dispersion are investigated using rheo-SAXS measurements carried out on the ID02 beamline at the European Synchrotron Radiation Facility (ESRF, Grenoble, France) \cite{Narayanan2022}. The incident X-ray beam of wavelength 0.1~nm is collimated to a vertical size of 50~$\mu$m and a horizontal size of 100~$\mu$m. The 2D scattering patterns were measured using an Eiger2 4M pixel array detector and the subsequent data reduction procedure is described elsewhere \cite{Panine2003}. The scattering intensity $I(q)$ is obtained by subtracting the two-dimensional scattering patterns of the carbon black gel and the mineral oil. The resulting scattering intensity presented in this article always remained isotropic (see Fig.~\ref{fig:isotropy} in Appendix~\ref{Apd:RheoSaxs}). Therefore, we radially averaged the normalized intensity pattern to obtain one dimensional $I(q)$. 
Note that measurements were performed in both radial and tangential configurations, and they turn out to be equivalent due to the isotropy of the gel microstructure.

\subsection{Rheological protocol} 
We apply the protocol sketched in Fig.~\ref{fig:protocol}, which is divided in two sequences: a flow cessation sequence to shape up the gel properties followed by a sequence of characterization of the gel mechanical properties inherited from the flow cessation protocol.
\begin{figure}[t!]
\centering
\includegraphics[width=8 cm]{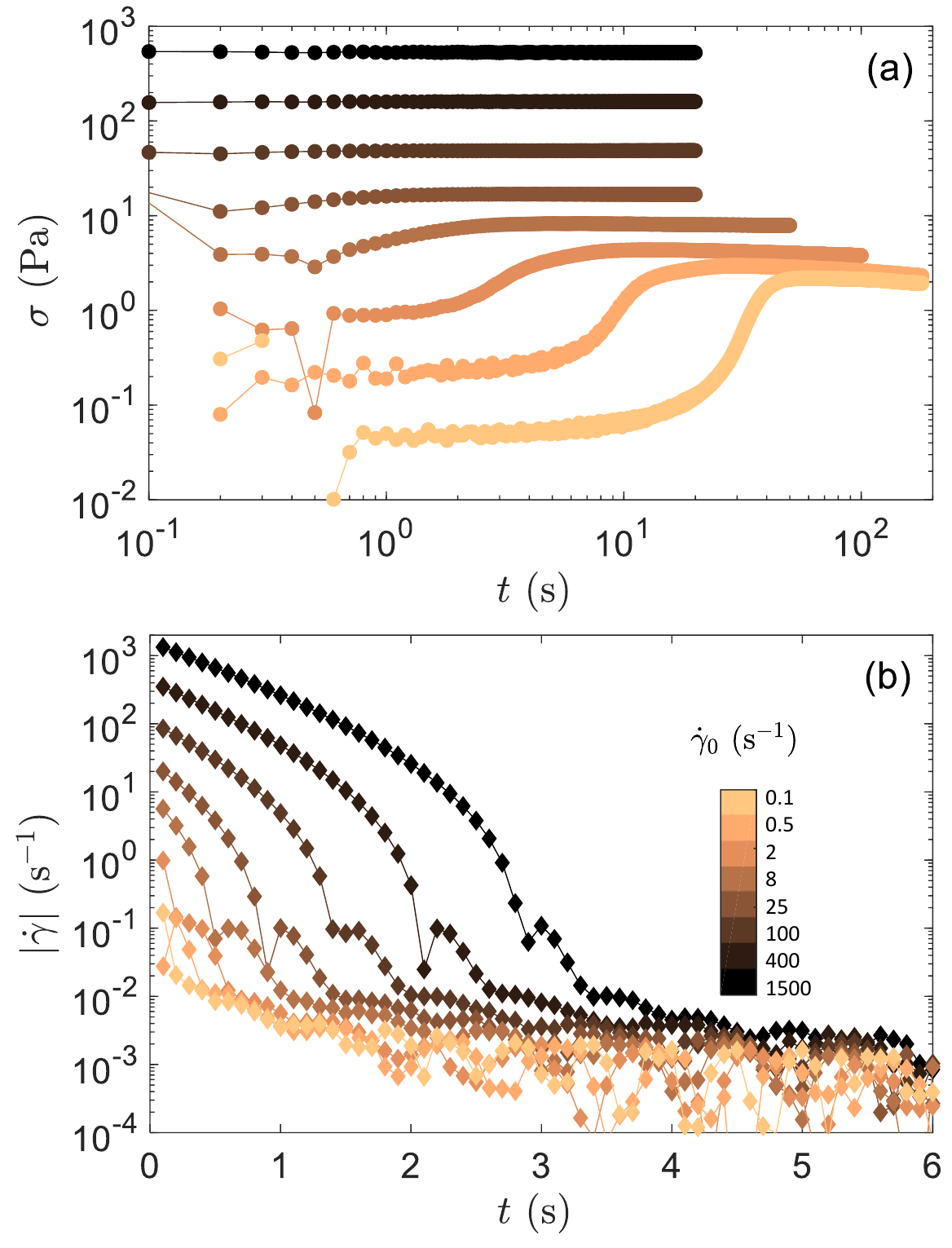}
\caption{Flow cessation sequence performed with different values of the shear rate $\dot\gamma_0$ in a $c_w=4$~\% carbon black dispersions. The sequence is composed of two steps: a quench in shear rate from 1000~s$^{-1}$ to $\dot\gamma_0$ yielding a stress response $\sigma(t)$ pictured in (a), followed by a complete flow cessation which results in (b) the relaxation of $\vert\dot{\gamma} \vert(t)$ when imposing $\sigma=0$~Pa. In (a) and (b) colors encode the value of $\dot\gamma_0$ ranging from 1500 (black) to 0.1~s$^{-1}$ (yellow), see legend in (b). }
\label{fig:cessation}
\end{figure}

\begin{figure*}[!ht]
\centering
\includegraphics[width=16.6 cm]{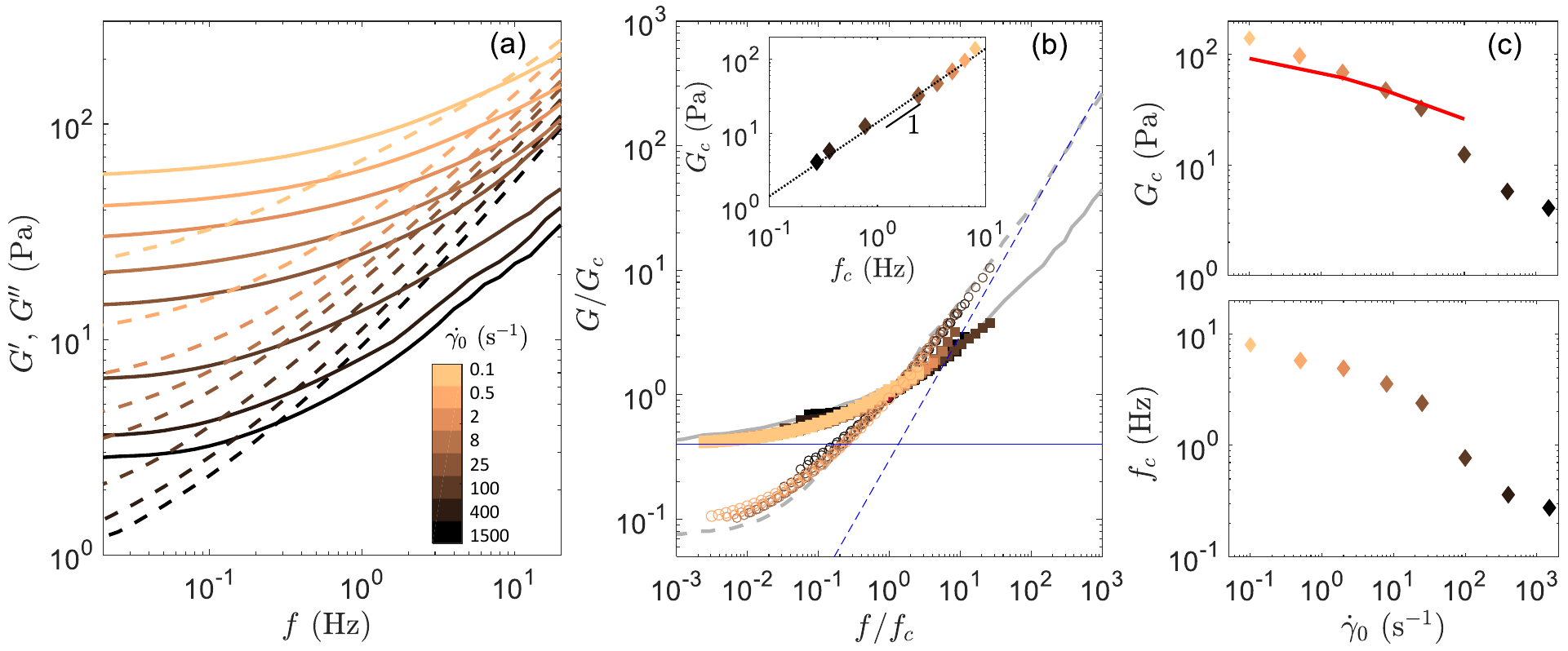}
\caption{(a) Elastic $G'$ and viscous modulus $G''$ vs.~the frequency $f$ in a $c_w=4$~\% carbon black dispersions. The color codes for the shear rate intensity $\dot\gamma_0$ applied before flow cessation. (b)~Normalized viscoelastic spectrum: $G'/G_c$ and $G''/G_c$ vs.~the normalized frequency $f/f_c$, where $G_c$ and $f_c$ denote respectively the modulus and the frequency at which $G'=G''$ in (a). The blue curves correspond to a Kelvin-Voigt model (see Appendix~\ref{Appendix:model}). The grey curves correspond to the master curves obtained for $G'$ (solid line) and $G''$ (dashed line) by rescaling viscoelastic spectra of CB dispersions in oil for various volume fractions [data extracted from Ref.~\cite{trappe2000}]. Inset: $G_c$ vs.~$f_c$. The red dotted line is the best linear fit of the data: $G_c=14.5 f_c$. (c)~Evolution of  $G_c$ (top) and $f_c$ (bottom) vs.~$\dot\gamma_0$. The red line is the best fit of the data using Eq.~\eqref{eq:Model} and the structural information reported in Fig.~\ref{fig:structure}. The best fit is obtained with a single adjustable parameter, namely the prefactor $G_{CB}= 9$~Pa. 
} 
\label{fig:scaling}
\end{figure*}

In practice, the flow cessation protocol is divided into three steps. First, we carry out a rejuvenation step during which the sample is sheared at $\dot{\gamma}=1000$~s$^{-1}$ for $\Delta t=60$~s to erase any shear history that would influence the gel mechanical properties later on. 
Second, we modify the shear intensity by imposing a quench from $\dot{\gamma}=1000$~s$^{-1}$ to a constant shear rate $\dot{\gamma}_0$ $\in [0.1 , 1500]$~s$^{-1}$ for a duration  $\Delta t_0 \in [20 , 200]$~s. Figure~\ref{fig:cessation}(a) shows the stress response $\sigma(t)$ of the CB gel resulting from quenches to various values of $\dot{\gamma}_0$. For high shear rates $\dot{\gamma}_0$, a duration of $\Delta t_0=20$~s is sufficient to reach a steady state. However, for $\dot{\gamma}_0<10$~s$^{-1}$, we must impose $\dot{\gamma}_0$ for longer durations, as $\sigma$ increases significantly, before reaching a maximum and then slowly decreases. The increase of $\sigma$ at short time scales corresponds to a transient regime necessary for the system to adapt to the new shear rate $\dot{\gamma}_0$~\cite{dullaert2005} (see also Appendix~\ref{Apd:HB}). The slow decrease at longer time scales might be due to some slippage of the dispersion at the walls of the shear cell~\cite{meeker2004}.
Third, we apply a flow cessation  by imposing $\sigma=0$~Pa for 30~s, while recording the shear-rate response $\dot{\gamma}(t)$ as displayed in Fig.~\ref{fig:cessation}(b). We observe that $\dot{\gamma}$ decreases to values beneath 10$^{-3}$~s$^{-1}$ within a few seconds indicating that the rotor is immobile and that flow cessation is complete. 
At short time scales, the shear rate decreases exponentially as expected for a simple, viscous fluid. At intermediate timescales, $\dot{\gamma}$ drops faster than exponentially and displays oscillations typical of the viscoelastic ringing observed in soft solids during creep tests~\cite{Zolzer:1993,Baravian:1998,Ewoldt:2007,Benmouffok:2010}. This indicates that gels reforms within a few seconds.

Finally, the characterization sequence following flow cessation consists in three steps. First, we let the system rest for 360~s, while measuring the elastic $G'$ and viscous $G''$ modulus using oscillations of small amplitude $\gamma=0.1$~\% at a frequency $f=1$~Hz. As shown in the Appendix~\ref{Apd:Rest} Fig.~\ref{fig:GpGpptime}, the viscoelastic moduli of the dispersion rapidly reach a regime where aging is weak. Second we perform a frequency sweep at $\gamma=0.3$~\% with 10 points per decade for frequencies $f$ ranging from 0.02 to 20~Hz. To gain some insights on the gel microstructure during these two sequences, the entire protocol was carried out in the rheo-SAXS setup for four distinct shear intensity $\dot\gamma_0$. The scattered intensity $I(q)$ of the gel obtained after flow cessation is discussed in the next section.

Based on reference~\cite{radhakrishnan2017}, we estimate that through out the rheological protocol the gel is homogeneously sheared and does not display shear banding. Indeed shear banding may appear in carbon black suspensions. When going from large to low value of the shear rate, shear banding only happens below a critical shear rate of $\dot{\gamma}_{SB}$. $\dot{\gamma}_{SB}$ is easily identifiable on the flow curve by a drop of the shear stress $\sigma$ at low shear rates. From Fig.~\ref{fig:flowcurve} we determine $\dot{\gamma}_{SB}\sim 0.1$~s$^{-1}$. This value justify carrying out preshear of intensity $\dot{\gamma}_0$ no lower than 0.1~s$^{-1}$ to garantie the homogeneity of the flow profile in the rheometer.

\begin{figure*}[!t] 
\centering
\includegraphics[width=16.6 cm]{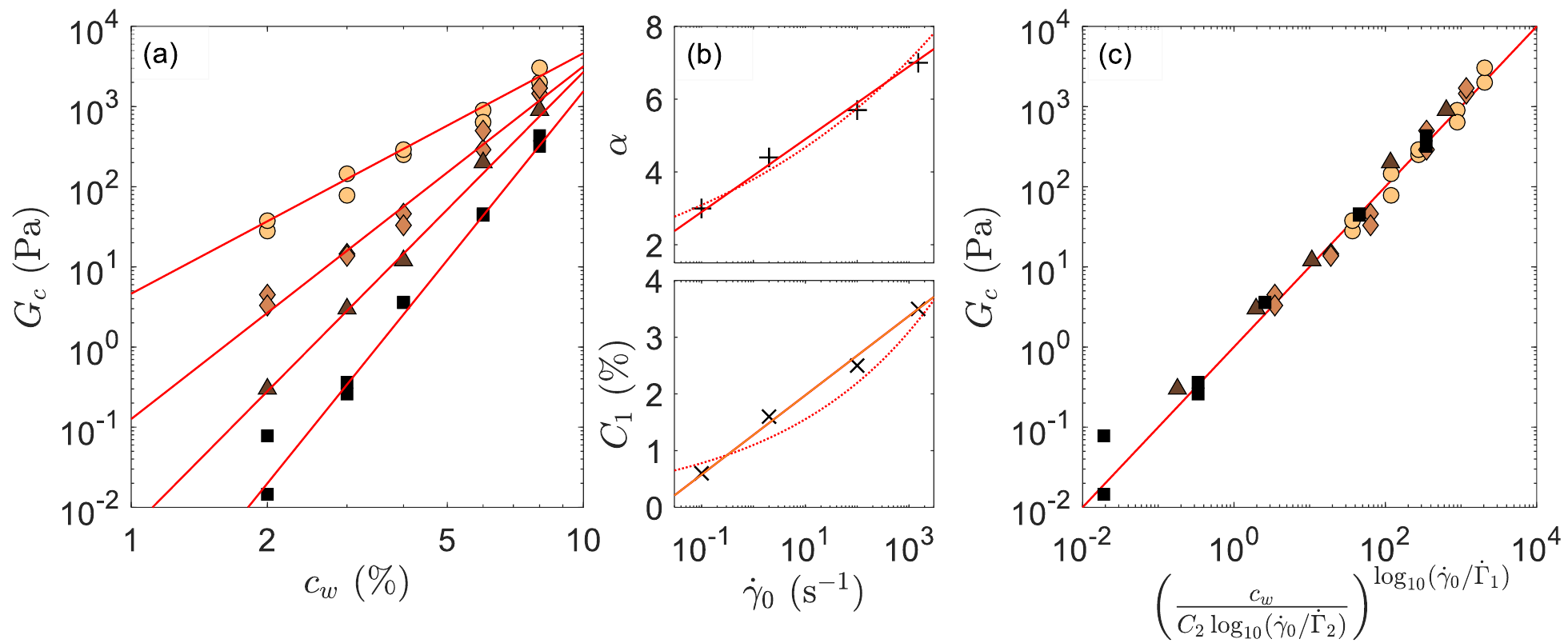}
\caption{(a) Evolution of $G_c$ as a function of the weight concentration in carbon black particles $c_w$ for various shear rates applied prior to flow cessation: $\dot{\gamma}_0=0.1$~s$^{-1}$ (circle), 2~s$^{-1}$ (diamond), 100~s$^{-1}$ (triangle) and 1500~s$^{-1}$ (square). Red lines are the best power-law fit of the data $G_c = (c_w/C_1)^{\alpha}$. Both $\alpha$ and $C_1$ depend on $\dot{\gamma}_0$. (b) $\alpha$ and $C_1$ vs. the shear rate $\dot{\gamma}_0$. Red lines are the best logarithmic fit of the data: $\alpha=\log_{10}(\dot{\gamma}_0/\dot{\Gamma}_1)$ with $\dot{\Gamma}_1$=10$^{-3.9}$~s$^{-1}$ and $C_1=C_2\log_{10}(\dot{\gamma}_0/\dot{\Gamma}_2)$ with $C_2=0.7$~\% and $\dot{\Gamma}_2=0.015$~s$^{-1}$. Dotted lines are the best power law fits. (c) Evolution of $G_c$ as function of the dimensionless concentration and shear rate. 
}
\label{fig:sensitivity}
\end{figure*}

\section{Results} \label{sec:Re}

\subsection{Impact of the shear rate $\dot{\gamma}_0$ on the gel linear viscoelastic spectrum} \label{Mastercurve}

We first focus on the impact of $\dot \gamma_0$ on the linear viscoelastic properties of the gel formed upon flow cessation (see ``Frequency sweep" in the characterization sequence sketched in Fig.~\ref{fig:protocol}).
The gel frequency spectrum  is reported in Fig.~\ref{fig:scaling}(a) for various values of the  shear rate intensity $\dot{\gamma}_0$ spanning over four decades. Overall, we observe that low $\dot \gamma_0$ produce more elastic gels. More precisely, whatever the shear rate intensity $\dot{\gamma}_0$, the elastic and viscous moduli are increasing functions of the frequency and cross at a frequency $f_c$ that shifts towards larger values for decreasing $\dot \gamma_0$. Moreover, in the limit of low frequencies, all spectra show a plateau in elasticity with $G'>G''$, which confirms the solid-like behavior of the sample, regardless of the shear rate intensity applied prior to flow cessation. The shape of the viscoelastic spectrum is robust, and appears merely shifted, which prompts us to construct a master curve from the data in Fig.~\ref{fig:scaling}(a). By normalizing each spectra by the coordinate  ($f_c$, $G_c$) defined by the crossover of $G'$ and $G''$, we obtain the master curve reported in Fig.~\ref{fig:scaling}(b). This scaling behavior is also clearly visible in the tan($\delta$)$=G''/G'$ representation. Since $G'$ and $G''$ are scaled by the same factor, building a master curve with tan($\delta$) from different $\dot{\gamma}_0$ only requires to scale the frequency axis as shown in Fig.~\ref{fig:tand}.
The asymptotic behavior of the master curve corresponds to a Kelvin-Voigt model (see Appendix.\ref{Appendix:model}) displayed as blue lines in Fig.~\ref{fig:scaling}(b). At low frequencies, i.e., $f\ll f_c$, the elastic modulus $G'$ tends towards a plateau value much larger than $G''$, which is the hallmark of a solid-like behavior at rest.
At high frequencies, i.e., $f\gg f_c$, the viscous modulus $G''$ dominates and increases linearly with the frequency and the solvent viscosity $\eta$, such that $G''=2\pi \eta f$. In this range of frequencies, the variations of $G''$ correspond to the viscous dissipation due to the thermal fluctuations of the gel network in the background solvent. We note that the master curve can be fully fitted by a fractional Kelvin-Voigt model (see Appendix.\ref{Appendix:model}).
Such a master curve is strongly reminiscent of that obtained on fractal gels by varying the particle volume fraction, and the interparticle potential  \cite{trappe2000,prasad2003, won2005}. In contrast, here, the master curve is generated by varying the shear history on a sample of {\it fixed} composition. Yet, rescaled data extracted from Ref.~\cite{trappe2000} and obtained with different CB particles suspended in another solvent [see gray curves in Fig.~\ref{fig:scaling}(b)], fall remarkably well on our master curve. This suggests that various shear histories allows generating gels, whose microstructure shares some similarity with that generated by varying the colloid volume fraction.


The high sensitivity of CB gels to shear history is encoded in the dependence of the locus of $G'$ and $G''$ crossing point ($f_c$, $G_c$) with $\dot \gamma_0$. As shown in Fig.~\ref{fig:scaling}(c), both $f_c$ and $G_c$ decrease by almost two orders of magnitude when increasing $\dot{\gamma}_0$ from 0.1~s$^{-1}$ to 1500~s$^{-1}$.
Such influence of shear history is not obvious, for it shows a trend similar to that observed in boehmite  gels~\cite{Sudreau:2022} and silica sphere and rods gels~\cite{das2022} but opposite to that reported in depletion gels, where a strong shear yields a more homogeneous and more elastic structure upon flow cessation \cite{koumakis2015}. 

\subsection{Influence of the carbon black weight concentration}



The rescaling and the master curve introduced in Section~\ref{Mastercurve} are robust to changes in the CB weight concentration, from $c_w=2$ to 8~\%. For instance, the viscoelastic spectrum of a $c_w=2$\% carbon black dispersion obtained with various shear rate $\dot \gamma_0$ applied before flow cessation can be rescaled on the same master curve as that displayed in Fig.~\ref{fig:scaling}(b) (see Fig.~\ref{fig:concentration} in appendix~\ref{Apd:Concentration}). Moreover, for a fixed shear intensity $\dot \gamma_0$, the modulus $G_c$ increases as a power law of $c_w$, $c_w=(c_w/C_1)^{\alpha}$ with a concentration $C_1$ and an exponent $\alpha$ that depends on $\dot \gamma_0$ [Fig.~\ref{fig:sensitivity}(b)]. 
While a power-law increase of the gel elasticity for increasing particle weight concentration or volume fraction is classically reported in colloidal gels with an exponent $\alpha$ ranging between 2 and 4.5 depending on the range of the interparticle potential and the nature of the particles \cite{buscall1988,trappe2001,prasad2003}, the sensitivity of $\alpha$ to shear history is a key result of the present study. Here, in Fig.~\ref{fig:sensitivity}(b), we show that $\alpha$ increases for increasing shear rate intensity applied before flow cessation, varying between $\alpha\simeq 3$ for $\dot\gamma_0=0.1$~s$^{-1}$ to surprisingly high values, i.e. $\alpha\simeq 7$ for $\dot\gamma_0=1500$~s$^{-1}$. Finally, in Fig.~\ref{fig:sensitivity}(c), we show that $G_c$ follows a master curve driven by a dimensionless concentration and shear rate intensity. Although this dependence remains empirical, this master curve highlights the fact that there are many ways to obtain gels with identical $G_c$. For instance to get $G_c~\simeq 3$~Pa one can either prepare a gel at ($c_w=4$~\%, $\dot{\gamma}_0=1500$~s$^{-1}$), ($c_w=3$~\%, $\dot{\gamma}_0=100$~s$^{-1}$) or ($c_w=2$~\%, $\dot{\gamma}_0=2$~s$^{-1}$).



To connect these results to the gel microstructure, one can be tempted to 
combine the power-law exponent $\alpha$ with the scaling theories developed for fractal gels \cite{shih1990}, in order to estimate the cluster fractal dimension $d_f$. The theory developed in the context of Brownian aggregation  distinguishes between two types of network, depending on the relative value of the elastic constant of the inter-cluster links to that of the cluster. In the case of weak links $\alpha= 1/(3-d_f)$, which yields $2.6<d_f<2.9$, whereas in the case of strong links, $\alpha= (3+x)/(3-d_f)$ with $x<d_f$ the fractal dimension of the gel backbone, and $1.5<d_f<2.5$ (see Fig.~\ref{fig:df} in appendix~\ref{Apd:Concentration}). These values motivate an experimental characterization of the gel microstructure, and especially of the cluster fractal dimension to test the relevance of such scaling theories.   

\begin{figure*}[ht]
\centering
\includegraphics[width=17 cm]{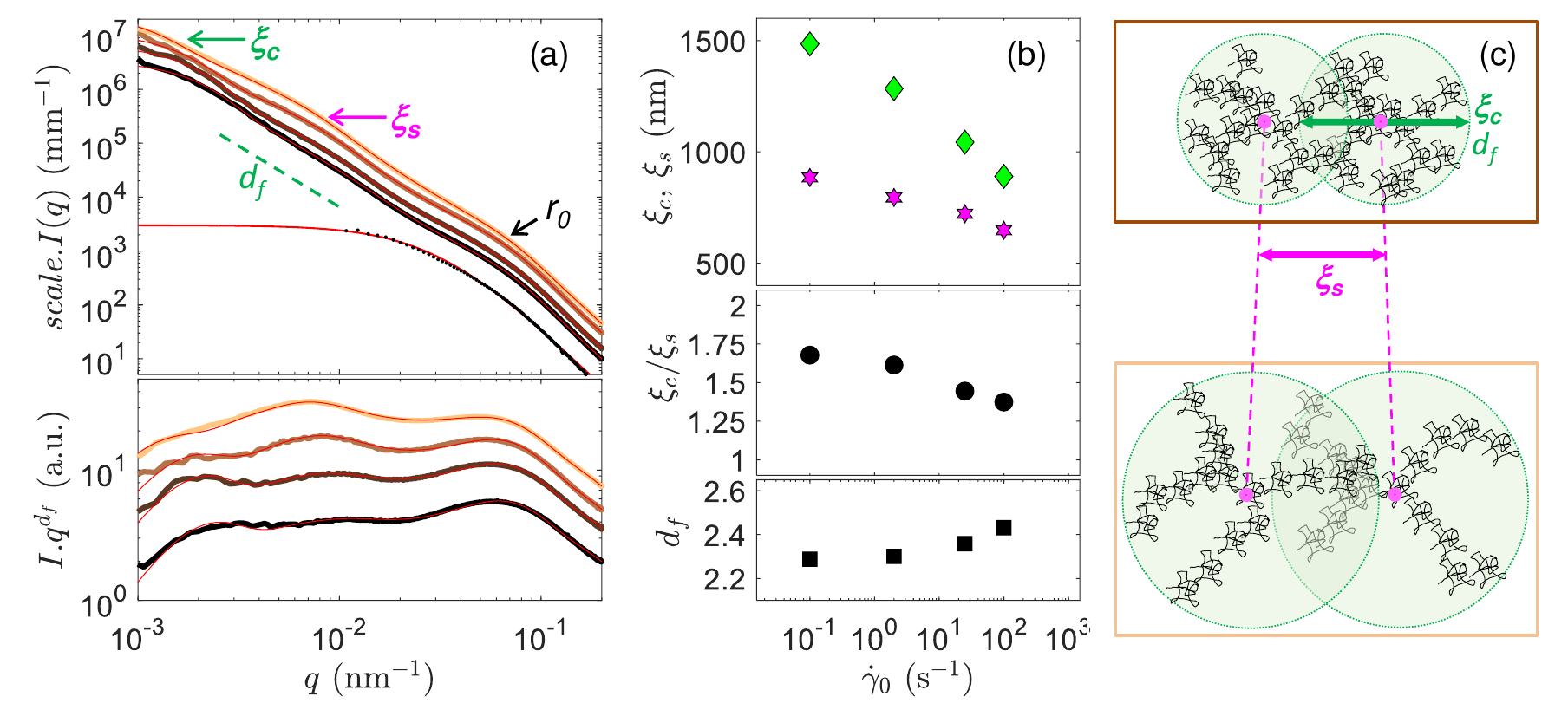}
\caption{Structure of the gel obtained after flow cessation. The scattering intensity $I(q)$ is measured 360~s after the end of the flow cessation test for different preshear $\dot\gamma_0$ in carbon black dispersions at $c_w=4$~\%. (a) The thick lines represent the evolution of $I(q)$ (top) and $I.q^{df}$ (bottom) as a function of the wave vector $q$. Color codes from pink to black for $\dot\gamma_0$=0.1, 2, 20 and 100~s$^{-1}$.  The dotted line is the form factor obtained at $c_w=0.1\%$. Red lines are fit to the experimental data by a  two-level Beaucage model composed of the CB particles of radius $r_0$ and clusters of size $\xi_c$ with fractal dimension $d_f$, modified by an intercluster structure factor that accounts for the center-to-center distance $\xi_s$ between the clusters. The scattering intensity $I(q)$ and $I.q^{d_f}$ are translated along the $y$-axis for better readability. (b) Evolution of the fit parameters $\xi_c$ (green diamond), $\xi_s$ (pink star), $\xi_c/\xi_s$ (circle) and $d_f$ (square) as a function of $\dot\gamma_0$. For the four fits, we obtained radius of gyration $r_0=28.5$~nm. (c) Schematic of the clusters evolution as the preshear $\dot\gamma_0$ decreases between the top and bottom panels.
} 
\label{fig:structure}
\end{figure*}

\subsection{Microstructure of the gel as function of the shear $\dot\gamma_0$ applied before flow cessation}


To better understand the interplay between shear history and the gel microstructural properties, we perform rheo-SAXS experiments using the protocol sketched in Fig.~\ref{fig:protocol}. 
The scattering intensity $I(q)$  measured 360~s after flow cessation protocol for four different shear rate intensities $\dot\gamma_0$ prior to flow cessation is reported in Fig.~\ref{fig:structure}(a), as a function of the wave number $q$. 
In all four cases, the scattered intensity $I(q)$ presents similar features. First, $I(q)$ is isotropic, i.e., tangential and radial measurements are equivalent (see Fig.~\ref{fig:isotropy} in Appendix~\ref{Apd:RheoSaxs}). Therefore, the gel displays an isotropic structure at all length scaled probed by SAXS, which is why we only report the azimuthally averaged $I(q)$. Second, $I(q)$ presents three characteristic bumps around 0.04, 0.01 and 0.002~nm$^{-1}$ characteristic of three length scales. This structure is atypical. Indeed, in gels driven solely by thermal agitation, $I(q)$ classically displays only two characteristic length scales: the particle size $r_0$ and the cluster size $\xi_c$ separated by a power-law regime, and the exponent of which is related to the cluster fractal dimension \cite{courtens1987}.


Here, we attribute the high-$q$ bump to the CB particles of size $r_0$, the low-$q$ bump to clusters of size $\xi_c$ and fractal dimension $d_f$, and the bump at intermediate $q$ to the  structural distance $\xi_s$ between the centers of two adjacent clusters. Those bumps appears more clearly in the Kratky-like representation $I.q^{d_f}$ reported in  Fig.~\ref{fig:structure}(a). In the Kratky-like representation we used the $d_f$ displayed in Fig.~\ref{fig:structure}(b). 

A description of the gel microstructure based on three characteristic length scales is implemented in a modified two-level Beaucage model. In short, the two-level Beaucage model~\cite{beaucage1995,beaucage1996} accounts for the scattering of clusters of size $\xi_c$ and fractal dimension $d_f$ composed of particles of size $r_0$. To account for the increase of scattering at $\xi_s$, we have multiplied the cluster scattering intensity term in the Beaucage model by an ad-hoc inter-cluster structure factor (see appendix~\ref{Apd:RheoSaxs} for more details). 
This modified Beaucage model provides an excellent fit to the experimental data obtained for different shear histories  [Fig.~\ref{fig:structure}(a)]. Moreover, since the gel weight concentration is identical in the four measurements, the fit parameters $r_0$, $\xi_c$, $d_f$ and $\xi_s$ should obey mass conservation. 
In practice, such a constraint can be expressed at the scale of the unit cell of the gel network, i.e., the minimum structural repeating unit necessary to construct the gel structure defined by the correlation length $\xi_s$. 
The number of particles $N=(\xi_c/r_0)^{d_f}$ in a unit cell corresponds to the number of particles in the cluster of size $\xi_c$ and fractal dimension $d_f$. Based on the values of the fit parameters obtained from adjusting the modified two-level Beaucage model to the SAXS data, we check that, indeed, $\rho=(\xi_c/r_0)^{d_f}/\xi_s^3$ remains constant across the four measurements with $\rho=7250 \pm 230$~particles/$\mu$m$^3$. This value is also in agreement with the carbon black weight concentration $c_w=4$~\% (see Fig.~\ref{fig:rho} in appendix~\ref{Apd:RheoSaxs}), which confirms that 
our analysis of the scattering data is self-consistent. 

Figure~\ref{fig:structure}(b) shows the dependence of the fit parameters $\xi_c$, $\xi_s$, and $d_f$ with the shear rate intensity $\dot{\gamma}_0$ applied before flow cessation. A gel prepared with a lower shear intensity shows a larger and looser microstructure since $\xi_c$ increases and $d_f$ decreases for decreasing $\dot \gamma_0$. Considering the evolution of only those two parameters suggests a decrease of the gel elasticity for decreasing values of $\dot{\gamma}_0$, in stark contrast with our observations.
However, $\xi_c$ and $d_f$ are not the only parameters, and the correlation length $\xi_s$, which corresponds to the cluster to cluster center distance, plays an important role. In particular, $\xi_s$ is smaller than the cluster size $\xi_c$, indicating that the clusters interpenetrate each other. Such cluster interpenetration has recently been suggested in carbon black gels to interpret step down shear rate rheology experiments~\cite{wang2022}. In our case, for lower shear rate intensity $\dot{\gamma}_0$ prior to flow cessation, the ratio $\xi_c/\xi_s$ increases, i.e., the clusters become more interpenetrated, accounting for the reinforcement of the gel elasticity. We therefore hypothesize that the gel elasticity is related to the cluster interpenetration, increasing the elasticity of the gel network as compared to the case where clusters would be packed in a random close-packing configuration [Fig.~\ref{fig:structure}(c)]. 

\subsection{A structure based model to account for the gel elasticity}
There are numerous models accounting for the elastic properties of fractal gels derived from microscopic considerations, namely the $\phi$-power law models, especially in the context of diffusion-limited cluster aggregation (DLCA) and reaction-limited cluster aggregation (RLCA)~\cite{wessel1992,shih1990,Kantor1984a,Kantor1984b,Mewis2012,Marangoni2000,Marangoni2021,Roldughin2003,Mellema2002,Wu2001}. However, these models do not take into consideration the case where shear history interferes with the gelation pathway activated by thermal energy. Therefore, unsurprisingly, such models cannot capture our observations (see Fig.~\ref{fig:df} in Appendix~\ref{Apd:RheoSaxs}). 
These models notably predict that 
the cluster size $\xi_c$ is set by its fractal dimension $d_f$, the particle volume fraction $\phi$ and the particle size $r_0$ in stark contrast with our observations where the shear applied prior to flow cessation appears as an additional key parameter that act on the gel structure.
Moreover, in these models, the elastic properties of colloidal gels are either connected to the local bending cost of the particle network, or to the cluster connectivity. However, to the best of our knowledge, none of these approaches accounts for an overlap, or equivalently for an interpenetration, of two neighboring clusters.   


Here, we introduce the interpenetration $\phi$-power law model. This model is an implementation of the $\phi$-power law models proposed in \cite{shih1990,Mellema2002,Wu2001}. In practice, we assume that the gel is composed of particles of size $r_0$ that form clusters of size $\xi_c$ and fractal dimension $d_f$ separated by a center-to-center distance $\xi_s$. If $\xi_s>\xi_c$, the clusters are independent and the dispersion is a fluid. However, if $\xi_s<\xi_c$, clusters interpenetrate each other and form a gel. We have mostly replaced the factor from the $\phi$-power law models that accounts for the elasticity of two adjacent clusters by an elongation elasticity due to the interpenetration. We assumed a decomposition of the microscopic stiffnesses in three contributions as springs in series, namely the intra-cluster  the intermicroscopic and the interpenetration respectively. We additionally assumed the that the interpenetration stiffness is negligible at the microscopic scale, thus dominates the macroscopic rheological behaviour.
The gel elastic modulus $G'_{\infty}$ at low frequencies is calculated in Appendix~\ref{Apd:model} and yields the following expression:
\begin{equation}
    G'_\infty=\underset{G_{\mathrm{CB}}}{\underbrace{\frac{U}{r_{0}\delta^{2}}}}\underset{g_{\mathrm{Interp}}}{\underbrace{\frac{1}{2}\left(\frac{\xi_c}{r_0}\right)^{d_f}\left(1+\frac{\xi_s}{2\xi_c}\right)^{\frac{d_f}{3}}\left(1-\frac{\xi_s}{\xi_c}\right)^{\frac{2d_f}{3}}}}\underset{g_{\mathrm{Net}}}{\underbrace{\phi\left(\frac{\xi_{s}}{r_{0}}\right)^{2-d_{f}}}}
    \label{eq:Model}
\end{equation}
This expression displays an elasticity that follow the hierarchical structural properties of the gel.
$G_{\mathrm{CB}}$ is the elasticity arising from colloid-colloid interactions where $U$ and $\delta$ are respectively the depth and the range of the carbon black attraction. $g_{\mathrm{Interp}}$ corresponds to the scaling that accounts for the cluster-cluster interpenetration and $g_{\mathrm{Net}}$ is the scaling attributed to the network formed by the clusters at the macroscopic scale.


To test the relevance of interpenetration $\phi$-power law model, we report in  Fig.~\ref{fig:scaling}(c) the best fit of $G_c$ as a function of the $\dot\gamma_0$ using Eq.~\eqref{eq:Model} with the values of the structural parameters inferred from Fig.~\ref{fig:structure}(b), the fact that $G'_{\infty}=0.3 G_c$ (see Appendix~\ref{Apd:model}), and the sole adjustable parameter $G_{\mathrm{CB}}= 9$Pa. The model correctly captures the decrease of the elasticity of the gel network as $\dot\gamma_0$ increases. However, taking $U=10~k_BT$ and $\delta=0.2 r_0$ we obtain $G_{CB}\sim 2000$~Pa a value much larger that the fit value: the model fails to capture the absolute value of the gel elasticity. 

The interpenetration $\phi$-power law model thus shows that cluster interpenetration accounts for the scaling of the mechanical properties of the gels and allows to rationalize the counter-intuitive observation that lower shear rate intensities before flow cessation yield stronger gels. Such results raise open questions, which are listed below.

\section{Discussion} \label{sec:Disc}

We have used mechanical shear to explore various configurations of carbon black gels. Starting from a fluidized state under an applied shear rate $\dot\gamma_0$, we use an abrupt flow cessation to trigger a liquid-to-solid transition. Varying $\dot\gamma_0$ allows us to tune the gel viscoelastic properties, whose 
spectrum can be mapped on a single master curve asymptotically defined at low frequencies by the elasticity of the gel network $G_{\infty}$ and at high frequencies by the viscosity $\eta$ of the background solvent. Coupling rheometry and SAXS, we have shown that the gel microstructure is composed of fractal clusters that interpenetrate each other, and the degree of interpenetration appears to be a key parameter contributing to the gel elasticity. We have validated this hypothesis developing an interpenetration $\phi$-power law model that account for the decrease of elasticity as $\dot\gamma_0$ increases.

\subsection{Is the scaling behavior of  the viscoelastic spectrum a consequence of an underlying superposition principle?}
In light of the scaling behavior of the viscoelastic spectrum, it is tempting to interpret the master curve obtained by varying $\dot \gamma_0$ as the result of some shear-frequency superposition principle. Superposition principles in soft matter mechanics rely on the idea that dynamical processes in soft materials can be accessed equivalently using time or frequency and another well-chosen variable. For example, time-temperature superposition in polymer melts~\cite{van1998} relies on the acceleration of all activated processes at high temperatures, enabling probing of longer effective time scales at high temperatures. In other words, the average relaxation time of the material changes with temperature without affecting the shape of its viscoelastic spectrum. 

This is not what we observe here, for the rescaling of the viscoelastic spectrum requires a shift along the frequency axis and a shift along the viscoelastic moduli. Such behavior has, however, been observed in different systems and still attributed to a superposition principle such as in colloidal low-methoxyl pectin~\cite{huang2021} in the context of gelling time/relaxation time superposition, protein condensates~\cite{jawerth2020} in the context of aging Maxwell fluids, in triblock copolymer solutions~\cite{krishnan2010} in the context of time-composition superposition, and in soft colloidal glasses~\cite{wen2015,wen2014} in the context of time-concentration superposition. In the case of carbon black gels subject to various shear rate intensities before flow cessation, the viscoelastic spectrum scaling is attributed to deep structural changes, such as the cluster fractal dimension $d_f$ (the gel does not have self-similar structures) rather than changes in the dynamics. Such results rule out a superposition principle. In other words, it is possible to form carbon black gels with the same value of the elastic plateau $G_{\infty}$ using different gels structures through shear history and concentration as shown in Fig.~\ref{fig:sensitivity}(c).


\subsection{Physical origin of the gel structure}
The multiple metastable gels formed following various shear preparations belong to the category of fractal gels. This is probably why our results do not match the trends observed in Ref.~\cite{koumakis2015} which belong to gels formed through arrested phase separation.  We find clusters of fractal dimension $d_f\in[2.3 , 2.5]$ larger than the prediction from diffusion-limited cluster aggregation DLCA where $d_f\sim1.8$~\cite{weitz1984} or reaction-limited cluster aggregation RLCA where $d_f\sim2.1$~\cite{schaefer1984} but smaller than the value obtained for sheared fractal aggregates where $d_f\sim2.6$~\cite{wessel1992}.

In our system, the Brownian time $\tau_B=R^3 6\pi\eta / k_BT \in [0.07 ; 1600]$~s is set by the diffusion of particles of size $R$ ranging from the CB dimension $r_0=35$~nm to the cluster size $\xi_c \sim 1$~$\mu$m. $\tau_B$ compares to the time necessary for flow cessation to take place $\tau_{fc}<4~$s (Fig.~\ref{fig:cessation}) supporting again the fact that  aggregation and flow cessation are coupled. 

The interplay between flow and structure in gels has been tackled mainly in the flow regime~\cite{hipp2021,nabizadeh2021} but has not been formalized to model multiple metastable gel states induced by a shear protocol. This is an important challenge in the pursue of memory materials~\cite{keim2019} which aim to encode, access, and erase signatures of past history in the state of a system.

The flow cessation protocol inducing the sol-gel transition could be addressed qualitatively through the use the Mason number $\mathrm{Mn}$~\cite{varga2018,Jamali2020}, which is here in the range $\left[3.5\times 10^{-4},5.2\right]$, and agrees rather well with the values in \cite{Jamali2019,Jamali2020}, or the adhesion number $\mathrm{Ad}$ \cite{Eggersdorfer2010,Marshall2014,Kimbonguila2014}, which is here in the range $\left[3.6,5.5\times10^4\right]$ weighting the relative importance of adhesion forces compared to shear forces and quantitatively using coagulation-fragmentation equation~\cite{Banasiak2020a,Stadnichuk2015,Sorensen1987} which embrace the competition between different aggregation mechanism and fragmentation through shear or collisions. 

Finally, we suggest another way to apprehend the multiple metastable gels formed through flow cessation. Indeed, carbon black gels as many other systems display delayed yielding~\cite{gibaud2010,sprakel2011,grenard2014}, i.e., when pushed at a constant shear stress $\sigma$, the gel initially at rest will start flowing on time scales that decrease exponentially with increasing value of $\sigma$. Our results prompt us to revisit delayed yielding phenomena and answer the following questions: How does gel prepared through flow cessation impact the delayed yielding mechanism? Is delayed yielding still characterized by an Arrhenius law? If so, is the energy barrier necessary to flow in the delayed yielding experiment related to the energy barrier to form the gel state induced by flow cessation?

\section{Conclusion}

In conclusion, we have quantified the impact of shear history on the viscoelastic properties of carbon black gels. We observe that, for a fixed content in nanoparticles, low shear intensities yield strong gels upon flow cessation, whereas larger shear intensities yield weaker gels. Such a variation in the gel strength was linked to the degree of interpenetration of the clusters that compose the gel microstructure. In that framework, we have introduced a mechanical model, th interpenetration $\phi$-power law model, that captures the impact of shear history on the gel elasticity, yielding a prediction for the scaling exponent that links the gel elasticity to the gel structural properties. These results highlight the power of shear history as an experimental tool to tune the viscoelastic properties of colloidal gels without changing the content in nanoparticles or their interactions. 

The viscoelastic spectrum~\cite{prasad2003} and the non-linear gel properties such as delayed yielding~\cite{sprakel2011} observed in the carbon black gels are also observed in many other colloidal gels, we therefore believe that cluster interpenetration could also be quite a general concept applicable in attractive colloidal systems. 
In addition, our work raises several fundamental questions, such as predicting for any type of colloids, the respective contributions of cluster interpenetration, cluster connectivity, and bending to a colloidal gel elasticity. Finally, future work could focus on determining the role of cluster interpenetration into the gel non-linear mechanical response, which might be a versatile parameter to tune the failure scenario of soft viscoelastic gels.




\section{appendix}
\subsection{Carbon black particles} \label{Apd:caracCB} 
\begin{figure}[ht]
\centering
\includegraphics[width=8 cm]{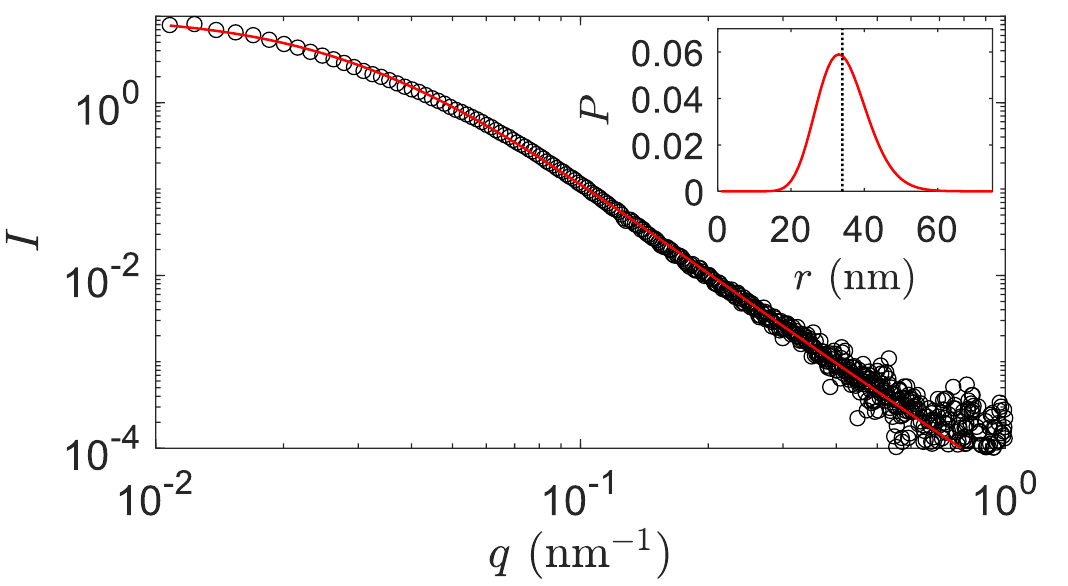}
\caption{Scattering intensity $I$ vs.~scattering wave number $q$ of dilute dispersion of carbon black Vulcan PF nanoparticles in the RTM17 mineral oil ($c_w=0.01$~\%). The red line is a mass fractal fit of such a form factor using a fractal dimension $d_{f0}=2.9$ and a Schulz particle radius distribution $P(r)$ centered on a radius of gyration $r_g=35$~nm (dashed line) with a polydispersity of 20~\% as shown in the inset.}
\label{fig:ffsaxs}
\end{figure}

Fig.~\ref{fig:ffsaxs} shows the scattering intensity vs.~wave vector $q$ for a dilute dispersion of CB particles (Vulcan PF, Cabot). Individual CB particle are fractal-like particles composed of fused nanoparticles of carbon~\cite{martinez2017,sztucki2007}, which motivates the use of a mass fractal model~\cite{teixeira1988} to fit $I(q)$. The fit yields a radius of gyration $r_g=35$~nm with 20\% polydispersity and a fractal dimension $d_{f0}=2.9$. Density measurements of CB powder were performed by helium pycnometer (AccuPyc II 1340, Micromeritics). Before the measurements, the powder was dried in an oven at 80$^{\circ}$C during 72 hours (2~\% in weight was lost). Two sets of measurements were done with 5 measures in a row for the first sample ($m=0.8619$~g) and 10 measurements in a row for the second sample ($m= 0.779$~g). The samples were poured in a 10~cm$^3$ aluminum vessel. We obtained a  density of the carbon black particles $d_{cb}=2.26\pm 0.03$ 

\subsection{Flow properties of Carbon black gels} \label{Apd:HB}
\begin{figure}[ht]
\centering
\includegraphics[width=8 cm]{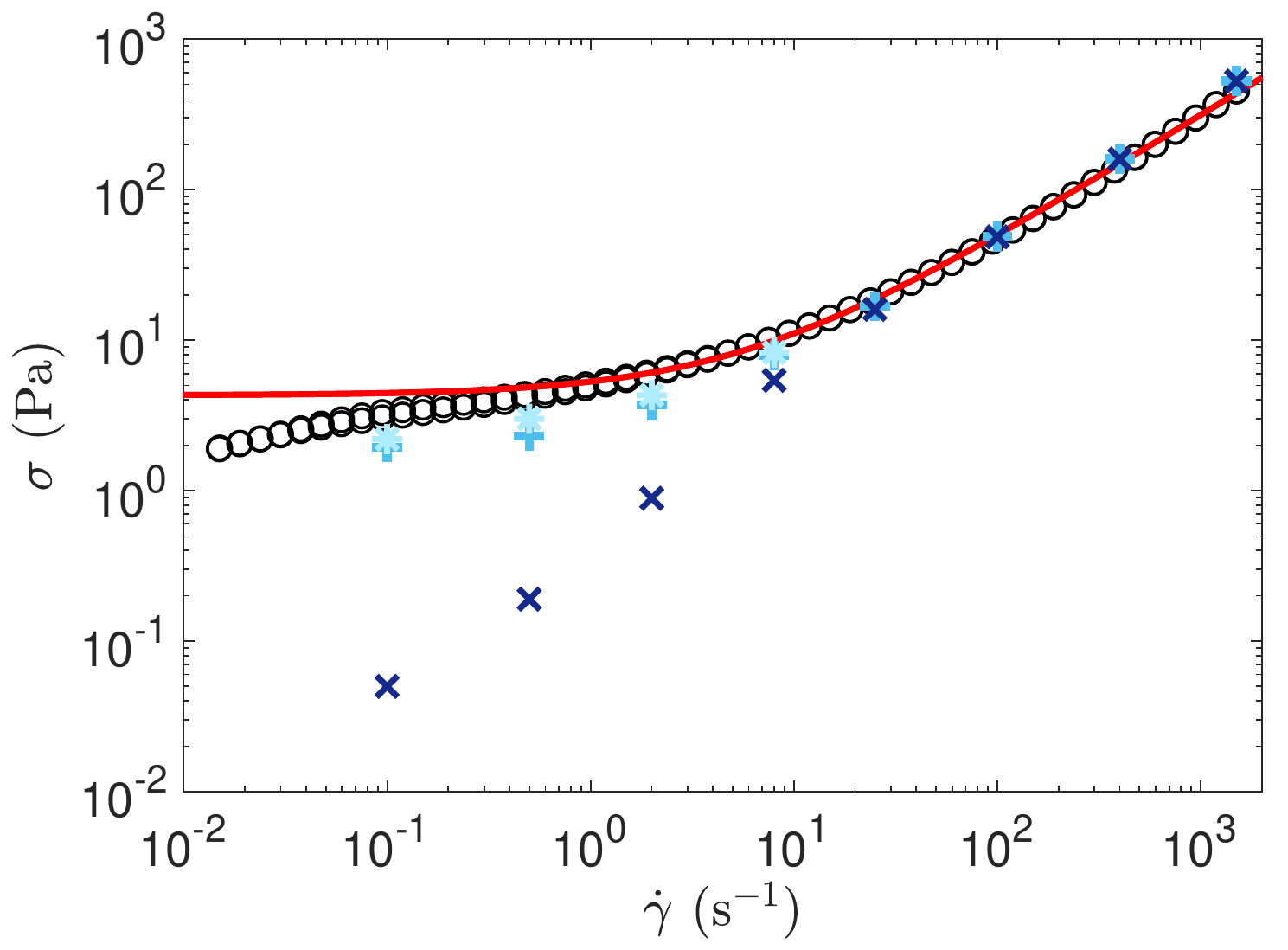}
\caption{Flow curve of the carbon black dispersion at $c_w=4\%w$ showing the evolution of the measured stress $\sigma$ as a function of the imposed shear rate $\dot{\gamma}$. $\dot{\gamma}$ is ramped down then up between 1500~s$^{-1}$ to 0.015 s$^{-1}$ with 10 points per decades at a rate of one point every 1~s. The red line is an Herschel-Bulkley fit. Unstationnary flow curve extracted from the preashear at $\dot\gamma_0$ in Fig.~\ref{fig:cessation}b at $t=1$~s (x), at the maximum of $\sigma$ (+) and at the end of the protocol (*). }
\label{fig:flowcurve}
\end{figure}

Figure~\ref{fig:flowcurve} reports the flow curve $\sigma(\dot \gamma)$ of a $c_w=4\%$w CB dispersion obtained by a decreasing ramp of shear rate. 
The flow curve  is fitted with a Herschel-Bulkley model, $\sigma=\sigma_y+K\dot{\gamma}^n$, and yields a dynamical yield stress $\sigma_y=4.5$~Pa, a fluidity index $n=0.83$, and a consistency index $K=1.0$~Pa.s$^{1/0.83}$.
In Fig.~\ref{fig:flowcurve}, we also show flow curve extracted from the preshear at $\dot\gamma_0$. We observe that those flow curve are not stationary. This is mostly due to the fact that the dispersion needs time to adapt from the jump in shear from 1000~s$^{-1}$ to $\dot\gamma_0$.

\subsection{Evolution of the viscoelastic modulus during the rest} \label{Apd:Rest}

\begin{figure}[ht]
\centering
\includegraphics[width=8 cm]{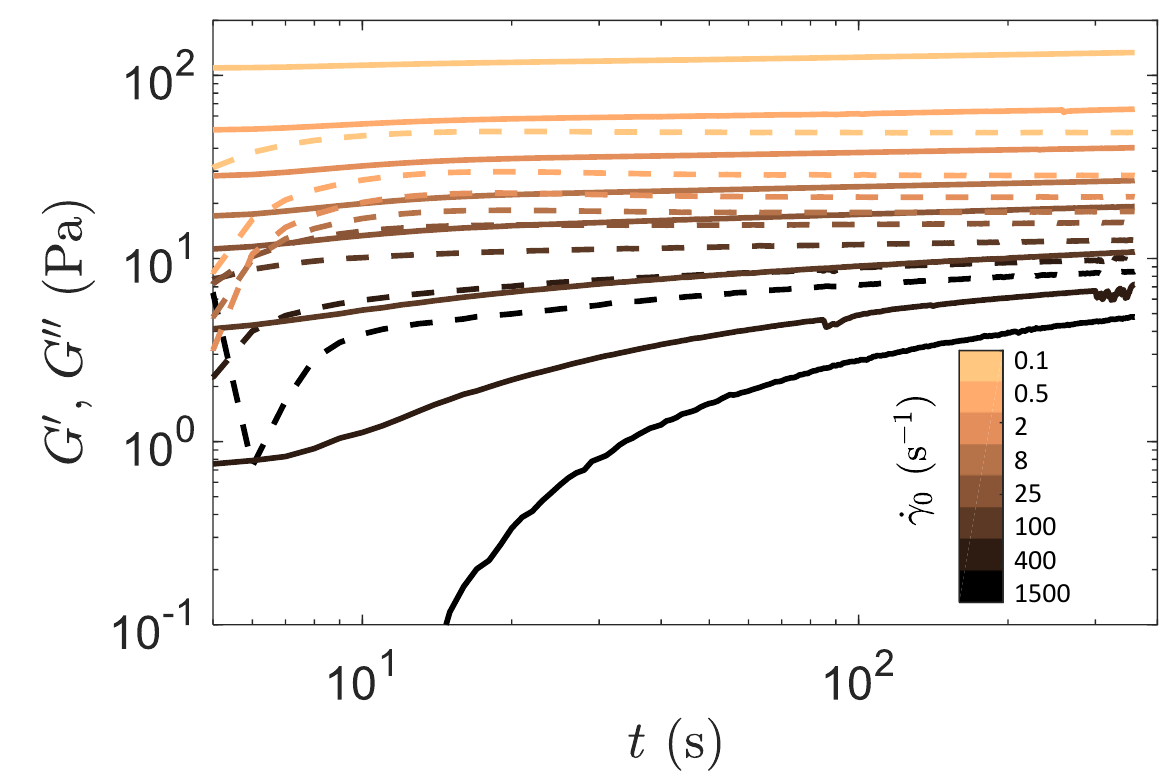}
\caption{Evolution of the viscoelastic moduli during the rest just after the Flow cessation protocol. The viscoelastic moduli $G'$ (line) and $G''$ (dash) are measured during 360~s at an oscillation amplitude of $\gamma=0.1$~\% and a frequency of $f=1$~Hz.}  
\label{fig:GpGpptime}
\end{figure}

During the rest period that fallows the flow cessation protocol, we measured the viscoelastic moduli of the dispersion in its linear regime, Fig.~\ref{fig:GpGpptime}. For high $\dot{\gamma}_0$ the gel reach within $\sim~100$~s a regime where aging becomes very slow. For low $\dot{\gamma}_0$ this slow aging regime is reached within a few seconds.

\subsection{tan($\delta$) representation of the viscoelastic spectrum} \label{Apd:Rest}

\begin{figure}[ht]
\centering
\includegraphics[width=8 cm]{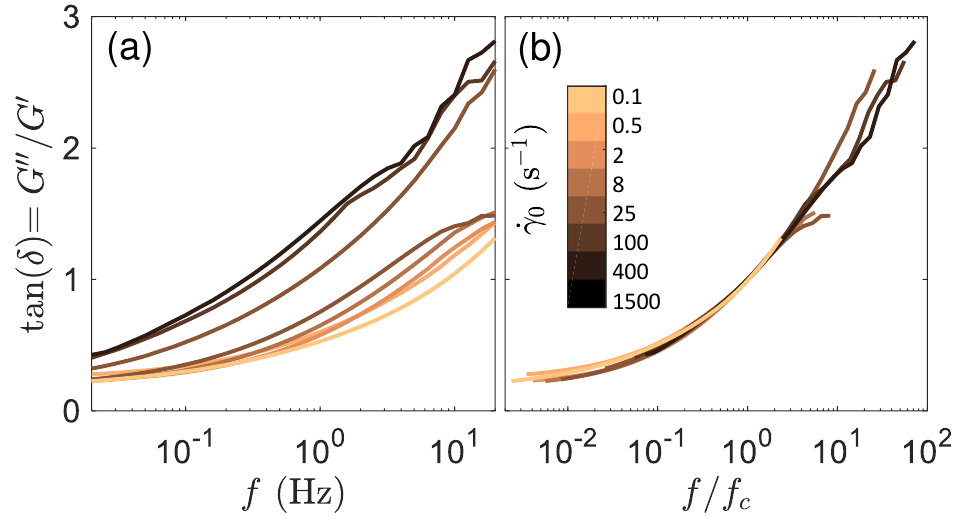}
\caption{tan($\delta$) representation of the viscoelastic spectrum measured in Fig.~\ref{fig:scaling}a-b. (a) tan($\delta$) as a function of the frequency $f$. (b) tan($\delta$) as a function of the normalized frequency $f/f_c$.} \label{fig:tand}
\end{figure}

Alternatively, the viscoelastic spectrums plotted in Fig.~\ref{fig:scaling}a may be represented by tan($\delta$)$=G''/G'$ as a function of the frequency $f$, Fig.~\ref{fig:tand}a. In this representation, tan($\delta$)$>1$ indicates that dissipation dominates the rheological behavior whereas tan($\delta$)$<1$ indicates a solid-like behavior at the corresponding frequency. In Fig.~\ref{fig:scaling}b, the viscoelastic spectrum is rescaled according to the coordinate ($f_c, G_c$). In the tan($\delta$) representation as $G'$ and $G''$ are rescaled by the same factor $G_c$ only the frequency axis needs to be rescaled. Doing so the tan($\delta$) measured for different $\dot{\gamma}_0$ scale on a master curve as displayed in Fig.~\ref{fig:tand}b.

\subsection{Model for the normalized viscoelastic spectrum of carbon black gels} \label{Appendix:model}

\begin{figure}[ht]
\centering
\includegraphics[width=8 cm]{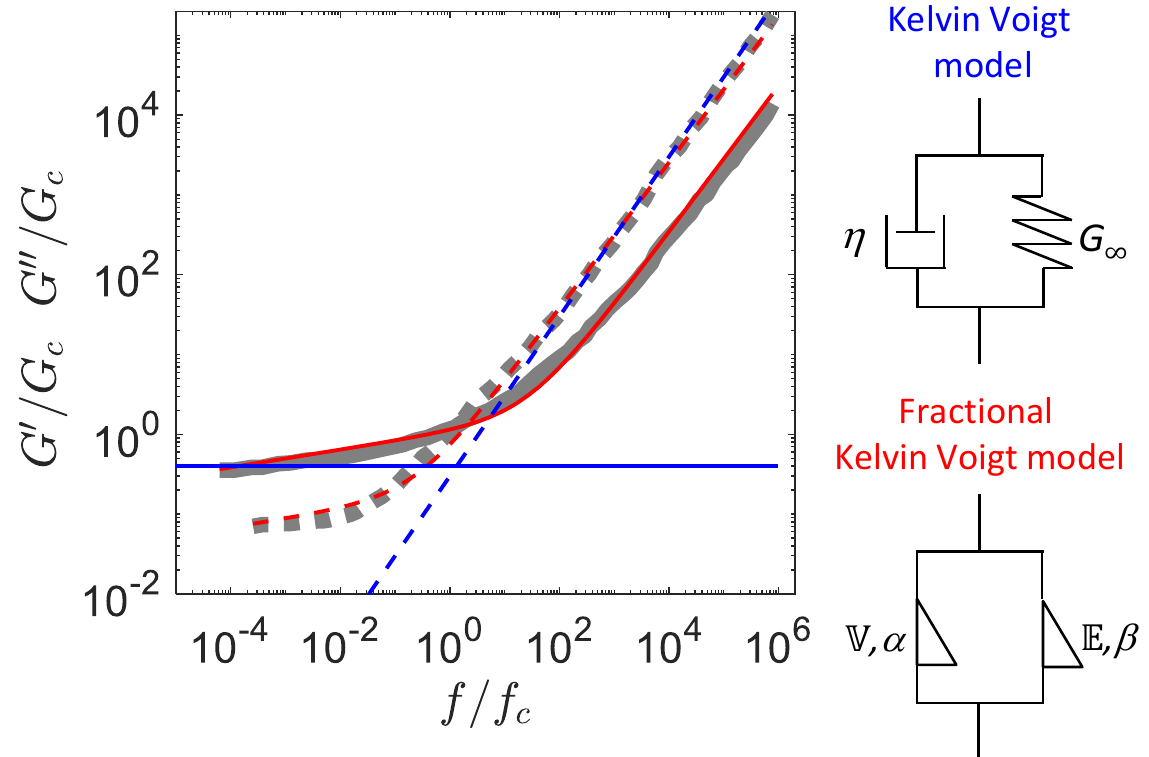}
\caption{Model of the normalized viscoelastic spectrum. Evolution of the normalized elastic $G'/G_c$ (grey line) and viscous $G''/G_c$ (dashed grey line) moduli as a function of the frequency $f/f_c$, extracted from~\cite{trappe2000}. ($f_c$, $G_c$) are the coordinates of the crossover between $G'$ and $G''$. The Kelvin-Voigt model is pictured in blue and the fractional Kelvin-Voigt model in red.}
\label{fig:rheomodel}
\end{figure}

The master curve reported in Fig.~\ref{fig:structure} is fitted using a Kelvin-Voigt model and a fractional Kelvin-Voigt model~\cite{schiessel1995}, as illustrated in Fig.~\ref{fig:rheomodel}. The fractional Kelvin-Voigt model consists of two springpots in parallel,  defined by their quasi-properties ($\mathds{V}$, $\mathds{E}$) and their dimensionless exponents ($\alpha$, $\beta$). Each springpot can be understood as a mechanical element having intermediate properties between that of a spring when its exponent is 0 and a dashpot when its exponent is 1. The resolution of the fractional Kelvin-Voigt model leads to
\begin{equation}
\bigg\{
\begin{array}{ll}
G' = \mathds{E} f^{\beta} \cos(\beta \pi/2)  + \mathds{V} f^{\alpha} \cos(\alpha \pi/2) \\
G'' = \mathds{E} f^{\beta} \sin(\beta \pi/2)  + \mathds{V} f^{\alpha} \sin(\alpha \pi/2)
\end{array}
\end{equation}
In the limit where $\alpha=1$ and $\beta=0$ we recover the classical Kelvin-Voigt model, which is defined by a dashpot of viscosity $\eta$ in parallel with a spring of elasticity $G_{\infty}$.

\begin{equation}
\bigg\{
\begin{array}{ll}
G' = \mathds{E}=G_{\infty}  \\
G'' = \mathds{V} f=2\pi \eta f
\end{array}
\end{equation}

As shown in Fig.~\ref{fig:rheomodel} the fractional model fits relatively well the normalized viscoelastic spectrum of carbon black gels in mineral oil using ($\mathds{V}/G_c=0.56$, $\alpha=0.91$) and ($\mathds{E}/G_c=1.09 $, $\beta=0.11$). 
Its classical counterpart can only capture the asymptotic behavior of the viscoelastic spectrum, i.e., the network elasticity $G_{\infty}= G' (f \ll f_c)$ and the background viscosity $\eta=G'' (f \gg f_c)/(2\pi f)$. We note that $G_{\infty}=0.3 G_c$. 



\subsection{Influence of the concentration $c_w$} \label{Apd:Concentration}
\begin{figure}[ht]
\centering
\includegraphics[width=8.5cm]{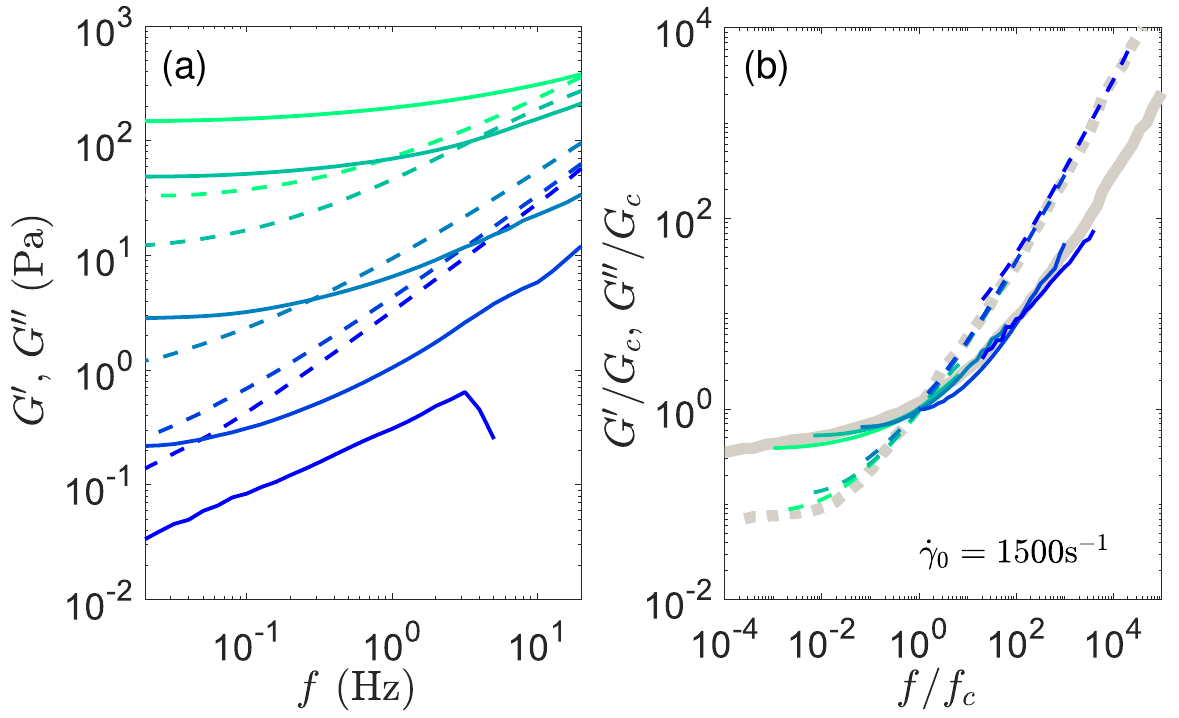}
\caption{Evolution of the viscoelastic spectrum for different gel concentration $c_w$ at $\dot{\gamma_0}=1500$~s$^{-1}$. (a) $G'$ (line) and $G''$ (dash) as a function of $f$. The concentration is varied from $c_w=2$ (blue) to 8\% (green). (b) Rescaled viscoelastic spectrum. The grey curves are taken from \cite{trappe2000} and corresponds to the master curve obtained by scaling a concentration series of carbon black dipersion in oil
}
\label{fig:concentration}
\end{figure}

Following the protocol displays in Fig.~\ref{fig:protocol} we test the influence of the concentration for $c_w=2$, 3, 6 and 8~\% in addition to $c_w=4$~\%. Such concentration series are presented in Fig.~\ref{fig:concentration} for $\dot{\gamma_0}=1500$~s$^{-1}$. We observe that high concentration shift the viscoelastic spectrum to higher elasticities. The cross over point is not always reachable within the frequency window. We scale the viscoelastic spectrum $G', G''$ on the master curve displayed in Fig.~\ref{fig:scaling}b to determine ($f_c, G_c$) and plot the results in Fig.~\ref{fig:sensitivity}.

\begin{figure}[ht]
\centering
\includegraphics[width=8.5cm]{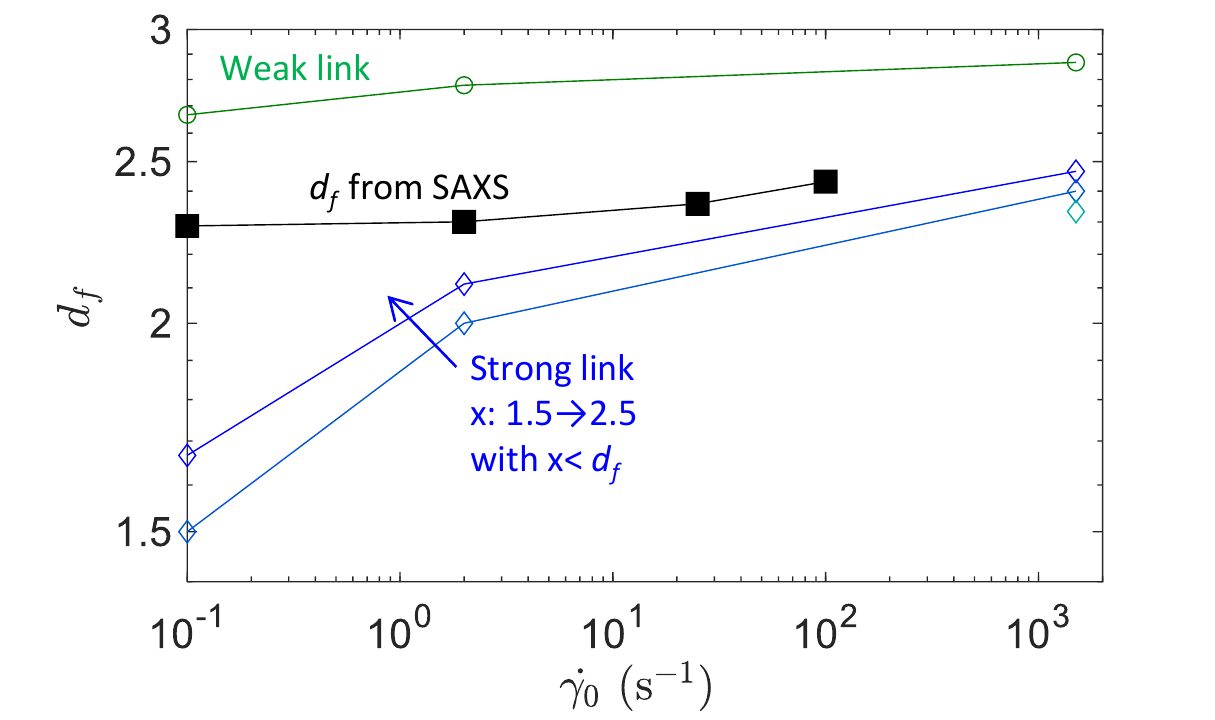}
\caption{Fractal dimension $d_f$ as a function of the shear step $\dot\gamma_0$. Measurements of $d_f$ extracted from the SAXS fits in Fig.\ref{fig:structure} (square) and evaluated from Fig.~\ref{fig:sensitivity} using the weak link model (circle) and the strong link model (diamond) from \cite{shih1990}. The fractal dimension $x$ of the gel backbone has been varied from x=1.5 (light blue) to x=2.5 (dark blue) in the strong link model.
}
\label{fig:df}
\end{figure}

\subsection{Analysis of the rheo-SAXS data} \label{Apd:RheoSaxs} 

\begin{figure}
\centering
\includegraphics[width=8 cm]{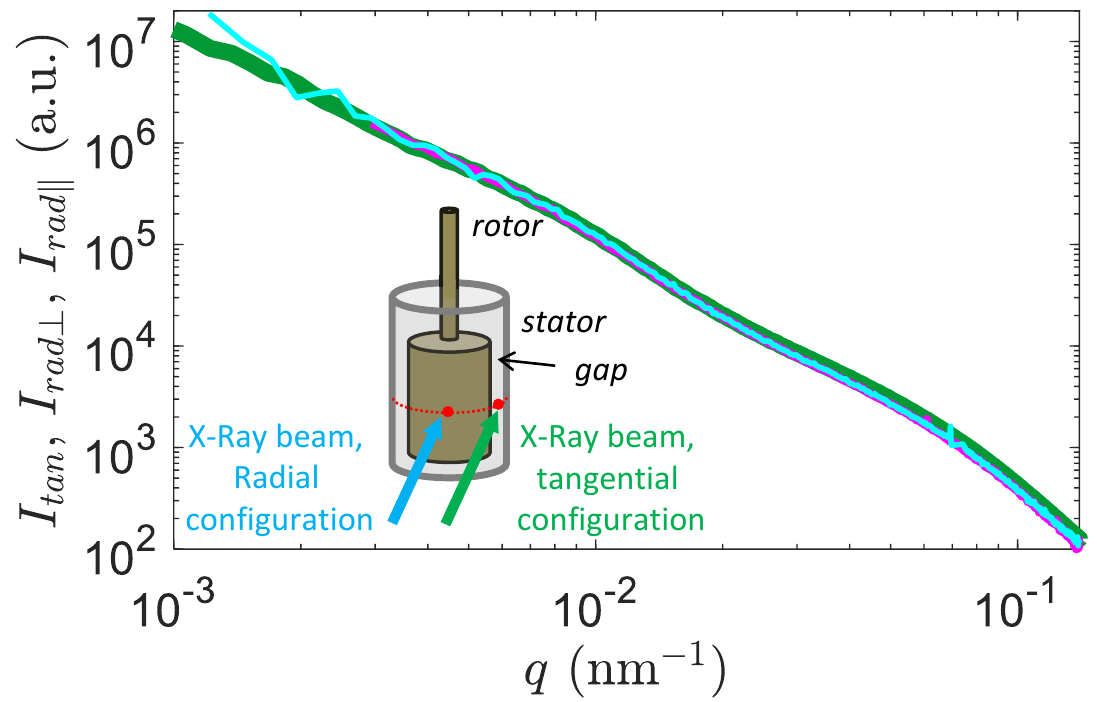}
\caption{Comparison of the scattering intensity between the tangential $I_{tan}$ (green) and radial $I_{rad \perp}$ (cyan) $I_{rad \parallel}$ (magenta) configuration. 
}
\label{fig:isotropy}
\end{figure}

The CB gel scattered intensities $I_{rad}(q)$ and $I_{tan}(q)$ are respectively measured in the radial and the tangential configuration. In the radial configuration, the X-ray beam probe the microstructure in the flow-vorticity plane, while in the tangential configuration, the X-ray beam probe the sample along velocity gradient-vorticity plane  as sketch in Fig.~\ref{fig:isotropy}. After the rest in the protocol from Fig.~\ref{fig:protocol}, the radial and tangential scattered intensity are compared. More precisely the radial scattering $I_{rad}$ is decomposed in its perpendicular $I_{rad \perp}$ and parallel $I_{rad \parallel}$ components. In Fig.~\ref{fig:isotropy}, we observe that $I_{tan}=I_{rad \perp}=I_{rad \parallel}$. The scattering intensity is isotropic and equal in all configurations. For better statistics, we focus on the tangential signal and radially average its 2D spectrum. In the paper, we note $I(q)=I_{tan}(q)$.

\begin{figure}
\centering
\includegraphics[width=8 cm]{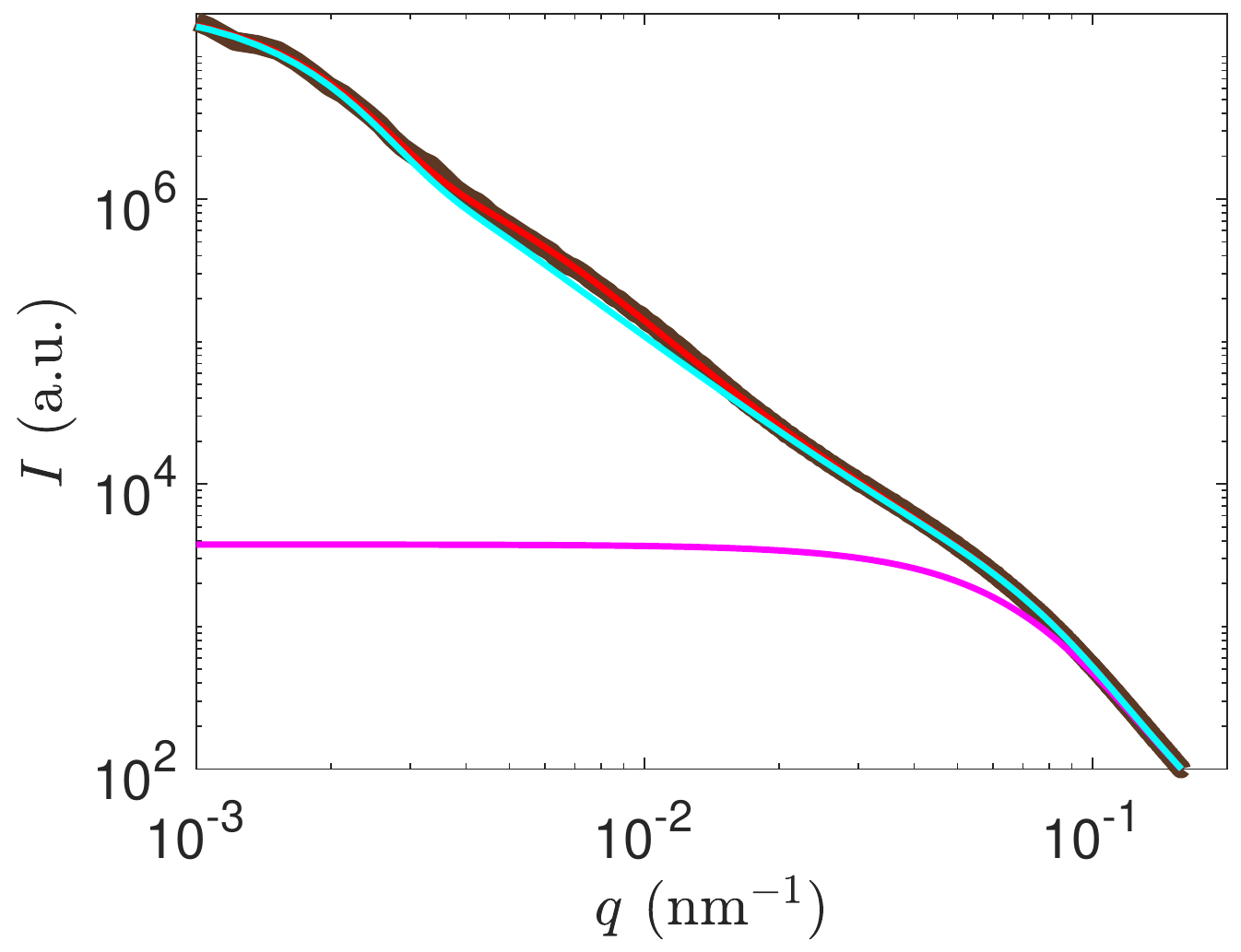}
\caption{Decomposition of the modified Beaucage model as written in Eq.~\ref{eq:beaucage}. Scattering intensity $I(q)$ as a function of $q$: experimental data (black), $I_2$ (pink), $I_1+I_2$ (cyan), $I_1.S_1+I_2$ (red).
}
\label{fig:beaucage}
\end{figure}

The intensity scattered by the carbon black is fitted in log scale using a modified Beaucage model \cite{beaucage1995,beaucage1996,keshavarz2021}, Fig.~\ref{fig:beaucage}:

\begin{multline}
I(q)=\underbrace{ \left[ G_1 \exp\left(-\frac{q^2 r_1^2}{3}\right) + B_1 \exp\left(-\frac{q^2 r_2^2}{3}\right)   q_1^{*p_1} \right]}_\text{Beaucage, Cluster level, $I_1$} . \\
\underbrace{\left[1+C_0 \left( \left(\frac{q}{q_s}\right)^2 + \left(\frac{q_s}{q}\right)^2 \right)^{-1} \right]}_\text{Inter-Cluster structure, $S_1$} + \underbrace{\left[G_2 \exp\left(-\frac{q^2 r_2^2}{3}\right) + B_2q_2^{*p_2}\right]}_\text{Beaucage, CB particle level, $I_2$},
\\\textrm{with}\quad q_{i=1,2}^*=q\left(\mathrm{erf}\left(\frac{q r_i}{\sqrt6}\right) \right)^{-3}
\label{eq:beaucage}
\end{multline}

In Eq.~\ref{eq:beaucage}, $I_2(q)$ refers to scattering due to the CB particle of size $r_0=r_2$ and fractal dimension $d_{f0}=-p_2$. The scattering due to the clusters of size $\xi_c=r_1$ and fractal dimension $d_f=-p_1$ is contained in the term $I_1(q)$. The modification of the two level Beaucage model consists in introducing an inter cluster structure factor $S_1(q)$ that accounts for the center-to-center distance between adjacent clusters, $\xi_s=2\pi/q_s$. $S_1(q)$ is a function that peaks at $q_s$ to a maximum  value $S_1(q_s)=1+C_0/2$ and that converge to 1 away from $q_s$. 

\begin{figure}
\centering
\includegraphics[width=7.5 cm]{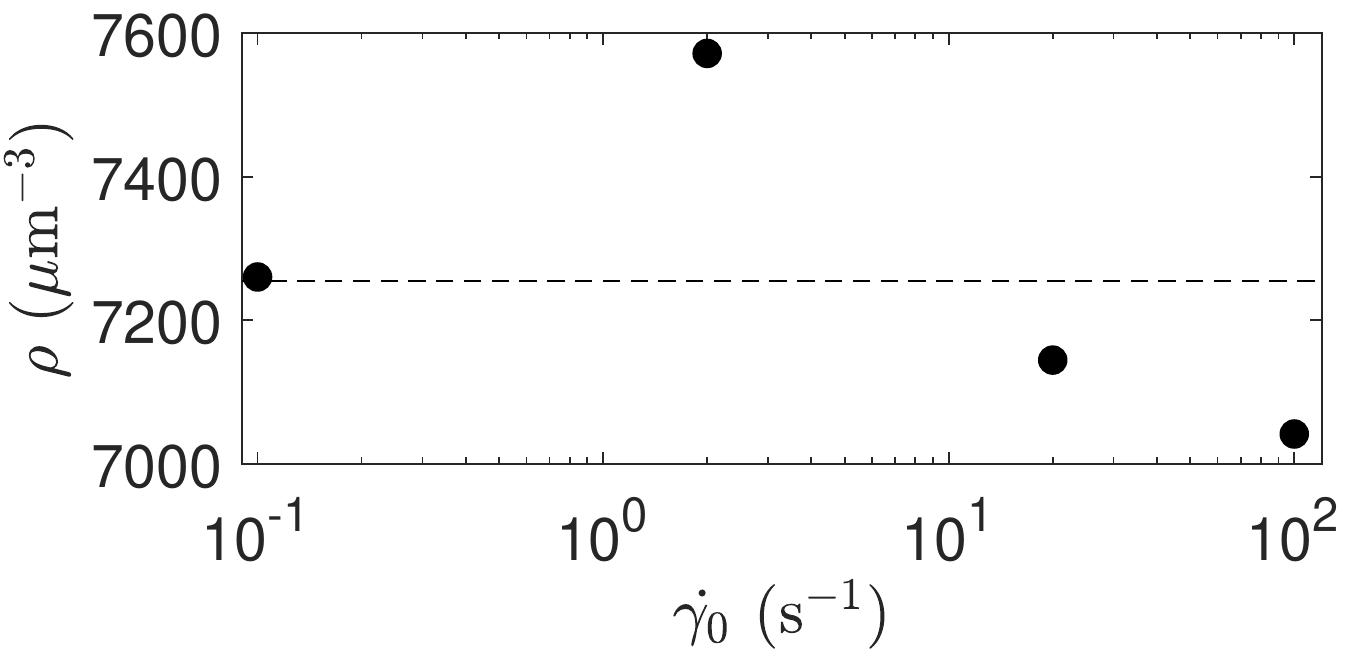}
\caption{Density of particles per unit cell $\rho$ as the function of the step shear $\dot\gamma_0$. $\rho$ is calculated from the values $r_0$, $\xi_c$, $d_f$ and $\xi_s$ obtained fitting the scattering intensity $I(q)$ plotted in Fig.\ref{fig:structure}a with Eq.~\ref{eq:beaucage}. The dash line is the average. 
}
\label{fig:rho}
\end{figure}

This fit is constrained by mass conservation. Indeed, as the multi-stable gels results from the same carbon black dispersion, all the gels have the exact same average number of particles at the macroscopic scale and in their unit cell. The unit cell is the minimum structural repeating unit necessary to construct the gel. In our case, the gel network unit cell is defined by the structural length $\xi_s$. The number of particles $N=(\xi_c/r_0)^{d_f}$ in a unit cell corresponds to the number of particles in the cluster of size $\xi_c$ and fractal dimension $d_f$. This gives a particle density $\rho=\frac{(\xi_c/r_0)^{d_f}}{\xi_s^3}=cste$. This constrain is well verified based on parameters obtained by fitting the SAXS data. As shown in Fig.~\ref{fig:rho}, $\rho=7250 \pm 230$~particles/$\mu$m$^3$ is constant within a margin of error of 3\%. Moreover, $\rho$ can be related to the particle concentration $c_w$ provided a good knowledge of the background oil density $d_{bck}=0.871$ ($T=20~^{\circ}$C), the individual carbon black particle density $d_{cb}=2.26 \pm 0.03$ ($T=20~^{\circ}$C) and the volume of a carbon black particle $v_{cb}$. The carbon black density $\rho$ in the gel is then

\begin{equation}
\rho=\frac{\phi}{v_{cb}},
\text{ with } \phi=\frac{c_w}{\left(c_w+\frac{d_{cb}}{d_{bck}}(1-c_w)\right)}
\label{eq:conservation}
\end{equation}

As it is hard to measure $v_{cb}$ given the particle fractal nature, $v_{cb}$ was evaluated from $\rho$. Using $v_{cb}=4/3\pi r_{\rho}^3$ we obtain $r_{\rho}=8.1$~nm a value lower than the radius of gyration of 35~nm extracted from the form factor measured in SAXS~Fig.~\ref{fig:ffsaxs}. This value is a bit smaller because the CB particles are fractal and polydisperse. Moreover, SAXS measurements tend to overestimate the particle size distribution as SAXS is more sensitive to larger particles. Finally, we might overestimate $\rho$ as we assumed that the clusters pack the space homogeneously. Despite the fact that $r_{\rho}$ is a bit small, we find $\rho=cste$ which tells us that the model is self-consistent.

\subsection{The interpenetration$\phi$-power law model, a model of the gel elasticity based on scaling arguments}
\label{Apd:model} 
\begin{figure}[ht] 
\centering
\includegraphics[width=\columnwidth]{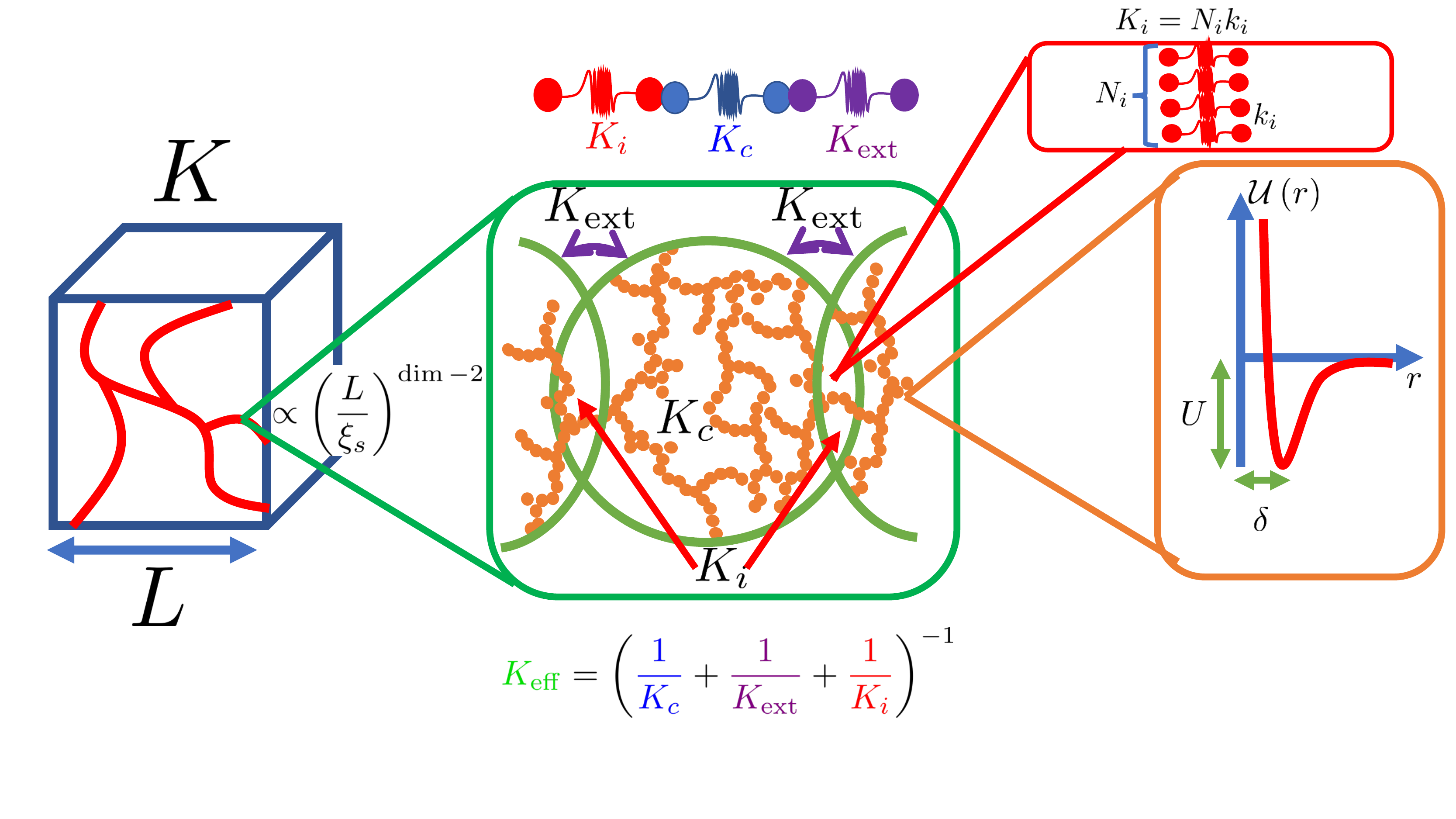}
\caption{Sketch of the contributions to the gel elasticity associated with the hierarchical structure of the gel.
}
\label{fig:ModelSketch}
\end{figure}

We assume that the gel is composed of particles of size $r_0$ that form clusters of size $\xi_c$ and fractal dimension $d_f$ separated by a center to center distance $\xi_s$. If $\xi_s>\xi_c$, the clusters are independent and the dispersion is a fluid. However, if $\xi_s<\xi_c$, clusters interpenetrate each other and form a gel. To model the elastic constant of the gel network, we follow the demonstrations of the $\phi$-power law models proposed in \cite{shih1990,Mellema2002,Wu2001} which allows to write
\begin{align}
    G'_\infty&=\frac{K}{L}\\
    K&=\left(\frac{L}{\xi_{s}}\right)^{\dim-2}K_\mathrm{eff}\\
    \phi\left(\frac{\xi_s}{r_0}\right)^{3-d_f}&=\left(\frac{\xi_s}{L}\right)^{3-\dim}
\end{align}
 with $G'_\infty$ the linear storage modulus, $\phi$ the particle volume fraction, $K$ the macroscopic stiffeness of the gel, $L$ the macroscopic size, $K_\mathrm{eff}$ the elementary effective stiffness of clusters and $\dim$ the dimension of the network which can be equal to the dimension of the euclidean space. These $\phi$-power law models are based on building the relationship between a microscopic stiffness due to the interaction potential between the colloids and a macroscopic scale through different extrapolation by mean of scaling laws. Then, to demonstrate our interpenetration $\phi$-power law model, we assume that $K_\mathrm{eff}$ may be written 
\begin{equation}
    \frac{1}{K_\mathrm{eff}}=\frac{1}{K_c}+\frac{1}{K_\mathrm{ext}}+\frac{1}{K_i}
\label{eq:general}
\end{equation}
with $K_c$, $K_\mathrm{ext}$ and $K_i$ the elastic stiffness related to the inside of the cluster, the intermicroscopic (see Ref\cite{Wu2001}) and the interpenetration of the cluster, respectively. These different stiffnesses are considered as spring in series as sketch on Fig. \ref{fig:ModelSketch} and as commonly assumed in the literature \cite{Wu2001}.

Let us now quantify $K_i$. $K_i$ is assumed to be proportional to the number of contact $N_i$ between interpenetrating clusters: $K_i=N_i k_i$, where $k_i$ is a reference interpenetration stiffness. The last expression comes from the fact we assumed that the contacts in the interpenetration zone are parallel springs: this justifies the additivity of the stiffnesses. This point has never been highlighted in the literature and is part of our proposition for the interpenetration $\phi$-power law model. Also, the reference interpenetration stiffness $k_i$ is a hard point in the model. Indeed, referring ourselves to \cite{Kantor1984a,Kantor1984b}, there may be a competition between elongation stiffness and bending stiffness. Without any true experimental insights about the interpenetration zone, it is hard to disentangle the contribution of each and we will assume that it is directly related to the depth of the interaction potential $U$ and the distance of interaction $\delta$ through $k_i=U/\delta^2$. Due to the fractal nature of the clusters, there are $N_i=\left(6V_i/\pi r_0^3\right)^\frac{d_f}{3}$ particles inside the intersection volume $V_i$ between two clusters. Geometrically assimilating clusters to spheres leads to an intersection volume \cite[p. 97]{Polyanin2007,Kern1967}

\begin{equation}
    V_i=\frac{\pi}{12}\xi_{c}^3\left(2+\frac{\xi_s}{\xi_c}\right)
    \left(1-\frac{\xi_s}{\xi_c}\right)^{2} \textbf{1}_{\left\{\xi_s<\xi_c\right\}}.
    \label{eq:vi}
\end{equation}
We assume that each particle brought by each cluster in $V_i$ form a contact adding rigidity to the whole system. Therefore, the numbers of contact is roughly $N_i$. Putting together the last expressions, we get

\begin{equation}
    K_i=\frac{U}{2\delta^2}\left(\frac{\xi_{c}}{r_{0}}\right)^{d_{f}}\left(1+\frac{\xi_s}{2\xi_c}\right)^{\frac{d_{f}}{3}}\left(1-\frac{\xi_s}{\xi_c}\right)^{\frac{2d_{f}}{3}}\textbf{1}_{\left\{\xi_s<\xi_c\right\}}.
    \label{eq:Model1}
\end{equation}

Let us now compare $K_i$ with $K_c$ and $K_\mathrm{ext}$. There are different ways to consider that $K_i\ll \min\left(K_c,K_\mathrm{ext}\right)$. To simplify the comparison, following \cite{Wu2001}, we will write 
\begin{equation}
    \frac{1}{K_c}+\frac{1}{K_\mathrm{ext}}=\frac{1}{K_c}\left(\frac{K_c}{K_\mathrm{ext}}\right)^\alpha
\end{equation}
with $\alpha\in\left[0,1\right]$ allowing to make a transition between the weak-link and the strong-link regime.
A first way to compare $K_i$ with $K_c\left(K_\mathrm{ext}/K_c\right)^\alpha$ is to say that the system is in the regime $\xi_c/\xi_s\gtrsim1$. Thus, one can re-write Eq.~\ref{eq:Model1} as

\begin{equation}
    K_i\underset {\xi_c/\xi_s\gtrsim1}{\propto} \frac{U}{2\delta^2} \left(1-\frac{\xi_s}{\xi_c}\right)^{\frac{2d_{f}}{3}}
\end{equation}

telling us that $K_i$ depends strongly on the distance of $\xi_c/\xi_s$ from unity. Therefore, $K_i$ is negligible when $\xi_c/\xi_s\gtrsim1$ compared to $K_c\left(K_\mathrm{ext}/K_c\right)^\alpha$ and Eq.~(\ref{eq:general}), we get $K_\mathrm{eff}\approx K_i$.

The other way to consider the system is, following previous approaches in \cite{Kantor1984a,Kantor1984b,shih1990,Wu2001,Mellema2002,wessel1992}, estimating $K_c\left(K_\mathrm{ext}/K_c\right)^\alpha\propto\xi_s^{-\mu}$ with $\mu\in\left[1,5\right]$ function of the fractal dimension $d_f$, the dimension of the elastic backbone and the regime of strong-link or weak-link because $\xi_s$ is similar to a cluster size with contact. Recalling that $\left(\xi_c/r_0\right)^{d_f}\propto\xi_s^3$, one gets in this case

\begin{equation}
    \frac{K_i}{K_c}\left(\frac{K_c}{K_\mathrm{ext}}\right)^\alpha\propto\xi_s^{3+\mu}\left(1+\frac{\xi_s}{2\xi_c}\right)^{\frac{d_{f}}{3}}\left(1-\frac{\xi_s}{\xi_c}\right)^{\frac{2d_{f}}{3}}.
\end{equation}

\begin{figure}[th!] 
\centering
\includegraphics[width=0.9\linewidth]{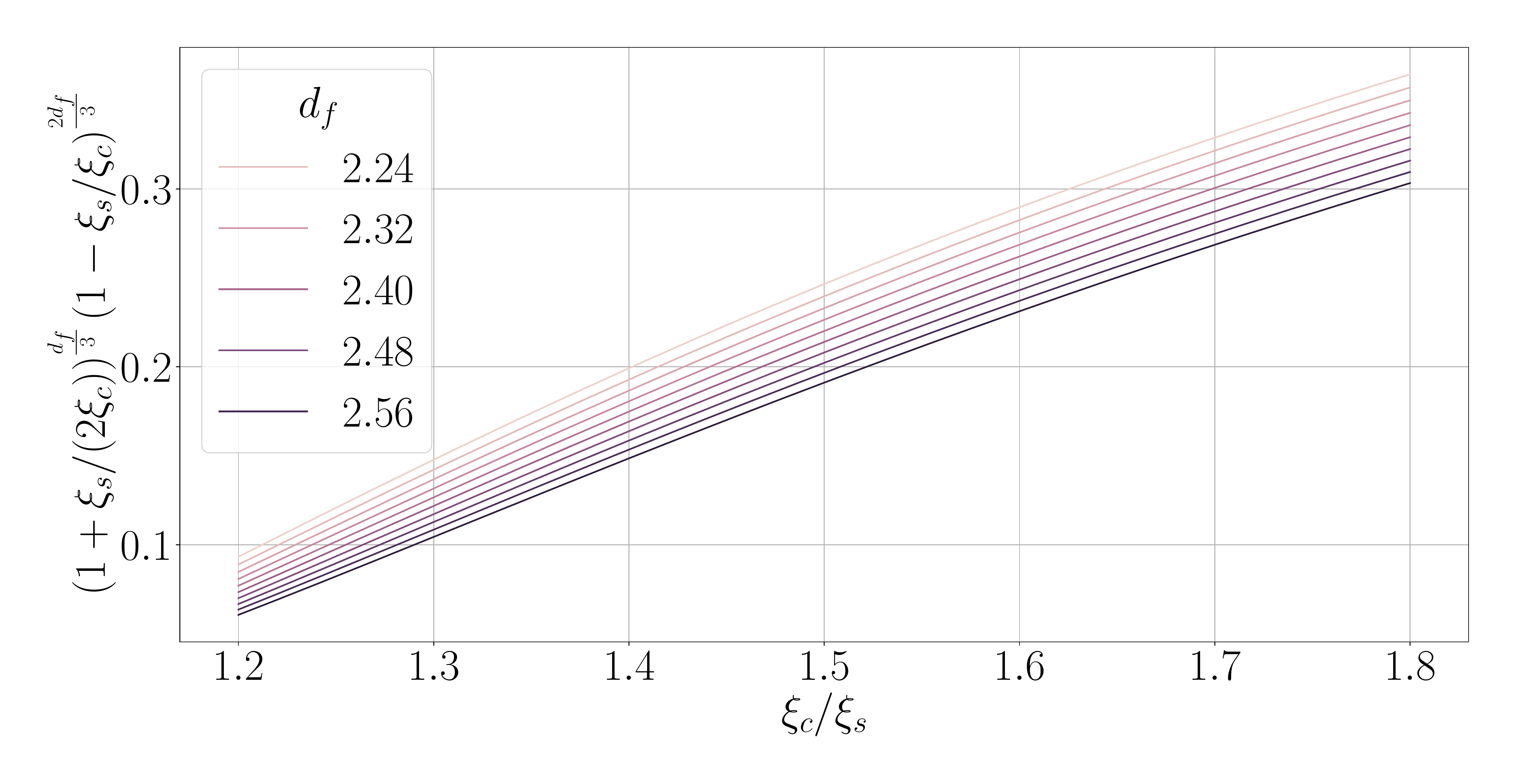}
\caption{Evolution of $\xi_c/\xi_s\mapsto\left(1+\xi_s/(2\xi_c)\right)^{\frac{d_{f}}{3}}\left(1-\xi_s/\xi_c\right)^{\frac{2d_{f}}{3}}$ for different values of $d_f$ on $\left[1.2,1.8\right]$
}
\label{fig:FunctionScaling}
\end{figure}

Assuming that $\xi_s$ does not vary much, $K_i/K_c\left(K_c/K_\mathrm{ext}\right)^\alpha$ is governed by the values of $\xi_c/\xi_s\mapsto\left(1+\xi_s/(2\xi_c)\right)^{\frac{d_{f}}{3}}\left(1-\xi_s/\xi_c\right)^{\frac{2d_{f}}{3}}$ on $\left[1.2,1.8\right]$. Referring to Fig. \ref{fig:FunctionScaling}, $K_i/K_c\left(K_c/K_\mathrm{ext}\right)^\alpha$ is between 0.1 and 0.3. Therefore, one can assume that $K_i\ll K_c\left(K_\mathrm{ext}/K_c\right)^\alpha$, at least for the first values, and following Eq.~(\ref{eq:general}), we get $K\approx K_i$.

Generally, the more $K_i$ is getting closer to $K_c\left(K_\mathrm{ext}/K_c\right)^\alpha$, the more difficult it is to consider that only one phenomenon prevails. If someone wants to completely understand the balance between the different contributions, she or he needs to model both phenomena and their coupling. This is not the goal of this model which tries to give some orders of magnitude without exhaustively modeling the system.

This final expression of the interpenetration $\phi$-power law model is then
\begin{equation}
    G'_\infty=\underset{G_{\mathrm{CB}}}{\underbrace{\frac{U}{r_{0}\delta^{2}}}}\underset{g_{\mathrm{Interp}}}{\underbrace{\frac{1}{2}\left(\frac{\xi_c}{r_0}\right)^{d_f}\left(1+\frac{\xi_s}{2\xi_c}\right)^{\frac{d_f}{3}}\left(1-\frac{\xi_s}{\xi_c}\right)^{\frac{2d_f}{3}}}}\underset{g_{\mathrm{Net}}}{\underbrace{\phi\left(\frac{\xi_{s}}{r_{0}}\right)^{2-d_{f}}}}
\end{equation}
expliciting that $G_{\mathrm{CB}}$ corresponds to the elasticity arising from colloid-colloid interaction with $U$ and $\delta$ respectively the depth and the range of the attraction, $g_{\mathrm{Interp}}$ is the elementary scaling for elasticity to account for interpenetration and may be changed according to \cite{shih1990} and $g_{\mathrm{Net}}$ is the network contribution from the element of elasticity to the macroscopic scale. The dimension of the network $\dim$ is not required in the final expression due to the contribution of the effective volume fraction through the particle volume fraction and the fractal dimension related to $\xi_s$.

The difference with the usual $\phi$-power law model\cite{shih1990} relies on the fact that $K_i=N_ik_i$ for our interpenetration $\phi$-power law model and $K_i=\left(\xi_s/r_0\right)^{-2-x}k_i$ with $x$ the chemical dimension or the dimension of the elastic backbone for the usual $\phi$-power law model\cite{shih1990}. If we assume $\dim=3$, it is possible to find a weak-link like regime\cite{shih1990} with $G'_\infty\propto\phi^{\frac{1}{3-d_f}}$. 

To summarize the approach and the assumptions :
\begin{itemize}
    \item Most of the ingredients come from the previous $\phi$-power law models building elasticity from microscopic quantities\cite{shih1990,Wu2001,Mellema2002}. The scaling between the macroscopic stiffness and the effective microscopic stiffness is conserved, the colloid-colloid interaction is conserved without expliciting the relations with bending or elongation, the construction of the effective microscopic stiffness as spring in series is also conserved.
    \item We assumed the domination in the behaviour of $K_i$ instead of the other stiffenesses, at least at the beginning of the interpenetration. As discussed above, the more pronounced the interpenetration, the more questionable this hypothesis. In a fully interpenetrated case, the contribution of each phenomenon may be of the same order of magnitude and the previous demonstration does not hold anymore. The derivation of an exhaustive model allowing the transition from one case to the other will be out of the scope of this paper. However, we encourage future research to dig into this model to make it complete and exhaustive against the literature and the present results.
\end{itemize}


\section{Erratum}

There is a mistake in the derivation of the interpenetration $\phi$-power law model in this article~\cite{dages2022}. This mistake appears in Eq. (8) and has repercussion on the final equation giving the gel elasticity $G'_{\infty}$ in Eq. (1) and (15). Here, we correct this mistake. The corrected calculation of $G'_{\infty}$ corroborate the experimental data presented in Fig.~4c~\cite{dages2022}. The paper conclusion remains therefore unchanged.
\\

In the article~\cite{dages2022}, in Eq. (8), the conservation of mass is expressed as 
\begin{equation}
    \phi\left(\frac{\xi_s}{r_0}\right)^{3-d_f}=\left(\frac{\xi_s}{L}\right)^{3-\dim}
\end{equation}
This is true if the size $\xi_s$ is the only length scale playing a role in the clusters and the network. However, in our case, there are two length scales $\xi_s$ and $\xi_c$ which brings 
\begin{equation}
    \phi=\left(\frac{r_{0}}{\xi_{s}}\right)^{3}\left(\frac{\xi_{c}}{r_{0}}\right)^{d_{f}}\left(\frac{\xi_{s}}{L}\right)^{3-\dim}
\end{equation}

This has an impact on Eq.~(15) in ~\cite{dages2022} which becomes 
\begin{equation}
\begin{split}
    G'_{\infty}=\underset{G_{\mathrm{CB}}}{\underbrace{\frac{U}{r_0\delta^{2}}}}\underset{g_{\mathrm{Interp}}}{\underbrace{\frac{1}{2}\left(\frac{\xi_{c}}{r_{0}}\right)^{d_{f}}\left(1+\frac{\xi_{s}}{2\xi_{c}}\right)^{\frac{d_{f}}{3}}\left(1-\frac{\xi_{s}}{\xi_{c}}\right)^{\frac{2d_{f}}{3}}}}
    \\ 
    \underset{g_{\mathrm{Net}}}{\underbrace{\phi\left(\frac{\xi_{s}}{r_{0}}\right)^{2}\left(\frac{\xi_{c}}{r_{0}}\right)^{-d_{f}}}}
    \end{split}
\end{equation}
or in a compact form 
\begin{equation}
    G'_{\infty}=\frac{U}{a\delta^{2}}\frac{1}{2}\left(1+\frac{\xi_{s}}{2\xi_{c}}\right)^{\frac{d_{f}}{3}}\left(1-\frac{\xi_{s}}{\xi_{c}}\right)^{\frac{2d_{f}}{3}}\phi\left(\frac{\xi_{s}}{r_{0}}\right)^{2}.
    \label{eq:Model}
\end{equation}

\begin{figure}
    \centering
    \includegraphics[width=\columnwidth]{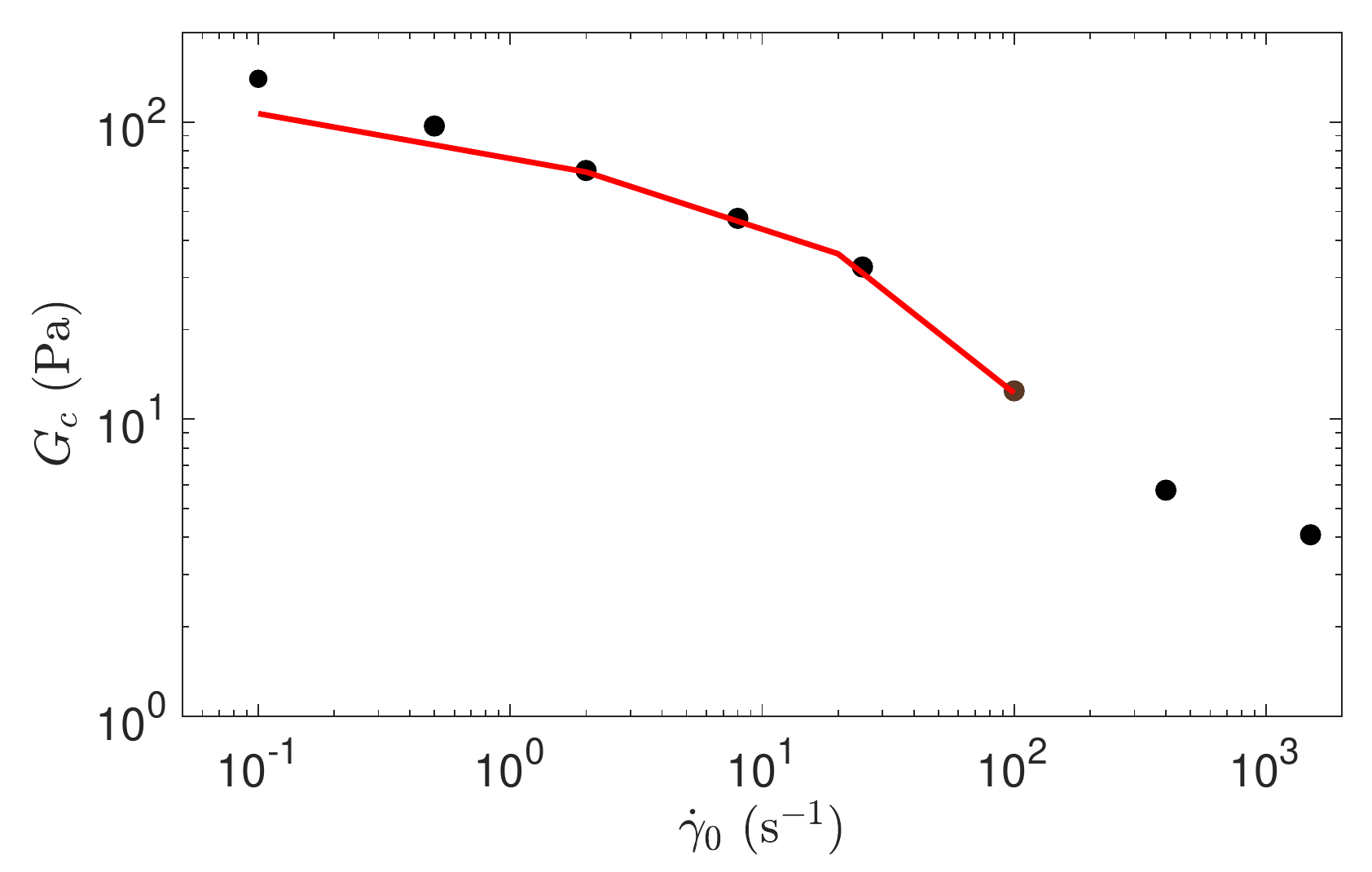}
    \caption{Evolution of  $G_c$  vs.~$\dot\gamma_0$. The red line is the best fit of the data using Eq.~\eqref{eq:Model} and the structural information reported in Fig.~(6a) of the paper~\cite{dages2022}. The best fit is obtained with a single adjustable parameter, namely the prefactor $G_{CB}= 9$~Pa.}
    \label{fig:rheo}
\end{figure}

The Fig.~(4c) in~\cite{dages2022} displays the variations of the gel crossover elasticity $G_c$ as a function of the shear rate intensity $\dot\gamma_0$ applied before flow cessation and the fit using the interpenetration $\phi$-power law model. We note that $G'_{\infty}=0.3 G_c$. The fit of $G_c$ in this figure needs to be modified accordingly to Eq.~\ref{eq:Model} and replaced by Fig.~\ref{fig:rheo}. In Fig.~\ref{fig:rheo}, the corrected model fits well the data. The conclusion of the paper remains therefore unchanged.

\section*{Author Contributions}
T.G. supervised the project. T.G., S.M. and T.D. conceived and designed the experiments. N.D., L.M. and T.G. carried out the measurements. N.D., A.P. and T.G. analyzed the experiments. L.V.B. developed the rheology model. T.G. wrote the paper. T.D. edited the manuscript.

\section*{Conflicts of interest}
There are no conflicts to declare.

\section*{Acknowledgements}
We thank the ESRF for beamtime at the beamline ID02 (proposal SC5099).
We are especially grateful to Theyencheri Narayanan for the discussions and technical support for the USAXS measurements. We are thankful to Wilbert Smit and Gauthier Legrand for their help in data acquisition and data analysis. This work was supported by the Région Auvergne-Rhône-Alpes ``Pack Ambition Recherche", the LABEX iMUST (ANR-10-LABX-0064) of Université de Lyon, within the program "Investissements d'Avenir" (ANR-11-IDEX-0007), the ANR grants (ANR-18-CE06-0013 and ANR-21-CE06-0020-01). This work benefited from meetings within the French working group GDR CNRS 2019 ``Solliciter LA Matière Molle" (SLAMM).


\begin{thebibliography}{91}%
\makeatletter
\providecommand \@ifxundefined [1]{%
 \@ifx{#1\undefined}
}%
\providecommand \@ifnum [1]{%
 \ifnum #1\expandafter \@firstoftwo
 \else \expandafter \@secondoftwo
 \fi
}%
\providecommand \@ifx [1]{%
 \ifx #1\expandafter \@firstoftwo
 \else \expandafter \@secondoftwo
 \fi
}%
\providecommand \natexlab [1]{#1}%
\providecommand \enquote  [1]{``#1''}%
\providecommand \bibnamefont  [1]{#1}%
\providecommand \bibfnamefont [1]{#1}%
\providecommand \citenamefont [1]{#1}%
\providecommand \href@noop [0]{\@secondoftwo}%
\providecommand \href [0]{\begingroup \@sanitize@url \@href}%
\providecommand \@href[1]{\@@startlink{#1}\@@href}%
\providecommand \@@href[1]{\endgroup#1\@@endlink}%
\providecommand \@sanitize@url [0]{\catcode `\\12\catcode `\$12\catcode
  `\&12\catcode `\#12\catcode `\^12\catcode `\_12\catcode `\%12\relax}%
\providecommand \@@startlink[1]{}%
\providecommand \@@endlink[0]{}%
\providecommand \url  [0]{\begingroup\@sanitize@url \@url }%
\providecommand \@url [1]{\endgroup\@href {#1}{\urlprefix }}%
\providecommand \urlprefix  [0]{URL }%
\providecommand \Eprint [0]{\href }%
\providecommand \doibase [0]{https://doi.org/}%
\providecommand \selectlanguage [0]{\@gobble}%
\providecommand \bibinfo  [0]{\@secondoftwo}%
\providecommand \bibfield  [0]{\@secondoftwo}%
\providecommand \translation [1]{[#1]}%
\providecommand \BibitemOpen [0]{}%
\providecommand \bibitemStop [0]{}%
\providecommand \bibitemNoStop [0]{.\EOS\space}%
\providecommand \EOS [0]{\spacefactor3000\relax}%
\providecommand \BibitemShut  [1]{\csname bibitem#1\endcsname}%
\let\auto@bib@innerbib\@empty
\bibitem [{\citenamefont {Trappe}\ \emph {et~al.}(2001)\citenamefont {Trappe},
  \citenamefont {Prasad}, \citenamefont {Cipelletti}, \citenamefont {Segre},\
  and\ \citenamefont {Weitz}}]{trappe2001}%
  \BibitemOpen
  \bibfield  {author} {\bibinfo {author} {\bibfnamefont {V.}~\bibnamefont
  {Trappe}}, \bibinfo {author} {\bibfnamefont {V.}~\bibnamefont {Prasad}},
  \bibinfo {author} {\bibfnamefont {L.}~\bibnamefont {Cipelletti}}, \bibinfo
  {author} {\bibfnamefont {P.}~\bibnamefont {Segre}},\ and\ \bibinfo {author}
  {\bibfnamefont {D.~A.}\ \bibnamefont {Weitz}},\ }\bibfield  {title} {\bibinfo
  {title} {Jamming phase diagram for attractive particles},\ }\href@noop {}
  {\bibfield  {journal} {\bibinfo  {journal} {Nature}\ }\textbf {\bibinfo
  {volume} {411}},\ \bibinfo {pages} {772} (\bibinfo {year}
  {2001})}\BibitemShut {NoStop}%
\bibitem [{\citenamefont {Lu}\ and\ \citenamefont {Weitz}(2013)}]{Lu:2013}%
  \BibitemOpen
  \bibfield  {author} {\bibinfo {author} {\bibfnamefont {P.}~\bibnamefont
  {Lu}}\ and\ \bibinfo {author} {\bibfnamefont {D.}~\bibnamefont {Weitz}},\
  }\bibfield  {title} {\bibinfo {title} {Colloidal particles: Crystals,
  glasses, and gels},\ }\href@noop {} {\bibfield  {journal} {\bibinfo
  {journal} {Annu. Rev. Condens. Matter Phys.}\ }\textbf {\bibinfo {volume}
  {4}},\ \bibinfo {pages} {217} (\bibinfo {year} {2013})}\BibitemShut {NoStop}%
\bibitem [{\citenamefont {Ioannidou}\ \emph {et~al.}(2016)\citenamefont
  {Ioannidou}, \citenamefont {Kanduc}, \citenamefont {Li}, \citenamefont
  {Frenkel}, \citenamefont {Dobnikar},\ and\ \citenamefont
  {Gado}}]{Ioannidou:2016}%
  \BibitemOpen
  \bibfield  {author} {\bibinfo {author} {\bibfnamefont {K.}~\bibnamefont
  {Ioannidou}}, \bibinfo {author} {\bibfnamefont {M.}~\bibnamefont {Kanduc}},
  \bibinfo {author} {\bibfnamefont {L.}~\bibnamefont {Li}}, \bibinfo {author}
  {\bibfnamefont {D.}~\bibnamefont {Frenkel}}, \bibinfo {author} {\bibfnamefont
  {J.}~\bibnamefont {Dobnikar}},\ and\ \bibinfo {author} {\bibfnamefont
  {E.~D.}\ \bibnamefont {Gado}},\ }\bibfield  {title} {\bibinfo {title} {The
  crucial effect of early-stage gelation on the mechanical properties of cement
  hydrates},\ }\href@noop {} {\bibfield  {journal} {\bibinfo  {journal} {Nat.
  Commun.}\ }\textbf {\bibinfo {volume} {7}},\ \bibinfo {pages} {12106}
  (\bibinfo {year} {2016})}\BibitemShut {NoStop}%
\bibitem [{\citenamefont {Parant}\ \emph {et~al.}(2017)\citenamefont {Parant},
  \citenamefont {Muller}, \citenamefont {{Le Mercier}}, \citenamefont
  {Tarascon}, \citenamefont {Poulin},\ and\ \citenamefont
  {Colin}}]{Parant:2017}%
  \BibitemOpen
  \bibfield  {author} {\bibinfo {author} {\bibfnamefont {H.}~\bibnamefont
  {Parant}}, \bibinfo {author} {\bibfnamefont {G.}~\bibnamefont {Muller}},
  \bibinfo {author} {\bibfnamefont {T.}~\bibnamefont {{Le Mercier}}}, \bibinfo
  {author} {\bibfnamefont {J.}~\bibnamefont {Tarascon}}, \bibinfo {author}
  {\bibfnamefont {P.}~\bibnamefont {Poulin}},\ and\ \bibinfo {author}
  {\bibfnamefont {A.}~\bibnamefont {Colin}},\ }\bibfield  {title} {\bibinfo
  {title} {Flowing suspensions of carbon black with high electronic
  conductivity for flow applications: Comparison between carbons black and
  exhibition of specific aggregation of carbon particles},\ }\href@noop {}
  {\bibfield  {journal} {\bibinfo  {journal} {Carbon}\ }\textbf {\bibinfo
  {volume} {119}},\ \bibinfo {pages} {10} (\bibinfo {year} {2017})}\BibitemShut
  {NoStop}%
\bibitem [{\citenamefont {Cao}\ and\ \citenamefont
  {Mezzenga}(2020)}]{Cao:2020}%
  \BibitemOpen
  \bibfield  {author} {\bibinfo {author} {\bibfnamefont {Y.}~\bibnamefont
  {Cao}}\ and\ \bibinfo {author} {\bibfnamefont {R.}~\bibnamefont {Mezzenga}},\
  }\bibfield  {title} {\bibinfo {title} {Design principles of food gels},\
  }\href {https://doi.org/10.1038/s43016-019-0009-x} {\bibfield  {journal}
  {\bibinfo  {journal} {Nature Food}\ }\textbf {\bibinfo {volume} {1}},\
  \bibinfo {pages} {106} (\bibinfo {year} {2020})}\BibitemShut {NoStop}%
\bibitem [{\citenamefont {Trappe}\ and\ \citenamefont
  {Sandk{\"u}hler}(2004)}]{trappe2004}%
  \BibitemOpen
  \bibfield  {author} {\bibinfo {author} {\bibfnamefont {V.}~\bibnamefont
  {Trappe}}\ and\ \bibinfo {author} {\bibfnamefont {P.}~\bibnamefont
  {Sandk{\"u}hler}},\ }\bibfield  {title} {\bibinfo {title} {Colloidal
  gels—low-density disordered solid-like states},\ }\href@noop {} {\bibfield
  {journal} {\bibinfo  {journal} {Current opinion in colloid \& interface
  science}\ }\textbf {\bibinfo {volume} {8}},\ \bibinfo {pages} {494} (\bibinfo
  {year} {2004})}\BibitemShut {NoStop}%
\bibitem [{\citenamefont {Zaccarelli}(2007)}]{Zaccarelli:2007}%
  \BibitemOpen
  \bibfield  {author} {\bibinfo {author} {\bibfnamefont {E.}~\bibnamefont
  {Zaccarelli}},\ }\bibfield  {title} {\bibinfo {title} {Colloidal gels:
  equilibrium and non-equilibrium routes},\ }\href@noop {} {\bibfield
  {journal} {\bibinfo  {journal} {Journal of Physics: Condensed Matter}\
  }\textbf {\bibinfo {volume} {19}},\ \bibinfo {pages} {323101} (\bibinfo
  {year} {2007})}\BibitemShut {NoStop}%
\bibitem [{\citenamefont {Weitz}\ and\ \citenamefont
  {Oliveria}(1984)}]{weitz1984}%
  \BibitemOpen
  \bibfield  {author} {\bibinfo {author} {\bibfnamefont {D.}~\bibnamefont
  {Weitz}}\ and\ \bibinfo {author} {\bibfnamefont {M.}~\bibnamefont
  {Oliveria}},\ }\bibfield  {title} {\bibinfo {title} {Fractal structures
  formed by kinetic aggregation of aqueous gold colloids},\ }\href@noop {}
  {\bibfield  {journal} {\bibinfo  {journal} {Physical review letters}\
  }\textbf {\bibinfo {volume} {52}},\ \bibinfo {pages} {1433} (\bibinfo {year}
  {1984})}\BibitemShut {NoStop}%
\bibitem [{\citenamefont {Schaefer}\ \emph {et~al.}(1984)\citenamefont
  {Schaefer}, \citenamefont {Martin}, \citenamefont {Wiltzius},\ and\
  \citenamefont {Cannell}}]{schaefer1984}%
  \BibitemOpen
  \bibfield  {author} {\bibinfo {author} {\bibfnamefont {D.~W.}\ \bibnamefont
  {Schaefer}}, \bibinfo {author} {\bibfnamefont {J.~E.}\ \bibnamefont
  {Martin}}, \bibinfo {author} {\bibfnamefont {P.}~\bibnamefont {Wiltzius}},\
  and\ \bibinfo {author} {\bibfnamefont {D.~S.}\ \bibnamefont {Cannell}},\
  }\bibfield  {title} {\bibinfo {title} {Fractal geometry of colloidal
  aggregates},\ }\href@noop {} {\bibfield  {journal} {\bibinfo  {journal}
  {Physical Review Letters}\ }\textbf {\bibinfo {volume} {52}},\ \bibinfo
  {pages} {2371} (\bibinfo {year} {1984})}\BibitemShut {NoStop}%
\bibitem [{\citenamefont {Cardinaux}\ \emph {et~al.}(2007)\citenamefont
  {Cardinaux}, \citenamefont {Gibaud}, \citenamefont {Stradner},\ and\
  \citenamefont {Schurtenberger}}]{cardinaux2007}%
  \BibitemOpen
  \bibfield  {author} {\bibinfo {author} {\bibfnamefont {F.}~\bibnamefont
  {Cardinaux}}, \bibinfo {author} {\bibfnamefont {T.}~\bibnamefont {Gibaud}},
  \bibinfo {author} {\bibfnamefont {A.}~\bibnamefont {Stradner}},\ and\
  \bibinfo {author} {\bibfnamefont {P.}~\bibnamefont {Schurtenberger}},\
  }\bibfield  {title} {\bibinfo {title} {Interplay between spinodal
  decomposition and glass formation in proteins exhibiting short-range
  attractions},\ }\href@noop {} {\bibfield  {journal} {\bibinfo  {journal}
  {Physical Review Letters}\ }\textbf {\bibinfo {volume} {99}},\ \bibinfo
  {pages} {118301} (\bibinfo {year} {2007})}\BibitemShut {NoStop}%
\bibitem [{\citenamefont {Lu}\ \emph {et~al.}(2008)\citenamefont {Lu},
  \citenamefont {Zaccarelli}, \citenamefont {Ciulla}, \citenamefont
  {Schofield}, \citenamefont {Sciortino},\ and\ \citenamefont
  {Weitz}}]{lu2008}%
  \BibitemOpen
  \bibfield  {author} {\bibinfo {author} {\bibfnamefont {P.~J.}\ \bibnamefont
  {Lu}}, \bibinfo {author} {\bibfnamefont {E.}~\bibnamefont {Zaccarelli}},
  \bibinfo {author} {\bibfnamefont {F.}~\bibnamefont {Ciulla}}, \bibinfo
  {author} {\bibfnamefont {A.~B.}\ \bibnamefont {Schofield}}, \bibinfo {author}
  {\bibfnamefont {F.}~\bibnamefont {Sciortino}},\ and\ \bibinfo {author}
  {\bibfnamefont {D.~A.}\ \bibnamefont {Weitz}},\ }\bibfield  {title} {\bibinfo
  {title} {Gelation of particles with short-range attraction},\ }\href@noop {}
  {\bibfield  {journal} {\bibinfo  {journal} {Nature}\ }\textbf {\bibinfo
  {volume} {453}},\ \bibinfo {pages} {499} (\bibinfo {year}
  {2008})}\BibitemShut {NoStop}%
\bibitem [{\citenamefont {Zia}\ \emph {et~al.}(2014)\citenamefont {Zia},
  \citenamefont {Landrum},\ and\ \citenamefont {Russel}}]{Zia:2014}%
  \BibitemOpen
  \bibfield  {author} {\bibinfo {author} {\bibfnamefont {R.~N.}\ \bibnamefont
  {Zia}}, \bibinfo {author} {\bibfnamefont {B.~J.}\ \bibnamefont {Landrum}},\
  and\ \bibinfo {author} {\bibfnamefont {W.~B.}\ \bibnamefont {Russel}},\
  }\bibfield  {title} {\bibinfo {title} {A micro-mechanical study of coarsening
  and rheology of colloidal gels: Cage building, cage hopping, and
  smoluchowski's ratchet},\ }\href@noop {} {\bibfield  {journal} {\bibinfo
  {journal} {J. Rheol.}\ }\textbf {\bibinfo {volume} {58}},\ \bibinfo {pages}
  {1121} (\bibinfo {year} {2014})}\BibitemShut {NoStop}%
\bibitem [{\citenamefont {Shih}\ \emph {et~al.}(1990)\citenamefont {Shih},
  \citenamefont {Shih}, \citenamefont {Kim}, \citenamefont {Liu},\ and\
  \citenamefont {Aksay}}]{shih1990}%
  \BibitemOpen
  \bibfield  {author} {\bibinfo {author} {\bibfnamefont {W.-H.}\ \bibnamefont
  {Shih}}, \bibinfo {author} {\bibfnamefont {W.~Y.}\ \bibnamefont {Shih}},
  \bibinfo {author} {\bibfnamefont {S.-I.}\ \bibnamefont {Kim}}, \bibinfo
  {author} {\bibfnamefont {J.}~\bibnamefont {Liu}},\ and\ \bibinfo {author}
  {\bibfnamefont {I.~A.}\ \bibnamefont {Aksay}},\ }\bibfield  {title} {\bibinfo
  {title} {Scaling behavior of the elastic properties of colloidal gels},\
  }\href@noop {} {\bibfield  {journal} {\bibinfo  {journal} {Physical review
  A}\ }\textbf {\bibinfo {volume} {42}},\ \bibinfo {pages} {4772} (\bibinfo
  {year} {1990})}\BibitemShut {NoStop}%
\bibitem [{\citenamefont {Krall}\ and\ \citenamefont
  {Weitz}(1998)}]{Krall:1998}%
  \BibitemOpen
  \bibfield  {author} {\bibinfo {author} {\bibfnamefont {A.~H.}\ \bibnamefont
  {Krall}}\ and\ \bibinfo {author} {\bibfnamefont {D.~A.}\ \bibnamefont
  {Weitz}},\ }\bibfield  {title} {\bibinfo {title} {Internal dynamics and
  elasticity of fractal colloidal gels},\ }\href@noop {} {\bibfield  {journal}
  {\bibinfo  {journal} {Phys. Rev. Lett.}\ }\textbf {\bibinfo {volume} {80}},\
  \bibinfo {pages} {778} (\bibinfo {year} {1998})}\BibitemShut {NoStop}%
\bibitem [{\citenamefont {Zaccone}\ \emph {et~al.}(2009)\citenamefont
  {Zaccone}, \citenamefont {Wu},\ and\ \citenamefont {Del~Gado}}]{zaccone2009}%
  \BibitemOpen
  \bibfield  {author} {\bibinfo {author} {\bibfnamefont {A.}~\bibnamefont
  {Zaccone}}, \bibinfo {author} {\bibfnamefont {H.}~\bibnamefont {Wu}},\ and\
  \bibinfo {author} {\bibfnamefont {E.}~\bibnamefont {Del~Gado}},\ }\bibfield
  {title} {\bibinfo {title} {Elasticity of arrested short-ranged attractive
  colloids: Homogeneous and heterogeneous glasses},\ }\href@noop {} {\bibfield
  {journal} {\bibinfo  {journal} {Physical review letters}\ }\textbf {\bibinfo
  {volume} {103}},\ \bibinfo {pages} {208301} (\bibinfo {year}
  {2009})}\BibitemShut {NoStop}%
\bibitem [{\citenamefont {Whitaker}\ \emph {et~al.}(2019)\citenamefont
  {Whitaker}, \citenamefont {Varga}, \citenamefont {Hsiao}, \citenamefont
  {Solomon}, \citenamefont {Swan},\ and\ \citenamefont {Furst}}]{whitaker2019}%
  \BibitemOpen
  \bibfield  {author} {\bibinfo {author} {\bibfnamefont {K.~A.}\ \bibnamefont
  {Whitaker}}, \bibinfo {author} {\bibfnamefont {Z.}~\bibnamefont {Varga}},
  \bibinfo {author} {\bibfnamefont {L.~C.}\ \bibnamefont {Hsiao}}, \bibinfo
  {author} {\bibfnamefont {M.~J.}\ \bibnamefont {Solomon}}, \bibinfo {author}
  {\bibfnamefont {J.~W.}\ \bibnamefont {Swan}},\ and\ \bibinfo {author}
  {\bibfnamefont {E.~M.}\ \bibnamefont {Furst}},\ }\bibfield  {title} {\bibinfo
  {title} {Colloidal gel elasticity arises from the packing of locally glassy
  clusters},\ }\href@noop {} {\bibfield  {journal} {\bibinfo  {journal} {Nature
  communications}\ }\textbf {\bibinfo {volume} {10}},\ \bibinfo {pages} {1}
  (\bibinfo {year} {2019})}\BibitemShut {NoStop}%
\bibitem [{\citenamefont {Gibaud}\ \emph {et~al.}(2013)\citenamefont {Gibaud},
  \citenamefont {Zaccone}, \citenamefont {Del~Gado}, \citenamefont {Trappe},\
  and\ \citenamefont {Schurtenberger}}]{gibaud2013}%
  \BibitemOpen
  \bibfield  {author} {\bibinfo {author} {\bibfnamefont {T.}~\bibnamefont
  {Gibaud}}, \bibinfo {author} {\bibfnamefont {A.}~\bibnamefont {Zaccone}},
  \bibinfo {author} {\bibfnamefont {E.}~\bibnamefont {Del~Gado}}, \bibinfo
  {author} {\bibfnamefont {V.}~\bibnamefont {Trappe}},\ and\ \bibinfo {author}
  {\bibfnamefont {P.}~\bibnamefont {Schurtenberger}},\ }\bibfield  {title}
  {\bibinfo {title} {Unexpected decoupling of stretching and bending modes in
  protein gels},\ }\href@noop {} {\bibfield  {journal} {\bibinfo  {journal}
  {Physical review letters}\ }\textbf {\bibinfo {volume} {110}},\ \bibinfo
  {pages} {058303} (\bibinfo {year} {2013})}\BibitemShut {NoStop}%
\bibitem [{\citenamefont {Gibaud}\ and\ \citenamefont
  {Schurtenberger}(2009)}]{gibaud2009}%
  \BibitemOpen
  \bibfield  {author} {\bibinfo {author} {\bibfnamefont {T.}~\bibnamefont
  {Gibaud}}\ and\ \bibinfo {author} {\bibfnamefont {P.}~\bibnamefont
  {Schurtenberger}},\ }\bibfield  {title} {\bibinfo {title} {A closer look at
  arrested spinodal decomposition in protein solutions},\ }\href@noop {}
  {\bibfield  {journal} {\bibinfo  {journal} {Journal of Physics: Condensed
  Matter}\ }\textbf {\bibinfo {volume} {21}},\ \bibinfo {pages} {322201}
  (\bibinfo {year} {2009})}\BibitemShut {NoStop}%
\bibitem [{\citenamefont {Koumakis}\ \emph {et~al.}(2015)\citenamefont
  {Koumakis}, \citenamefont {Moghimi}, \citenamefont {Besseling}, \citenamefont
  {Poon}, \citenamefont {Brady},\ and\ \citenamefont
  {Petekidis}}]{koumakis2015}%
  \BibitemOpen
  \bibfield  {author} {\bibinfo {author} {\bibfnamefont {N.}~\bibnamefont
  {Koumakis}}, \bibinfo {author} {\bibfnamefont {E.}~\bibnamefont {Moghimi}},
  \bibinfo {author} {\bibfnamefont {R.}~\bibnamefont {Besseling}}, \bibinfo
  {author} {\bibfnamefont {W.~C.}\ \bibnamefont {Poon}}, \bibinfo {author}
  {\bibfnamefont {J.~F.}\ \bibnamefont {Brady}},\ and\ \bibinfo {author}
  {\bibfnamefont {G.}~\bibnamefont {Petekidis}},\ }\bibfield  {title} {\bibinfo
  {title} {Tuning colloidal gels by shear},\ }\href@noop {} {\bibfield
  {journal} {\bibinfo  {journal} {Soft Matter}\ }\textbf {\bibinfo {volume}
  {11}},\ \bibinfo {pages} {4640} (\bibinfo {year} {2015})}\BibitemShut
  {NoStop}%
\bibitem [{\citenamefont {Helal}\ \emph {et~al.}(2016)\citenamefont {Helal},
  \citenamefont {Divoux},\ and\ \citenamefont {McKinley}}]{helal2016}%
  \BibitemOpen
  \bibfield  {author} {\bibinfo {author} {\bibfnamefont {A.}~\bibnamefont
  {Helal}}, \bibinfo {author} {\bibfnamefont {T.}~\bibnamefont {Divoux}},\ and\
  \bibinfo {author} {\bibfnamefont {G.~H.}\ \bibnamefont {McKinley}},\
  }\bibfield  {title} {\bibinfo {title} {Simultaneous rheoelectric measurements
  of strongly conductive complex fluids},\ }\href@noop {} {\bibfield  {journal}
  {\bibinfo  {journal} {Physical Review Applied}\ }\textbf {\bibinfo {volume}
  {6}},\ \bibinfo {pages} {064004} (\bibinfo {year} {2016})}\BibitemShut
  {NoStop}%
\bibitem [{\citenamefont {Utz}\ \emph {et~al.}(2000)\citenamefont {Utz},
  \citenamefont {Debenedetti},\ and\ \citenamefont {Stillinger}}]{Utz:2000}%
  \BibitemOpen
  \bibfield  {author} {\bibinfo {author} {\bibfnamefont {M.}~\bibnamefont
  {Utz}}, \bibinfo {author} {\bibfnamefont {P.~G.}\ \bibnamefont
  {Debenedetti}},\ and\ \bibinfo {author} {\bibfnamefont {F.~H.}\ \bibnamefont
  {Stillinger}},\ }\bibfield  {title} {\bibinfo {title} {Atomistic simulation
  of aging and rejuvenation in glasses},\ }\href@noop {} {\bibfield  {journal}
  {\bibinfo  {journal} {Phys. Rev. Lett.}\ }\textbf {\bibinfo {volume} {84}},\
  \bibinfo {pages} {1471} (\bibinfo {year} {2000})}\BibitemShut {NoStop}%
\bibitem [{\citenamefont {Bonn}\ \emph {et~al.}(2004)\citenamefont {Bonn},
  \citenamefont {Tanaka}, \citenamefont {Coussot},\ and\ \citenamefont
  {Meunier}}]{Bonn:2004}%
  \BibitemOpen
  \bibfield  {author} {\bibinfo {author} {\bibfnamefont {D.}~\bibnamefont
  {Bonn}}, \bibinfo {author} {\bibfnamefont {H.}~\bibnamefont {Tanaka}},
  \bibinfo {author} {\bibfnamefont {P.}~\bibnamefont {Coussot}},\ and\ \bibinfo
  {author} {\bibfnamefont {J.}~\bibnamefont {Meunier}},\ }\bibfield  {title}
  {\bibinfo {title} {Ageing, shear rejuvenation and avalanches in soft glassy
  materials},\ }\href@noop {} {\bibfield  {journal} {\bibinfo  {journal}
  {Journal of Physics: Condensed Matter}\ }\textbf {\bibinfo {volume} {16}},\
  \bibinfo {pages} {S4987} (\bibinfo {year} {2004})}\BibitemShut {NoStop}%
\bibitem [{\citenamefont {Moghimi}\ \emph {et~al.}(2017)\citenamefont
  {Moghimi}, \citenamefont {Jacob}, \citenamefont {Koumakis},\ and\
  \citenamefont {Petekidis}}]{moghimi2017}%
  \BibitemOpen
  \bibfield  {author} {\bibinfo {author} {\bibfnamefont {E.}~\bibnamefont
  {Moghimi}}, \bibinfo {author} {\bibfnamefont {A.~R.}\ \bibnamefont {Jacob}},
  \bibinfo {author} {\bibfnamefont {N.}~\bibnamefont {Koumakis}},\ and\
  \bibinfo {author} {\bibfnamefont {G.}~\bibnamefont {Petekidis}},\ }\bibfield
  {title} {\bibinfo {title} {Colloidal gels tuned by oscillatory shear},\
  }\href@noop {} {\bibfield  {journal} {\bibinfo  {journal} {Soft Matter}\
  }\textbf {\bibinfo {volume} {13}},\ \bibinfo {pages} {2371} (\bibinfo {year}
  {2017})}\BibitemShut {NoStop}%
\bibitem [{\citenamefont {Ovarlez}\ \emph {et~al.}(2013)\citenamefont
  {Ovarlez}, \citenamefont {Tocquer}, \citenamefont {Bertrand},\ and\
  \citenamefont {Coussot}}]{ovarlez2013}%
  \BibitemOpen
  \bibfield  {author} {\bibinfo {author} {\bibfnamefont {G.}~\bibnamefont
  {Ovarlez}}, \bibinfo {author} {\bibfnamefont {L.}~\bibnamefont {Tocquer}},
  \bibinfo {author} {\bibfnamefont {F.}~\bibnamefont {Bertrand}},\ and\
  \bibinfo {author} {\bibfnamefont {P.}~\bibnamefont {Coussot}},\ }\bibfield
  {title} {\bibinfo {title} {Rheopexy and tunable yield stress of carbon black
  suspensions},\ }\href@noop {} {\bibfield  {journal} {\bibinfo  {journal}
  {Soft Matter}\ }\textbf {\bibinfo {volume} {9}},\ \bibinfo {pages} {5540}
  (\bibinfo {year} {2013})}\BibitemShut {NoStop}%
\bibitem [{\citenamefont {Raney}\ \emph {et~al.}(2018)\citenamefont {Raney},
  \citenamefont {Compton}, \citenamefont {Mueller}, \citenamefont {Ober},
  \citenamefont {Shea},\ and\ \citenamefont {Lewis}}]{Raney:2018}%
  \BibitemOpen
  \bibfield  {author} {\bibinfo {author} {\bibfnamefont {J.~R.}\ \bibnamefont
  {Raney}}, \bibinfo {author} {\bibfnamefont {B.~G.}\ \bibnamefont {Compton}},
  \bibinfo {author} {\bibfnamefont {J.}~\bibnamefont {Mueller}}, \bibinfo
  {author} {\bibfnamefont {T.~J.}\ \bibnamefont {Ober}}, \bibinfo {author}
  {\bibfnamefont {K.}~\bibnamefont {Shea}},\ and\ \bibinfo {author}
  {\bibfnamefont {J.~A.}\ \bibnamefont {Lewis}},\ }\bibfield  {title} {\bibinfo
  {title} {Rotational 3d printing of damage-tolerant composites with
  programmable mechanics},\ }\href@noop {} {\bibfield  {journal} {\bibinfo
  {journal} {Proceedings of the National Academy of Sciences}\ }\textbf
  {\bibinfo {volume} {115}},\ \bibinfo {pages} {1198} (\bibinfo {year}
  {2018})}\BibitemShut {NoStop}%
\bibitem [{\citenamefont {Dag{\`e}s}\ \emph {et~al.}(2021)\citenamefont
  {Dag{\`e}s}, \citenamefont {Lidon}, \citenamefont {Jung}, \citenamefont
  {Pignon}, \citenamefont {Manneville},\ and\ \citenamefont
  {Gibaud}}]{dages2021}%
  \BibitemOpen
  \bibfield  {author} {\bibinfo {author} {\bibfnamefont {N.}~\bibnamefont
  {Dag{\`e}s}}, \bibinfo {author} {\bibfnamefont {P.}~\bibnamefont {Lidon}},
  \bibinfo {author} {\bibfnamefont {G.}~\bibnamefont {Jung}}, \bibinfo {author}
  {\bibfnamefont {F.}~\bibnamefont {Pignon}}, \bibinfo {author} {\bibfnamefont
  {S.}~\bibnamefont {Manneville}},\ and\ \bibinfo {author} {\bibfnamefont
  {T.}~\bibnamefont {Gibaud}},\ }\bibfield  {title} {\bibinfo {title}
  {Mechanics and structure of carbon black gels under high-power ultrasound},\
  }\href@noop {} {\bibfield  {journal} {\bibinfo  {journal} {Journal of
  Rheology}\ }\textbf {\bibinfo {volume} {65}},\ \bibinfo {pages} {477}
  (\bibinfo {year} {2021})}\BibitemShut {NoStop}%
\bibitem [{\citenamefont {Lahaye}\ and\ \citenamefont
  {Ehrburger-Dolle}(1994)}]{lahaye1994}%
  \BibitemOpen
  \bibfield  {author} {\bibinfo {author} {\bibfnamefont {J.}~\bibnamefont
  {Lahaye}}\ and\ \bibinfo {author} {\bibfnamefont {F.}~\bibnamefont
  {Ehrburger-Dolle}},\ }\bibfield  {title} {\bibinfo {title} {Mechanisms of
  carbon black formation. correlation with the morphology of aggregates},\
  }\href@noop {} {\bibfield  {journal} {\bibinfo  {journal} {Carbon}\ }\textbf
  {\bibinfo {volume} {32}},\ \bibinfo {pages} {1319} (\bibinfo {year}
  {1994})}\BibitemShut {NoStop}%
\bibitem [{\citenamefont {Xi}\ and\ \citenamefont {Zhong}(2006)}]{xi2006}%
  \BibitemOpen
  \bibfield  {author} {\bibinfo {author} {\bibfnamefont {J.}~\bibnamefont
  {Xi}}\ and\ \bibinfo {author} {\bibfnamefont {B.-J.}\ \bibnamefont {Zhong}},\
  }\bibfield  {title} {\bibinfo {title} {Soot in diesel combustion systems},\
  }\href@noop {} {\bibfield  {journal} {\bibinfo  {journal} {Chemical
  Engineering \& Technology: Industrial Chemistry-Plant Equipment-Process
  Engineering-Biotechnology}\ }\textbf {\bibinfo {volume} {29}},\ \bibinfo
  {pages} {665} (\bibinfo {year} {2006})}\BibitemShut {NoStop}%
\bibitem [{\citenamefont {Sztucki}\ \emph {et~al.}(2007)\citenamefont
  {Sztucki}, \citenamefont {Narayanan},\ and\ \citenamefont
  {Beaucage}}]{sztucki2007}%
  \BibitemOpen
  \bibfield  {author} {\bibinfo {author} {\bibfnamefont {M.}~\bibnamefont
  {Sztucki}}, \bibinfo {author} {\bibfnamefont {T.}~\bibnamefont {Narayanan}},\
  and\ \bibinfo {author} {\bibfnamefont {G.}~\bibnamefont {Beaucage}},\
  }\bibfield  {title} {\bibinfo {title} {In situ study of aggregation of soot
  particles in an acetylene flame by small-angle x-ray scattering},\
  }\href@noop {} {\bibfield  {journal} {\bibinfo  {journal} {Journal of applied
  physics}\ }\textbf {\bibinfo {volume} {101}},\ \bibinfo {pages} {114304}
  (\bibinfo {year} {2007})}\BibitemShut {NoStop}%
\bibitem [{\citenamefont {Wang}(2018)}]{wang2018}%
  \BibitemOpen
  \bibfield  {author} {\bibinfo {author} {\bibfnamefont {M.-J.}\ \bibnamefont
  {Wang}},\ }\href@noop {} {\emph {\bibinfo {title} {Carbon Black: Science and
  Technology}}}\ (\bibinfo  {publisher} {Routledge},\ \bibinfo {year}
  {2018})\BibitemShut {NoStop}%
\bibitem [{\citenamefont {Herd}\ \emph {et~al.}(1992)\citenamefont {Herd},
  \citenamefont {McDonald},\ and\ \citenamefont {Hess}}]{herd1992}%
  \BibitemOpen
  \bibfield  {author} {\bibinfo {author} {\bibfnamefont {C.~R.}\ \bibnamefont
  {Herd}}, \bibinfo {author} {\bibfnamefont {G.~C.}\ \bibnamefont {McDonald}},\
  and\ \bibinfo {author} {\bibfnamefont {W.~M.}\ \bibnamefont {Hess}},\
  }\bibfield  {title} {\bibinfo {title} {Morphology of carbon-black aggregates:
  fractal versus euclidean geometry},\ }\href@noop {} {\bibfield  {journal}
  {\bibinfo  {journal} {Rubber chemistry and technology}\ }\textbf {\bibinfo
  {volume} {65}},\ \bibinfo {pages} {107} (\bibinfo {year} {1992})}\BibitemShut
  {NoStop}%
\bibitem [{\citenamefont {Martinez}\ \emph {et~al.}(2017)\citenamefont
  {Martinez}, \citenamefont {Iturrondobeitia}, \citenamefont {Ibarretxe},\ and\
  \citenamefont {Guraya}}]{martinez2017}%
  \BibitemOpen
  \bibfield  {author} {\bibinfo {author} {\bibfnamefont {R.~F.}\ \bibnamefont
  {Martinez}}, \bibinfo {author} {\bibfnamefont {M.}~\bibnamefont
  {Iturrondobeitia}}, \bibinfo {author} {\bibfnamefont {J.}~\bibnamefont
  {Ibarretxe}},\ and\ \bibinfo {author} {\bibfnamefont {T.}~\bibnamefont
  {Guraya}},\ }\bibfield  {title} {\bibinfo {title} {Methodology to classify
  the shape of reinforcement fillers: optimization, evaluation, comparison, and
  selection of models},\ }\href@noop {} {\bibfield  {journal} {\bibinfo
  {journal} {Journal of Materials Science}\ }\textbf {\bibinfo {volume} {52}},\
  \bibinfo {pages} {569} (\bibinfo {year} {2017})}\BibitemShut {NoStop}%
\bibitem [{\citenamefont {Richards}\ \emph {et~al.}(2017)\citenamefont
  {Richards}, \citenamefont {Hipp}, \citenamefont {Riley}, \citenamefont
  {Wagner},\ and\ \citenamefont {Butler}}]{richards2017}%
  \BibitemOpen
  \bibfield  {author} {\bibinfo {author} {\bibfnamefont {J.~J.}\ \bibnamefont
  {Richards}}, \bibinfo {author} {\bibfnamefont {J.~B.}\ \bibnamefont {Hipp}},
  \bibinfo {author} {\bibfnamefont {J.~K.}\ \bibnamefont {Riley}}, \bibinfo
  {author} {\bibfnamefont {N.~J.}\ \bibnamefont {Wagner}},\ and\ \bibinfo
  {author} {\bibfnamefont {P.~D.}\ \bibnamefont {Butler}},\ }\bibfield  {title}
  {\bibinfo {title} {Clustering and percolation in suspensions of carbon
  black},\ }\href@noop {} {\bibfield  {journal} {\bibinfo  {journal}
  {Langmuir}\ }\textbf {\bibinfo {volume} {33}},\ \bibinfo {pages} {12260}
  (\bibinfo {year} {2017})}\BibitemShut {NoStop}%
\bibitem [{\citenamefont {Hipp}\ \emph {et~al.}(2021)\citenamefont {Hipp},
  \citenamefont {Richards},\ and\ \citenamefont {Wagner}}]{hipp2021}%
  \BibitemOpen
  \bibfield  {author} {\bibinfo {author} {\bibfnamefont {J.~B.}\ \bibnamefont
  {Hipp}}, \bibinfo {author} {\bibfnamefont {J.~J.}\ \bibnamefont {Richards}},\
  and\ \bibinfo {author} {\bibfnamefont {N.~J.}\ \bibnamefont {Wagner}},\
  }\bibfield  {title} {\bibinfo {title} {Direct measurements of the
  microstructural origin of shear-thinning in carbon black suspensions},\
  }\href@noop {} {\bibfield  {journal} {\bibinfo  {journal} {Journal of
  Rheology}\ }\textbf {\bibinfo {volume} {65}},\ \bibinfo {pages} {145}
  (\bibinfo {year} {2021})}\BibitemShut {NoStop}%
\bibitem [{\citenamefont {Trappe}\ \emph {et~al.}(2007)\citenamefont {Trappe},
  \citenamefont {Pitard}, \citenamefont {Ramos}, \citenamefont {Robert},
  \citenamefont {Bissig},\ and\ \citenamefont {Cipelletti}}]{trappe2007}%
  \BibitemOpen
  \bibfield  {author} {\bibinfo {author} {\bibfnamefont {V.}~\bibnamefont
  {Trappe}}, \bibinfo {author} {\bibfnamefont {E.}~\bibnamefont {Pitard}},
  \bibinfo {author} {\bibfnamefont {L.}~\bibnamefont {Ramos}}, \bibinfo
  {author} {\bibfnamefont {A.}~\bibnamefont {Robert}}, \bibinfo {author}
  {\bibfnamefont {H.}~\bibnamefont {Bissig}},\ and\ \bibinfo {author}
  {\bibfnamefont {L.}~\bibnamefont {Cipelletti}},\ }\bibfield  {title}
  {\bibinfo {title} {Investigation of q-dependent dynamical heterogeneity in a
  colloidal gel by x-ray photon correlation spectroscopy},\ }\href@noop {}
  {\bibfield  {journal} {\bibinfo  {journal} {Physical Review E}\ }\textbf
  {\bibinfo {volume} {76}},\ \bibinfo {pages} {051404} (\bibinfo {year}
  {2007})}\BibitemShut {NoStop}%
\bibitem [{\citenamefont {Prasad}\ \emph {et~al.}(2003)\citenamefont {Prasad},
  \citenamefont {Trappe}, \citenamefont {Dinsmore}, \citenamefont {Segre},
  \citenamefont {Cipelletti},\ and\ \citenamefont {Weitz}}]{prasad2003}%
  \BibitemOpen
  \bibfield  {author} {\bibinfo {author} {\bibfnamefont {V.}~\bibnamefont
  {Prasad}}, \bibinfo {author} {\bibfnamefont {V.}~\bibnamefont {Trappe}},
  \bibinfo {author} {\bibfnamefont {A.}~\bibnamefont {Dinsmore}}, \bibinfo
  {author} {\bibfnamefont {P.}~\bibnamefont {Segre}}, \bibinfo {author}
  {\bibfnamefont {L.}~\bibnamefont {Cipelletti}},\ and\ \bibinfo {author}
  {\bibfnamefont {D.}~\bibnamefont {Weitz}},\ }\bibfield  {title} {\bibinfo
  {title} {Rideal lecture universal features of the fluid to solid transition
  for attractive colloidal particles},\ }\href@noop {} {\bibfield  {journal}
  {\bibinfo  {journal} {Faraday discussions}\ }\textbf {\bibinfo {volume}
  {123}},\ \bibinfo {pages} {1} (\bibinfo {year} {2003})}\BibitemShut {NoStop}%
\bibitem [{\citenamefont {Herschel}\ and\ \citenamefont
  {Bulkley}(1926)}]{herschel1926}%
  \BibitemOpen
  \bibfield  {author} {\bibinfo {author} {\bibfnamefont {W.}~\bibnamefont
  {Herschel}}\ and\ \bibinfo {author} {\bibfnamefont {R.}~\bibnamefont
  {Bulkley}},\ }\bibfield  {title} {\bibinfo {title} {Measurement of
  consistency as applied to rubber-benzene solutions},\ }in\ \href@noop {}
  {\emph {\bibinfo {booktitle} {Am. Soc. Test Proc}}},\ Vol.~\bibinfo {volume}
  {26}\ (\bibinfo {year} {1926})\ pp.\ \bibinfo {pages} {621--633}\BibitemShut
  {NoStop}%
\bibitem [{\citenamefont {Fardin}\ \emph {et~al.}(2014)\citenamefont {Fardin},
  \citenamefont {Perge},\ and\ \citenamefont {Taberlet}}]{fardin2014}%
  \BibitemOpen
  \bibfield  {author} {\bibinfo {author} {\bibfnamefont {M.}~\bibnamefont
  {Fardin}}, \bibinfo {author} {\bibfnamefont {C.}~\bibnamefont {Perge}},\ and\
  \bibinfo {author} {\bibfnamefont {N.}~\bibnamefont {Taberlet}},\ }\bibfield
  {title} {\bibinfo {title} {“the hydrogen atom of fluid
  dynamics”--introduction to the taylor--couette flow for soft matter
  scientists},\ }\href@noop {} {\bibfield  {journal} {\bibinfo  {journal} {Soft
  Matter}\ }\textbf {\bibinfo {volume} {10}},\ \bibinfo {pages} {3523}
  (\bibinfo {year} {2014})}\BibitemShut {NoStop}%
\bibitem [{\citenamefont {Narayanan}\ \emph {et~al.}(2022)\citenamefont
  {Narayanan}, \citenamefont {Sztucki}, \citenamefont {Zinn}, \citenamefont
  {Kieffer}, \citenamefont {Homs-Puron}, \citenamefont {Gorini}, \citenamefont
  {Van~Vaerenbergh},\ and\ \citenamefont {Boesecke}}]{Narayanan2022}%
  \BibitemOpen
  \bibfield  {author} {\bibinfo {author} {\bibfnamefont {T.}~\bibnamefont
  {Narayanan}}, \bibinfo {author} {\bibfnamefont {M.}~\bibnamefont {Sztucki}},
  \bibinfo {author} {\bibfnamefont {T.}~\bibnamefont {Zinn}}, \bibinfo {author}
  {\bibfnamefont {J.}~\bibnamefont {Kieffer}}, \bibinfo {author} {\bibfnamefont
  {A.}~\bibnamefont {Homs-Puron}}, \bibinfo {author} {\bibfnamefont
  {J.}~\bibnamefont {Gorini}}, \bibinfo {author} {\bibfnamefont
  {P.}~\bibnamefont {Van~Vaerenbergh}},\ and\ \bibinfo {author} {\bibfnamefont
  {P.}~\bibnamefont {Boesecke}},\ }\bibfield  {title} {\bibinfo {title}
  {Performance of the time-resolved ultra-small-angle x-ray scattering beamline
  with the extremely brilliant source},\ }\href@noop {} {\bibfield  {journal}
  {\bibinfo  {journal} {Journal of Applied Crystallography}\ }\textbf {\bibinfo
  {volume} {55}},\ \bibinfo {pages} {98} (\bibinfo {year} {2022})}\BibitemShut
  {NoStop}%
\bibitem [{\citenamefont {Panine}\ \emph {et~al.}(2003)\citenamefont {Panine},
  \citenamefont {Gradzielski},\ and\ \citenamefont {Narayanan}}]{Panine2003}%
  \BibitemOpen
  \bibfield  {author} {\bibinfo {author} {\bibfnamefont {P.}~\bibnamefont
  {Panine}}, \bibinfo {author} {\bibfnamefont {M.}~\bibnamefont
  {Gradzielski}},\ and\ \bibinfo {author} {\bibfnamefont {T.}~\bibnamefont
  {Narayanan}},\ }\bibfield  {title} {\bibinfo {title} {Combined rheometry and
  small-angle x-ray scattering},\ }\href@noop {} {\bibfield  {journal}
  {\bibinfo  {journal} {Review of Scientific Instruments}\ }\textbf {\bibinfo
  {volume} {74}},\ \bibinfo {pages} {2451} (\bibinfo {year}
  {2003})}\BibitemShut {NoStop}%
\bibitem [{\citenamefont {Trappe}\ and\ \citenamefont
  {Weitz}(2000)}]{trappe2000}%
  \BibitemOpen
  \bibfield  {author} {\bibinfo {author} {\bibfnamefont {V.}~\bibnamefont
  {Trappe}}\ and\ \bibinfo {author} {\bibfnamefont {D.}~\bibnamefont {Weitz}},\
  }\bibfield  {title} {\bibinfo {title} {Scaling of the viscoelasticity of
  weakly attractive particles},\ }\href@noop {} {\bibfield  {journal} {\bibinfo
   {journal} {Physical review letters}\ }\textbf {\bibinfo {volume} {85}},\
  \bibinfo {pages} {449} (\bibinfo {year} {2000})}\BibitemShut {NoStop}%
\bibitem [{\citenamefont {Dullaert}\ and\ \citenamefont
  {Mewis}(2005)}]{dullaert2005}%
  \BibitemOpen
  \bibfield  {author} {\bibinfo {author} {\bibfnamefont {K.}~\bibnamefont
  {Dullaert}}\ and\ \bibinfo {author} {\bibfnamefont {J.}~\bibnamefont
  {Mewis}},\ }\bibfield  {title} {\bibinfo {title} {Thixotropy: Build-up and
  breakdown curves during flow},\ }\href@noop {} {\bibfield  {journal}
  {\bibinfo  {journal} {Journal of Rheology}\ }\textbf {\bibinfo {volume}
  {49}},\ \bibinfo {pages} {1213} (\bibinfo {year} {2005})}\BibitemShut
  {NoStop}%
\bibitem [{\citenamefont {Meeker}\ \emph {et~al.}(2004)\citenamefont {Meeker},
  \citenamefont {Bonnecaze},\ and\ \citenamefont {Cloitre}}]{meeker2004}%
  \BibitemOpen
  \bibfield  {author} {\bibinfo {author} {\bibfnamefont {S.~P.}\ \bibnamefont
  {Meeker}}, \bibinfo {author} {\bibfnamefont {R.~T.}\ \bibnamefont
  {Bonnecaze}},\ and\ \bibinfo {author} {\bibfnamefont {M.}~\bibnamefont
  {Cloitre}},\ }\bibfield  {title} {\bibinfo {title} {Slip and flow in pastes
  of soft particles: Direct observation and rheology},\ }\href@noop {}
  {\bibfield  {journal} {\bibinfo  {journal} {Journal of Rheology}\ }\textbf
  {\bibinfo {volume} {48}},\ \bibinfo {pages} {1295} (\bibinfo {year}
  {2004})}\BibitemShut {NoStop}%
\bibitem [{\citenamefont {Z{\"o}lzer}\ and\ \citenamefont
  {Eicke}(1993)}]{Zolzer:1993}%
  \BibitemOpen
  \bibfield  {author} {\bibinfo {author} {\bibfnamefont {U.}~\bibnamefont
  {Z{\"o}lzer}}\ and\ \bibinfo {author} {\bibfnamefont {H.-F.}\ \bibnamefont
  {Eicke}},\ }\bibfield  {title} {\bibinfo {title} {Free oscillatory shear
  measurements—an interesting application of constant stress rheometers in
  the creep mode},\ }\href@noop {} {\bibfield  {journal} {\bibinfo  {journal}
  {Rheol. acta}\ }\textbf {\bibinfo {volume} {32}},\ \bibinfo {pages} {104}
  (\bibinfo {year} {1993})}\BibitemShut {NoStop}%
\bibitem [{\citenamefont {Baravian}\ and\ \citenamefont
  {Quemada}(1998)}]{Baravian:1998}%
  \BibitemOpen
  \bibfield  {author} {\bibinfo {author} {\bibfnamefont {C.}~\bibnamefont
  {Baravian}}\ and\ \bibinfo {author} {\bibfnamefont {D.}~\bibnamefont
  {Quemada}},\ }\bibfield  {title} {\bibinfo {title} {Using instrumental
  inertia in controlled stress rheometry},\ }\href@noop {} {\bibfield
  {journal} {\bibinfo  {journal} {Rheol Acta}\ }\textbf {\bibinfo {volume}
  {37}},\ \bibinfo {pages} {223} (\bibinfo {year} {1998})}\BibitemShut
  {NoStop}%
\bibitem [{\citenamefont {Ewoldt}\ and\ \citenamefont
  {McKinley}(2007)}]{Ewoldt:2007}%
  \BibitemOpen
  \bibfield  {author} {\bibinfo {author} {\bibfnamefont {R.~H.}\ \bibnamefont
  {Ewoldt}}\ and\ \bibinfo {author} {\bibfnamefont {G.~H.}\ \bibnamefont
  {McKinley}},\ }\bibfield  {title} {\bibinfo {title} {Creep ringing in
  rheometry or how to deal with oft-discarded data in step stress tests!},\
  }\href@noop {} {\bibfield  {journal} {\bibinfo  {journal} {Rheology
  Bulletin}\ }\textbf {\bibinfo {volume} {76}},\ \bibinfo {pages} {4} (\bibinfo
  {year} {2007})}\BibitemShut {NoStop}%
\bibitem [{\citenamefont {Benmouffok-Benbelkacem}\ \emph
  {et~al.}(2010)\citenamefont {Benmouffok-Benbelkacem}, \citenamefont {Caton},
  \citenamefont {Baravian},\ and\ \citenamefont
  {Skali-Lami}}]{Benmouffok:2010}%
  \BibitemOpen
  \bibfield  {author} {\bibinfo {author} {\bibfnamefont {G.}~\bibnamefont
  {Benmouffok-Benbelkacem}}, \bibinfo {author} {\bibfnamefont {F.}~\bibnamefont
  {Caton}}, \bibinfo {author} {\bibfnamefont {C.}~\bibnamefont {Baravian}},\
  and\ \bibinfo {author} {\bibfnamefont {S.}~\bibnamefont {Skali-Lami}},\
  }\bibfield  {title} {\bibinfo {title} {Non-linear viscoelasticity and
  temporal behavior of typical yield stress fluids: Carbopol, xanthan
  andketchup},\ }\href@noop {} {\bibfield  {journal} {\bibinfo  {journal}
  {Rheol Acta}\ }\textbf {\bibinfo {volume} {49}},\ \bibinfo {pages} {305}
  (\bibinfo {year} {2010})}\BibitemShut {NoStop}%
\bibitem [{\citenamefont {Radhakrishnan}\ \emph {et~al.}(2017)\citenamefont
  {Radhakrishnan}, \citenamefont {Divoux}, \citenamefont {Manneville},\ and\
  \citenamefont {Fielding}}]{radhakrishnan2017}%
  \BibitemOpen
  \bibfield  {author} {\bibinfo {author} {\bibfnamefont {R.}~\bibnamefont
  {Radhakrishnan}}, \bibinfo {author} {\bibfnamefont {T.}~\bibnamefont
  {Divoux}}, \bibinfo {author} {\bibfnamefont {S.}~\bibnamefont {Manneville}},\
  and\ \bibinfo {author} {\bibfnamefont {S.~M.}\ \bibnamefont {Fielding}},\
  }\bibfield  {title} {\bibinfo {title} {Understanding rheological hysteresis
  in soft glassy materials},\ }\href@noop {} {\bibfield  {journal} {\bibinfo
  {journal} {Soft Matter}\ }\textbf {\bibinfo {volume} {13}},\ \bibinfo {pages}
  {1834} (\bibinfo {year} {2017})}\BibitemShut {NoStop}%
\bibitem [{\citenamefont {Won}\ \emph {et~al.}(2005)\citenamefont {Won},
  \citenamefont {Meeker}, \citenamefont {Trappe}, \citenamefont {Weitz},
  \citenamefont {Diggs},\ and\ \citenamefont {Emert}}]{won2005}%
  \BibitemOpen
  \bibfield  {author} {\bibinfo {author} {\bibfnamefont {Y.-Y.}\ \bibnamefont
  {Won}}, \bibinfo {author} {\bibfnamefont {S.~P.}\ \bibnamefont {Meeker}},
  \bibinfo {author} {\bibfnamefont {V.}~\bibnamefont {Trappe}}, \bibinfo
  {author} {\bibfnamefont {D.~A.}\ \bibnamefont {Weitz}}, \bibinfo {author}
  {\bibfnamefont {N.~Z.}\ \bibnamefont {Diggs}},\ and\ \bibinfo {author}
  {\bibfnamefont {J.~I.}\ \bibnamefont {Emert}},\ }\bibfield  {title} {\bibinfo
  {title} {Effect of temperature on carbon-black agglomeration in hydrocarbon
  liquid with adsorbed dispersant},\ }\href@noop {} {\bibfield  {journal}
  {\bibinfo  {journal} {Langmuir}\ }\textbf {\bibinfo {volume} {21}},\ \bibinfo
  {pages} {924} (\bibinfo {year} {2005})}\BibitemShut {NoStop}%
\bibitem [{\citenamefont {Sudreau}\ \emph {et~al.}(2022)\citenamefont
  {Sudreau}, \citenamefont {Manneville}, \citenamefont {Servel},\ and\
  \citenamefont {Divoux}}]{Sudreau:2022}%
  \BibitemOpen
  \bibfield  {author} {\bibinfo {author} {\bibfnamefont {I.}~\bibnamefont
  {Sudreau}}, \bibinfo {author} {\bibfnamefont {S.}~\bibnamefont {Manneville}},
  \bibinfo {author} {\bibfnamefont {M.}~\bibnamefont {Servel}},\ and\ \bibinfo
  {author} {\bibfnamefont {T.}~\bibnamefont {Divoux}},\ }\bibfield  {title}
  {\bibinfo {title} {Shear-induced memory effects in boehmite gels},\
  }\href@noop {} {\bibfield  {journal} {\bibinfo  {journal} {Journal of
  Rheology}\ }\textbf {\bibinfo {volume} {66}},\ \bibinfo {pages} {91}
  (\bibinfo {year} {2022})}\BibitemShut {NoStop}%
\bibitem [{\citenamefont {Das}\ and\ \citenamefont
  {Petekidis}(2022)}]{das2022}%
  \BibitemOpen
  \bibfield  {author} {\bibinfo {author} {\bibfnamefont {M.}~\bibnamefont
  {Das}}\ and\ \bibinfo {author} {\bibfnamefont {G.}~\bibnamefont
  {Petekidis}},\ }\bibfield  {title} {\bibinfo {title} {Shear induced tuning
  and memory effects in colloidal gels of rods and spheres},\ }\href@noop {}
  {\bibfield  {journal} {\bibinfo  {journal} {arXiv:2207.05185}\ } (\bibinfo
  {year} {2022})}\BibitemShut {NoStop}%
\bibitem [{\citenamefont {Buscall}\ \emph {et~al.}(1988)\citenamefont
  {Buscall}, \citenamefont {Mills}, \citenamefont {Goodwin},\ and\
  \citenamefont {Lawson}}]{buscall1988}%
  \BibitemOpen
  \bibfield  {author} {\bibinfo {author} {\bibfnamefont {R.}~\bibnamefont
  {Buscall}}, \bibinfo {author} {\bibfnamefont {P.~D.}\ \bibnamefont {Mills}},
  \bibinfo {author} {\bibfnamefont {J.~W.}\ \bibnamefont {Goodwin}},\ and\
  \bibinfo {author} {\bibfnamefont {D.}~\bibnamefont {Lawson}},\ }\bibfield
  {title} {\bibinfo {title} {Scaling behaviour of the rheology of aggregate
  networks formed from colloidal particles},\ }\href@noop {} {\bibfield
  {journal} {\bibinfo  {journal} {Journal of the Chemical Society, Faraday
  Transactions 1: Physical Chemistry in Condensed Phases}\ }\textbf {\bibinfo
  {volume} {84}},\ \bibinfo {pages} {4249} (\bibinfo {year}
  {1988})}\BibitemShut {NoStop}%
\bibitem [{\citenamefont {Courtens}\ and\ \citenamefont
  {Vacher}(1987)}]{courtens1987}%
  \BibitemOpen
  \bibfield  {author} {\bibinfo {author} {\bibfnamefont {E.}~\bibnamefont
  {Courtens}}\ and\ \bibinfo {author} {\bibfnamefont {R.}~\bibnamefont
  {Vacher}},\ }\bibfield  {title} {\bibinfo {title} {Structure and dynamics of
  fractal aerogels},\ }\href@noop {} {\bibfield  {journal} {\bibinfo  {journal}
  {Zeitschrift f{\"u}r Physik B Condensed Matter}\ }\textbf {\bibinfo {volume}
  {68}},\ \bibinfo {pages} {355} (\bibinfo {year} {1987})}\BibitemShut
  {NoStop}%
\bibitem [{\citenamefont {Beaucage}(1995)}]{beaucage1995}%
  \BibitemOpen
  \bibfield  {author} {\bibinfo {author} {\bibfnamefont {G.}~\bibnamefont
  {Beaucage}},\ }\bibfield  {title} {\bibinfo {title} {Approximations leading
  to a unified exponential/power-law approach to small-angle scattering},\
  }\href@noop {} {\bibfield  {journal} {\bibinfo  {journal} {Journal of Applied
  Crystallography}\ }\textbf {\bibinfo {volume} {28}},\ \bibinfo {pages} {717}
  (\bibinfo {year} {1995})}\BibitemShut {NoStop}%
\bibitem [{\citenamefont {Beaucage}(1996)}]{beaucage1996}%
  \BibitemOpen
  \bibfield  {author} {\bibinfo {author} {\bibfnamefont {G.}~\bibnamefont
  {Beaucage}},\ }\bibfield  {title} {\bibinfo {title} {Small-angle scattering
  from polymeric mass fractals of arbitrary mass-fractal dimension},\
  }\href@noop {} {\bibfield  {journal} {\bibinfo  {journal} {Journal of applied
  crystallography}\ }\textbf {\bibinfo {volume} {29}},\ \bibinfo {pages} {134}
  (\bibinfo {year} {1996})}\BibitemShut {NoStop}%
\bibitem [{\citenamefont {Wang}\ and\ \citenamefont {Ewoldt}(2022)}]{wang2022}%
  \BibitemOpen
  \bibfield  {author} {\bibinfo {author} {\bibfnamefont {Y.}~\bibnamefont
  {Wang}}\ and\ \bibinfo {author} {\bibfnamefont {R.~H.}\ \bibnamefont
  {Ewoldt}},\ }\bibfield  {title} {\bibinfo {title} {New insights on carbon
  black suspension rheology--anisotropic thixotropy and anti-thixotropy},\
  }\href@noop {} {\bibfield  {journal} {\bibinfo  {journal} {arXiv preprint
  arXiv:2202.05772}\ } (\bibinfo {year} {2022})}\BibitemShut {NoStop}%
\bibitem [{\citenamefont {Wessel}\ and\ \citenamefont
  {Ball}(1992)}]{wessel1992}%
  \BibitemOpen
  \bibfield  {author} {\bibinfo {author} {\bibfnamefont {R.}~\bibnamefont
  {Wessel}}\ and\ \bibinfo {author} {\bibfnamefont {R.}~\bibnamefont {Ball}},\
  }\bibfield  {title} {\bibinfo {title} {Fractal aggregates and gels in shear
  flow},\ }\href@noop {} {\bibfield  {journal} {\bibinfo  {journal} {Physical
  Review A}\ }\textbf {\bibinfo {volume} {46}},\ \bibinfo {pages} {R3008}
  (\bibinfo {year} {1992})}\BibitemShut {NoStop}%
\bibitem [{\citenamefont {Kantor}\ and\ \citenamefont
  {Witten}(1984)}]{Kantor1984a}%
  \BibitemOpen
  \bibfield  {author} {\bibinfo {author} {\bibfnamefont {Y.}~\bibnamefont
  {Kantor}}\ and\ \bibinfo {author} {\bibfnamefont {T.~A.}\ \bibnamefont
  {Witten}},\ }\bibfield  {title} {\bibinfo {title} {Mechanical stability of
  tenuous objects},\ }\href
  {https://doi.org/10.1051/jphyslet:019840045013067500} {\bibfield  {journal}
  {\bibinfo  {journal} {Journal de Physique Lettres}\ }\textbf {\bibinfo
  {volume} {45}},\ \bibinfo {pages} {675} (\bibinfo {year} {1984})}\BibitemShut
  {NoStop}%
\bibitem [{\citenamefont {Kantor}\ and\ \citenamefont
  {Webman}(1984)}]{Kantor1984b}%
  \BibitemOpen
  \bibfield  {author} {\bibinfo {author} {\bibfnamefont {Y.}~\bibnamefont
  {Kantor}}\ and\ \bibinfo {author} {\bibfnamefont {I.}~\bibnamefont
  {Webman}},\ }\bibfield  {title} {\bibinfo {title} {{Elastic Properties of
  Random Percolating Systems}},\ }\href
  {https://doi.org/10.1103/PhysRevLett.52.1891} {\bibfield  {journal} {\bibinfo
   {journal} {Physical Review letters}\ }\textbf {\bibinfo {volume} {52}},\
  \bibinfo {pages} {1891} (\bibinfo {year} {1984})}\BibitemShut {NoStop}%
\bibitem [{\citenamefont {Mewis}\ and\ \citenamefont
  {Wagner}(2012)}]{Mewis2012}%
  \BibitemOpen
  \bibfield  {author} {\bibinfo {author} {\bibfnamefont {J.}~\bibnamefont
  {Mewis}}\ and\ \bibinfo {author} {\bibfnamefont {N.~J.}\ \bibnamefont
  {Wagner}},\ }\href {https://doi.org/10.1017/CBO9780511977978} {\emph
  {\bibinfo {title} {{Colloidal Suspension Rheology}}}},\ edited by\ \bibinfo
  {editor} {\bibfnamefont {J.}~\bibnamefont {Mewis}}\ and\ \bibinfo {editor}
  {\bibfnamefont {N.~J.}\ \bibnamefont {Wagner}}\ (\bibinfo  {publisher}
  {Cambridge Univesity Press},\ \bibinfo {year} {2012})\BibitemShut {NoStop}%
\bibitem [{\citenamefont {Marangoni}(2000)}]{Marangoni2000}%
  \BibitemOpen
  \bibfield  {author} {\bibinfo {author} {\bibfnamefont {A.~G.}\ \bibnamefont
  {Marangoni}},\ }\bibfield  {title} {\bibinfo {title} {{Elasticity of
  high-volume-fraction fractal aggregate networks: A thermodynamic approach}},\
  }\href {https://doi.org/10.1103/PhysRevB.62.13951} {\bibfield  {journal}
  {\bibinfo  {journal} {Physical Review B}\ }\textbf {\bibinfo {volume} {62}},\
  \bibinfo {pages} {13951} (\bibinfo {year} {2000})}\BibitemShut {NoStop}%
\bibitem [{\citenamefont {Gravelle}\ and\ \citenamefont
  {Marangoni}(2021)}]{Marangoni2021}%
  \BibitemOpen
  \bibfield  {author} {\bibinfo {author} {\bibfnamefont {A.~J.}\ \bibnamefont
  {Gravelle}}\ and\ \bibinfo {author} {\bibfnamefont {A.~G.}\ \bibnamefont
  {Marangoni}},\ }\bibfield  {title} {\bibinfo {title} {A new fractal
  structural-mechanical theory of particle-filled colloidal networks with
  heterogeneous stress translation},\ }\href
  {https://doi.org/10.1016/j.jcis.2021.03.180} {\bibfield  {journal} {\bibinfo
  {journal} {Journal of Colloid and Interface Science}\ }\textbf {\bibinfo
  {volume} {598}},\ \bibinfo {pages} {56} (\bibinfo {year} {2021})}\BibitemShut
  {NoStop}%
\bibitem [{\citenamefont {Roldughin}(2003)}]{Roldughin2003}%
  \BibitemOpen
  \bibfield  {author} {\bibinfo {author} {\bibfnamefont {V.~I.}\ \bibnamefont
  {Roldughin}},\ }\bibfield  {title} {\bibinfo {title} {The characteristics of
  fractal disperse systems},\ }\href
  {https://doi.org/10.1070/RC2003v072n11ABEH000829} {\bibfield  {journal}
  {\bibinfo  {journal} {Russian Chemical Reviews}\ }\textbf {\bibinfo {volume}
  {72}},\ \bibinfo {pages} {913} (\bibinfo {year} {2003})}\BibitemShut
  {NoStop}%
\bibitem [{\citenamefont {Mellema}\ \emph {et~al.}(2002)\citenamefont
  {Mellema}, \citenamefont {van Opheusde},\ and\ \citenamefont {van
  Vliet}}]{Mellema2002}%
  \BibitemOpen
  \bibfield  {author} {\bibinfo {author} {\bibfnamefont {M.}~\bibnamefont
  {Mellema}}, \bibinfo {author} {\bibfnamefont {J.~H.~J.}\ \bibnamefont {van
  Opheusde}},\ and\ \bibinfo {author} {\bibfnamefont {T.}~\bibnamefont {van
  Vliet}},\ }\bibfield  {title} {\bibinfo {title} {Categorization of
  rheological scaling models for particlegels applied to casein gels},\ }\href
  {https://doi.org/10.1122/1.1423311} {\bibfield  {journal} {\bibinfo
  {journal} {Journal of Rheology}\ }\textbf {\bibinfo {volume} {46}},\ \bibinfo
  {pages} {11} (\bibinfo {year} {2002})}\BibitemShut {NoStop}%
\bibitem [{\citenamefont {Wu}\ and\ \citenamefont {Morbidelli}(2001)}]{Wu2001}%
  \BibitemOpen
  \bibfield  {author} {\bibinfo {author} {\bibfnamefont {H.}~\bibnamefont
  {Wu}}\ and\ \bibinfo {author} {\bibfnamefont {M.}~\bibnamefont
  {Morbidelli}},\ }\bibfield  {title} {\bibinfo {title} {{Model Relating
  Structure of Colloidal Gels to Their Elastic Properties}},\ }\href
  {https://doi.org/10.1021/la001121f} {\bibfield  {journal} {\bibinfo
  {journal} {Langmuir}\ }\textbf {\bibinfo {volume} {17}},\ \bibinfo {pages}
  {1030} (\bibinfo {year} {2001})}\BibitemShut {NoStop}%
\bibitem [{\citenamefont {Van~Gurp}\ and\ \citenamefont
  {Palmen}(1998)}]{van1998}%
  \BibitemOpen
  \bibfield  {author} {\bibinfo {author} {\bibfnamefont {M.}~\bibnamefont
  {Van~Gurp}}\ and\ \bibinfo {author} {\bibfnamefont {J.}~\bibnamefont
  {Palmen}},\ }\bibfield  {title} {\bibinfo {title} {Time-temperature
  superposition for polymeric blends},\ }\href@noop {} {\bibfield  {journal}
  {\bibinfo  {journal} {Rheol. Bull}\ }\textbf {\bibinfo {volume} {67}},\
  \bibinfo {pages} {5} (\bibinfo {year} {1998})}\BibitemShut {NoStop}%
\bibitem [{\citenamefont {Huang}\ \emph {et~al.}(2021)\citenamefont {Huang},
  \citenamefont {Yang}, \citenamefont {Lin}, \citenamefont {Su},\ and\
  \citenamefont {Hua}}]{huang2021}%
  \BibitemOpen
  \bibfield  {author} {\bibinfo {author} {\bibfnamefont {S.-T.}\ \bibnamefont
  {Huang}}, \bibinfo {author} {\bibfnamefont {C.-H.}\ \bibnamefont {Yang}},
  \bibinfo {author} {\bibfnamefont {P.-J.}\ \bibnamefont {Lin}}, \bibinfo
  {author} {\bibfnamefont {C.-Y.}\ \bibnamefont {Su}},\ and\ \bibinfo {author}
  {\bibfnamefont {C.-C.}\ \bibnamefont {Hua}},\ }\bibfield  {title} {\bibinfo
  {title} {Multiscale structural and rheological features of colloidal
  low-methoxyl pectin solutions and calcium-induced sol--gel transition},\
  }\href@noop {} {\bibfield  {journal} {\bibinfo  {journal} {Physical Chemistry
  Chemical Physics}\ }\textbf {\bibinfo {volume} {23}},\ \bibinfo {pages}
  {19269} (\bibinfo {year} {2021})}\BibitemShut {NoStop}%
\bibitem [{\citenamefont {Jawerth}\ \emph {et~al.}(2020)\citenamefont
  {Jawerth}, \citenamefont {Fischer-Friedrich}, \citenamefont {Saha},
  \citenamefont {Wang}, \citenamefont {Franzmann}, \citenamefont {Zhang},
  \citenamefont {Sachweh}, \citenamefont {Ruer}, \citenamefont {Ijavi},
  \citenamefont {Saha} \emph {et~al.}}]{jawerth2020}%
  \BibitemOpen
  \bibfield  {author} {\bibinfo {author} {\bibfnamefont {L.}~\bibnamefont
  {Jawerth}}, \bibinfo {author} {\bibfnamefont {E.}~\bibnamefont
  {Fischer-Friedrich}}, \bibinfo {author} {\bibfnamefont {S.}~\bibnamefont
  {Saha}}, \bibinfo {author} {\bibfnamefont {J.}~\bibnamefont {Wang}}, \bibinfo
  {author} {\bibfnamefont {T.}~\bibnamefont {Franzmann}}, \bibinfo {author}
  {\bibfnamefont {X.}~\bibnamefont {Zhang}}, \bibinfo {author} {\bibfnamefont
  {J.}~\bibnamefont {Sachweh}}, \bibinfo {author} {\bibfnamefont
  {M.}~\bibnamefont {Ruer}}, \bibinfo {author} {\bibfnamefont {M.}~\bibnamefont
  {Ijavi}}, \bibinfo {author} {\bibfnamefont {S.}~\bibnamefont {Saha}}, \emph
  {et~al.},\ }\bibfield  {title} {\bibinfo {title} {Protein condensates as
  aging maxwell fluids},\ }\href@noop {} {\bibfield  {journal} {\bibinfo
  {journal} {Science}\ }\textbf {\bibinfo {volume} {370}},\ \bibinfo {pages}
  {1317} (\bibinfo {year} {2020})}\BibitemShut {NoStop}%
\bibitem [{\citenamefont {Krishnan}\ \emph {et~al.}(2010)\citenamefont
  {Krishnan}, \citenamefont {Seifert}, \citenamefont {Lee}, \citenamefont
  {Khan},\ and\ \citenamefont {Spontak}}]{krishnan2010}%
  \BibitemOpen
  \bibfield  {author} {\bibinfo {author} {\bibfnamefont {A.~S.}\ \bibnamefont
  {Krishnan}}, \bibinfo {author} {\bibfnamefont {S.}~\bibnamefont {Seifert}},
  \bibinfo {author} {\bibfnamefont {B.}~\bibnamefont {Lee}}, \bibinfo {author}
  {\bibfnamefont {S.~A.}\ \bibnamefont {Khan}},\ and\ \bibinfo {author}
  {\bibfnamefont {R.~J.}\ \bibnamefont {Spontak}},\ }\bibfield  {title}
  {\bibinfo {title} {Cosolvent-regulated time--composition rheological
  equivalence in block copolymer solutions},\ }\href@noop {} {\bibfield
  {journal} {\bibinfo  {journal} {Soft Matter}\ }\textbf {\bibinfo {volume}
  {6}},\ \bibinfo {pages} {4331} (\bibinfo {year} {2010})}\BibitemShut
  {NoStop}%
\bibitem [{\citenamefont {Wen}\ \emph {et~al.}(2015)\citenamefont {Wen},
  \citenamefont {Schaefer},\ and\ \citenamefont {Archer}}]{wen2015}%
  \BibitemOpen
  \bibfield  {author} {\bibinfo {author} {\bibfnamefont {Y.~H.}\ \bibnamefont
  {Wen}}, \bibinfo {author} {\bibfnamefont {J.~L.}\ \bibnamefont {Schaefer}},\
  and\ \bibinfo {author} {\bibfnamefont {L.~A.}\ \bibnamefont {Archer}},\
  }\bibfield  {title} {\bibinfo {title} {Dynamics and rheology of soft
  colloidal glassestrappe2007},\ }\href@noop {} {\bibfield  {journal} {\bibinfo
   {journal} {Acs Macro Letters}\ }\textbf {\bibinfo {volume} {4}},\ \bibinfo
  {pages} {119} (\bibinfo {year} {2015})}\BibitemShut {NoStop}%
\bibitem [{\citenamefont {Wen}\ \emph {et~al.}(2014)\citenamefont {Wen},
  \citenamefont {Lu}, \citenamefont {Dobosz},\ and\ \citenamefont
  {Archer}}]{wen2014}%
  \BibitemOpen
  \bibfield  {author} {\bibinfo {author} {\bibfnamefont {Y.~H.}\ \bibnamefont
  {Wen}}, \bibinfo {author} {\bibfnamefont {Y.}~\bibnamefont {Lu}}, \bibinfo
  {author} {\bibfnamefont {K.~M.}\ \bibnamefont {Dobosz}},\ and\ \bibinfo
  {author} {\bibfnamefont {L.~A.}\ \bibnamefont {Archer}},\ }\bibfield  {title}
  {\bibinfo {title} {Structure, ion transport, and rheology of nanoparticle
  salts},\ }\href@noop {} {\bibfield  {journal} {\bibinfo  {journal}
  {Macromolecules}\ }\textbf {\bibinfo {volume} {47}},\ \bibinfo {pages} {4479}
  (\bibinfo {year} {2014})}\BibitemShut {NoStop}%
\bibitem [{\citenamefont {Nabizadeh}\ and\ \citenamefont
  {Jamali}(2021)}]{nabizadeh2021}%
  \BibitemOpen
  \bibfield  {author} {\bibinfo {author} {\bibfnamefont {M.}~\bibnamefont
  {Nabizadeh}}\ and\ \bibinfo {author} {\bibfnamefont {S.}~\bibnamefont
  {Jamali}},\ }\bibfield  {title} {\bibinfo {title} {Life and death of
  colloidal bonds control the rate-dependent rheology of gels},\ }\href@noop {}
  {\bibfield  {journal} {\bibinfo  {journal} {Nature Communications}\ }\textbf
  {\bibinfo {volume} {12}},\ \bibinfo {pages} {1} (\bibinfo {year}
  {2021})}\BibitemShut {NoStop}%
\bibitem [{\citenamefont {Keim}\ \emph {et~al.}(2019)\citenamefont {Keim},
  \citenamefont {Paulsen}, \citenamefont {Zeravcic}, \citenamefont {Sastry},\
  and\ \citenamefont {Nagel}}]{keim2019}%
  \BibitemOpen
  \bibfield  {author} {\bibinfo {author} {\bibfnamefont {N.~C.}\ \bibnamefont
  {Keim}}, \bibinfo {author} {\bibfnamefont {J.~D.}\ \bibnamefont {Paulsen}},
  \bibinfo {author} {\bibfnamefont {Z.}~\bibnamefont {Zeravcic}}, \bibinfo
  {author} {\bibfnamefont {S.}~\bibnamefont {Sastry}},\ and\ \bibinfo {author}
  {\bibfnamefont {S.~R.}\ \bibnamefont {Nagel}},\ }\bibfield  {title} {\bibinfo
  {title} {Memory formation in matter},\ }\href@noop {} {\bibfield  {journal}
  {\bibinfo  {journal} {Reviews of Modern Physics}\ }\textbf {\bibinfo {volume}
  {91}},\ \bibinfo {pages} {035002} (\bibinfo {year} {2019})}\BibitemShut
  {NoStop}%
\bibitem [{\citenamefont {Varga}\ and\ \citenamefont {Swan}(2018)}]{varga2018}%
  \BibitemOpen
  \bibfield  {author} {\bibinfo {author} {\bibfnamefont {Z.}~\bibnamefont
  {Varga}}\ and\ \bibinfo {author} {\bibfnamefont {J.~W.}\ \bibnamefont
  {Swan}},\ }\bibfield  {title} {\bibinfo {title} {Large scale anisotropies in
  sheared colloidal gels},\ }\href@noop {} {\bibfield  {journal} {\bibinfo
  {journal} {Journal of Rheology}\ }\textbf {\bibinfo {volume} {62}},\ \bibinfo
  {pages} {405} (\bibinfo {year} {2018})}\BibitemShut {NoStop}%
\bibitem [{\citenamefont {Jamali}\ \emph {et~al.}(2020)\citenamefont {Jamali},
  \citenamefont {Armstrong},\ and\ \citenamefont {McKinley}}]{Jamali2020}%
  \BibitemOpen
  \bibfield  {author} {\bibinfo {author} {\bibfnamefont {S.}~\bibnamefont
  {Jamali}}, \bibinfo {author} {\bibfnamefont {R.~C.}\ \bibnamefont
  {Armstrong}},\ and\ \bibinfo {author} {\bibfnamefont {G.~H.}\ \bibnamefont
  {McKinley}},\ }\bibfield  {title} {\bibinfo {title} {Time-rate-transformation
  framework for targeted assembly of short-range attractive colloidal
  suspensions},\ }\href {https://doi.org/10.1016/j.mtadv.2019.100026}
  {\bibfield  {journal} {\bibinfo  {journal} {Materials Today Advances}\
  }\textbf {\bibinfo {volume} {5}},\ \bibinfo {pages} {100026} (\bibinfo {year}
  {2020})}\BibitemShut {NoStop}%
\bibitem [{\citenamefont {Jamali}\ \emph {et~al.}(2019)\citenamefont {Jamali},
  \citenamefont {Armstrong},\ and\ \citenamefont {McKinley}}]{Jamali2019}%
  \BibitemOpen
  \bibfield  {author} {\bibinfo {author} {\bibfnamefont {S.}~\bibnamefont
  {Jamali}}, \bibinfo {author} {\bibfnamefont {R.~C.}\ \bibnamefont
  {Armstrong}},\ and\ \bibinfo {author} {\bibfnamefont {G.~H.}\ \bibnamefont
  {McKinley}},\ }\bibfield  {title} {\bibinfo {title} {Multiscale nature of
  thixotropy and rheological hysteresis in attractive colloidal suspensions
  under shear},\ }\href@noop {} {\bibfield  {journal} {\bibinfo  {journal}
  {Physical review letters}\ }\textbf {\bibinfo {volume} {123}},\ \bibinfo
  {pages} {248003} (\bibinfo {year} {2019})}\BibitemShut {NoStop}%
\bibitem [{\citenamefont {Eggersdorfer}\ \emph {et~al.}(2010)\citenamefont
  {Eggersdorfer}, \citenamefont {Kadau}, \citenamefont {Herrmann},\ and\
  \citenamefont {Pratsinis}}]{Eggersdorfer2010}%
  \BibitemOpen
  \bibfield  {author} {\bibinfo {author} {\bibfnamefont {M.~L.}\ \bibnamefont
  {Eggersdorfer}}, \bibinfo {author} {\bibfnamefont {D.}~\bibnamefont {Kadau}},
  \bibinfo {author} {\bibfnamefont {H.~J.}\ \bibnamefont {Herrmann}},\ and\
  \bibinfo {author} {\bibfnamefont {S.~E.}\ \bibnamefont {Pratsinis}},\
  }\bibfield  {title} {\bibinfo {title} {{Fragmentation and restructuring of
  soft-agglomerates under shear}},\ }\href
  {https://doi.org/10.1016/j.jcis.2009.10.062} {\bibfield  {journal} {\bibinfo
  {journal} {Journal of Colloid and Interface Science}\ }\textbf {\bibinfo
  {volume} {342}},\ \bibinfo {pages} {261} (\bibinfo {year}
  {2010})}\BibitemShut {NoStop}%
\bibitem [{\citenamefont {Marshall}\ and\ \citenamefont
  {Li}(2014)}]{Marshall2014}%
  \BibitemOpen
  \bibfield  {author} {\bibinfo {author} {\bibfnamefont {J.~S.}\ \bibnamefont
  {Marshall}}\ and\ \bibinfo {author} {\bibfnamefont {S.}~\bibnamefont {Li}},\
  }\href {https://doi.org/10.1017/CBO9781139424547} {\emph {\bibinfo {title}
  {{Adhesive Particle Flow: A Discrete-Element Approach}}}},\ \bibinfo
  {edition} {1st}\ ed.,\ edited by\ \bibinfo {editor} {\bibfnamefont {J.~S.}\
  \bibnamefont {Marshall}}\ and\ \bibinfo {editor} {\bibfnamefont
  {S.}~\bibnamefont {Li}}\ (\bibinfo  {publisher} {Cambridge University
  Press},\ \bibinfo {address} {32 Avenue of the Americas, NY 10013-2473, USA},\
  \bibinfo {year} {2014})\BibitemShut {NoStop}%
\bibitem [{\citenamefont {Kimbonguila~Manounou}\ and\ \citenamefont
  {R\'emond}(2014)}]{Kimbonguila2014}%
  \BibitemOpen
  \bibfield  {author} {\bibinfo {author} {\bibfnamefont {A.}~\bibnamefont
  {Kimbonguila~Manounou}}\ and\ \bibinfo {author} {\bibfnamefont
  {S.}~\bibnamefont {R\'emond}},\ }\bibfield  {title} {\bibinfo {title}
  {{Discrete element modeling of the microstructure of fine particle
  agglomerates in sheared dilute suspension}},\ }\href
  {https://doi.org/10.1016/j.physa.2014.06.023} {\bibfield  {journal} {\bibinfo
   {journal} {Physica A}\ }\textbf {\bibinfo {volume} {412}},\ \bibinfo {pages}
  {66} (\bibinfo {year} {2014})}\BibitemShut {NoStop}%
\bibitem [{\citenamefont {Banasiak}\ \emph {et~al.}(2020)\citenamefont
  {Banasiak}, \citenamefont {Lamb},\ and\ \citenamefont
  {Lauren\c{c}ot}}]{Banasiak2020a}%
  \BibitemOpen
  \bibfield  {author} {\bibinfo {author} {\bibfnamefont {J.}~\bibnamefont
  {Banasiak}}, \bibinfo {author} {\bibfnamefont {W.}~\bibnamefont {Lamb}},\
  and\ \bibinfo {author} {\bibfnamefont {P.}~\bibnamefont {Lauren\c{c}ot}},\
  }\href@noop {} {\emph {\bibinfo {title} {{Analytic Methods for
  Coagulation-Fragmentation Models, Volume I}}}},\ edited by\ \bibinfo {editor}
  {\bibfnamefont {J.}~\bibnamefont {Banasiak}}, \bibinfo {editor}
  {\bibfnamefont {W.}~\bibnamefont {Lamb}},\ and\ \bibinfo {editor}
  {\bibfnamefont {P.}~\bibnamefont {Lauren\c{c}ot}},\ \bibinfo {series}
  {Monographs and Research Notes in Mathematics}, Vol.~\bibinfo {volume} {1}\
  (\bibinfo  {publisher} {CRC Press},\ \bibinfo {address} {CRC PressTaylor \&
  Francis Group6000 Broken Sound Parkway NW, Suite 300Boca Raton, FL
  33487-2742},\ \bibinfo {year} {2020})\BibitemShut {NoStop}%
\bibitem [{\citenamefont {Stadnichuk}\ \emph {et~al.}(2015)\citenamefont
  {Stadnichuk}, \citenamefont {Bodrova},\ and\ \citenamefont
  {Brilliantov}}]{Stadnichuk2015}%
  \BibitemOpen
  \bibfield  {author} {\bibinfo {author} {\bibfnamefont {V.}~\bibnamefont
  {Stadnichuk}}, \bibinfo {author} {\bibfnamefont {A.}~\bibnamefont
  {Bodrova}},\ and\ \bibinfo {author} {\bibfnamefont {N.}~\bibnamefont
  {Brilliantov}},\ }\bibfield  {title} {\bibinfo {title} {{Smoluchowski
  aggregation fragmentation equations: Fast numerical method to find
  steady-state solutions}},\ }\href {https://doi.org/10.1142/S0217979215502082}
  {\bibfield  {journal} {\bibinfo  {journal} {International Journal of Modern
  Physics B}\ }\textbf {\bibinfo {volume} {29}},\ \bibinfo {pages} {1}
  (\bibinfo {year} {2015})}\BibitemShut {NoStop}%
\bibitem [{\citenamefont {Sorensen}\ \emph {et~al.}(1987)\citenamefont
  {Sorensen}, \citenamefont {Zhang},\ and\ \citenamefont
  {Taylor}}]{Sorensen1987}%
  \BibitemOpen
  \bibfield  {author} {\bibinfo {author} {\bibfnamefont {C.~M.}\ \bibnamefont
  {Sorensen}}, \bibinfo {author} {\bibfnamefont {H.~X.}\ \bibnamefont
  {Zhang}},\ and\ \bibinfo {author} {\bibfnamefont {T.~W.}\ \bibnamefont
  {Taylor}},\ }\bibfield  {title} {\bibinfo {title} {Cluster-size evolution in
  a coagulation-fragmentation system},\ }\href
  {https://doi.org/10.1103/PhysRevLett.59.363} {\bibfield  {journal} {\bibinfo
  {journal} {Physical Review Letters}\ }\textbf {\bibinfo {volume} {59}},\
  \bibinfo {pages} {363} (\bibinfo {year} {1987})}\BibitemShut {NoStop}%
\bibitem [{\citenamefont {Gibaud}\ \emph {et~al.}(2010)\citenamefont {Gibaud},
  \citenamefont {Frelat},\ and\ \citenamefont {Manneville}}]{gibaud2010}%
  \BibitemOpen
  \bibfield  {author} {\bibinfo {author} {\bibfnamefont {T.}~\bibnamefont
  {Gibaud}}, \bibinfo {author} {\bibfnamefont {D.}~\bibnamefont {Frelat}},\
  and\ \bibinfo {author} {\bibfnamefont {S.}~\bibnamefont {Manneville}},\
  }\bibfield  {title} {\bibinfo {title} {Heterogeneous yielding dynamics in a
  colloidal gel},\ }\href@noop {} {\bibfield  {journal} {\bibinfo  {journal}
  {Soft Matter}\ }\textbf {\bibinfo {volume} {6}},\ \bibinfo {pages} {3482}
  (\bibinfo {year} {2010})}\BibitemShut {NoStop}%
\bibitem [{\citenamefont {Sprakel}\ \emph {et~al.}(2011)\citenamefont
  {Sprakel}, \citenamefont {Lindstr{\"o}m}, \citenamefont {Kodger},\ and\
  \citenamefont {Weitz}}]{sprakel2011}%
  \BibitemOpen
  \bibfield  {author} {\bibinfo {author} {\bibfnamefont {J.}~\bibnamefont
  {Sprakel}}, \bibinfo {author} {\bibfnamefont {S.~B.}\ \bibnamefont
  {Lindstr{\"o}m}}, \bibinfo {author} {\bibfnamefont {T.~E.}\ \bibnamefont
  {Kodger}},\ and\ \bibinfo {author} {\bibfnamefont {D.~A.}\ \bibnamefont
  {Weitz}},\ }\bibfield  {title} {\bibinfo {title} {Stress enhancement in the
  delayed yielding of colloidal gels},\ }\href@noop {} {\bibfield  {journal}
  {\bibinfo  {journal} {Physical review letters}\ }\textbf {\bibinfo {volume}
  {106}},\ \bibinfo {pages} {248303} (\bibinfo {year} {2011})}\BibitemShut
  {NoStop}%
\bibitem [{\citenamefont {Grenard}\ \emph {et~al.}(2014)\citenamefont
  {Grenard}, \citenamefont {Divoux}, \citenamefont {Taberlet},\ and\
  \citenamefont {Manneville}}]{grenard2014}%
  \BibitemOpen
  \bibfield  {author} {\bibinfo {author} {\bibfnamefont {V.}~\bibnamefont
  {Grenard}}, \bibinfo {author} {\bibfnamefont {T.}~\bibnamefont {Divoux}},
  \bibinfo {author} {\bibfnamefont {N.}~\bibnamefont {Taberlet}},\ and\
  \bibinfo {author} {\bibfnamefont {S.}~\bibnamefont {Manneville}},\ }\bibfield
   {title} {\bibinfo {title} {Timescales in creep and yielding of attractive
  gels},\ }\href@noop {} {\bibfield  {journal} {\bibinfo  {journal} {Soft
  matter}\ }\textbf {\bibinfo {volume} {10}},\ \bibinfo {pages} {1555}
  (\bibinfo {year} {2014})}\BibitemShut {NoStop}%
\bibitem [{\citenamefont {Teixeira}(1988)}]{teixeira1988}%
  \BibitemOpen
  \bibfield  {author} {\bibinfo {author} {\bibfnamefont {J.}~\bibnamefont
  {Teixeira}},\ }\bibfield  {title} {\bibinfo {title} {Small-angle scattering
  by fractal systems},\ }\href@noop {} {\bibfield  {journal} {\bibinfo
  {journal} {Journal of Applied Crystallography}\ }\textbf {\bibinfo {volume}
  {21}},\ \bibinfo {pages} {781} (\bibinfo {year} {1988})}\BibitemShut
  {NoStop}%
\bibitem [{\citenamefont {Schiessel}\ \emph {et~al.}(1995)\citenamefont
  {Schiessel}, \citenamefont {Metzler}, \citenamefont {Blumen},\ and\
  \citenamefont {Nonnenmacher}}]{schiessel1995}%
  \BibitemOpen
  \bibfield  {author} {\bibinfo {author} {\bibfnamefont {H.}~\bibnamefont
  {Schiessel}}, \bibinfo {author} {\bibfnamefont {R.}~\bibnamefont {Metzler}},
  \bibinfo {author} {\bibfnamefont {A.}~\bibnamefont {Blumen}},\ and\ \bibinfo
  {author} {\bibfnamefont {T.}~\bibnamefont {Nonnenmacher}},\ }\bibfield
  {title} {\bibinfo {title} {Generalized viscoelastic models: their fractional
  equations with solutions},\ }\href@noop {} {\bibfield  {journal} {\bibinfo
  {journal} {Journal of physics A: Mathematical and General}\ }\textbf
  {\bibinfo {volume} {28}},\ \bibinfo {pages} {6567} (\bibinfo {year}
  {1995})}\BibitemShut {NoStop}%
\bibitem [{\citenamefont {Keshavarz}\ \emph {et~al.}(2021)\citenamefont
  {Keshavarz}, \citenamefont {Rodrigues}, \citenamefont {Champenois},
  \citenamefont {Frith}, \citenamefont {Ilavsky}, \citenamefont {Geri},
  \citenamefont {Divoux}, \citenamefont {McKinley},\ and\ \citenamefont
  {Poulesquen}}]{keshavarz2021}%
  \BibitemOpen
  \bibfield  {author} {\bibinfo {author} {\bibfnamefont {B.}~\bibnamefont
  {Keshavarz}}, \bibinfo {author} {\bibfnamefont {D.~G.}\ \bibnamefont
  {Rodrigues}}, \bibinfo {author} {\bibfnamefont {J.-B.}\ \bibnamefont
  {Champenois}}, \bibinfo {author} {\bibfnamefont {M.~G.}\ \bibnamefont
  {Frith}}, \bibinfo {author} {\bibfnamefont {J.}~\bibnamefont {Ilavsky}},
  \bibinfo {author} {\bibfnamefont {M.}~\bibnamefont {Geri}}, \bibinfo {author}
  {\bibfnamefont {T.}~\bibnamefont {Divoux}}, \bibinfo {author} {\bibfnamefont
  {G.~H.}\ \bibnamefont {McKinley}},\ and\ \bibinfo {author} {\bibfnamefont
  {A.}~\bibnamefont {Poulesquen}},\ }\bibfield  {title} {\bibinfo {title}
  {Time--connectivity superposition and the gel/glass duality of weak colloidal
  gels},\ }\href@noop {} {\bibfield  {journal} {\bibinfo  {journal}
  {Proceedings of the National Academy of Sciences}\ }\textbf {\bibinfo
  {volume} {118}} (\bibinfo {year} {2021})}\BibitemShut {NoStop}%
\bibitem [{\citenamefont {Polyanin}\ and\ \citenamefont
  {Manzhirov}(2007)}]{Polyanin2007}%
  \BibitemOpen
  \bibfield  {author} {\bibinfo {author} {\bibfnamefont {A.~D.}\ \bibnamefont
  {Polyanin}}\ and\ \bibinfo {author} {\bibfnamefont {A.~V.}\ \bibnamefont
  {Manzhirov}},\ }\href@noop {} {\emph {\bibinfo {title} {{Handbook of
  Mathematics for Engineers and Scientists}}}},\ edited by\ \bibinfo {editor}
  {\bibfnamefont {A.~D.}\ \bibnamefont {Polyanin}}\ and\ \bibinfo {editor}
  {\bibfnamefont {A.~V.}\ \bibnamefont {Manzhirov}}\ (\bibinfo  {publisher}
  {Chapman \& Hall},\ \bibinfo {address} {Chapman \& Hall/CRCTaylor \& Francis
  Group6000 Broken Sound Parkway NW, Suite 300Boca Raton, FL 33487‑2742},\
  \bibinfo {year} {2007})\BibitemShut {NoStop}%
\bibitem [{\citenamefont {Kern}\ and\ \citenamefont {Bland}(1967)}]{Kern1967}%
  \BibitemOpen
  \bibfield  {author} {\bibinfo {author} {\bibnamefont {Kern}}\ and\ \bibinfo
  {author} {\bibnamefont {Bland}},\ }\href@noop {} {\emph {\bibinfo {title}
  {{Solid Mensuration}}}},\ \bibinfo {edition} {2nd}\ ed.,\ edited by\ \bibinfo
  {editor} {\bibnamefont {Kern}}\ and\ \bibinfo {editor} {\bibnamefont
  {Bland}}\ (\bibinfo  {publisher} {John Wiley \& Sons},\ \bibinfo {year}
  {1967})\BibitemShut {NoStop}%
\bibitem [{\citenamefont {Dag{\`e}s}\ \emph {et~al.}(2022)\citenamefont
  {Dag{\`e}s}, \citenamefont {Bouthier}, \citenamefont {Matthews},
  \citenamefont {Manneville}, \citenamefont {Divoux}, \citenamefont
  {Poulesquen},\ and\ \citenamefont {Gibaud}}]{dages2022}%
  \BibitemOpen
  \bibfield  {author} {\bibinfo {author} {\bibfnamefont {N.}~\bibnamefont
  {Dag{\`e}s}}, \bibinfo {author} {\bibfnamefont {L.~V.}\ \bibnamefont
  {Bouthier}}, \bibinfo {author} {\bibfnamefont {L.}~\bibnamefont {Matthews}},
  \bibinfo {author} {\bibfnamefont {S.}~\bibnamefont {Manneville}}, \bibinfo
  {author} {\bibfnamefont {T.}~\bibnamefont {Divoux}}, \bibinfo {author}
  {\bibfnamefont {A.}~\bibnamefont {Poulesquen}},\ and\ \bibinfo {author}
  {\bibfnamefont {T.}~\bibnamefont {Gibaud}},\ }\bibfield  {title} {\bibinfo
  {title} {Interpenetration of fractal clusters drives elasticity in colloidal
  gels formed upon flow cessation},\ }\href@noop {} {\bibfield  {journal}
  {\bibinfo  {journal} {Soft Matter}\ }\textbf {\bibinfo {volume} {18}},\
  \bibinfo {pages} {6645} (\bibinfo {year} {2022})}\BibitemShut {NoStop}%
\end{thebibliography}

%

\end{document}